\documentclass[%
 reprint,
superscriptaddress,
 amsmath,amssymb,
 aps,
]{revtex4-2}

\usepackage{graphicx}
\usepackage{multirow}
\usepackage{array}

\newcommand{\degree}{$^{\circ}$}

\begin{document}

\title{Development of a prototype superconducting radio-frequency cavity for conduction-cooled accelerators}

\author{G. Ciovati}
 \email{gciovati@jlab.org}
 \affiliation{Thomas Jefferson National Accelerator Facility, Newport News, VA 23606, USA}
 \affiliation{Center for Accelerator Science, Department of Physics, Old Dominion University, \\Norfolk, Virginia 23529, USA}

\author{J. Anderson}
 \affiliation{General Atomics, San Diego, CA 92121, USA}
\author{S. Balachandran}
 \affiliation{Thomas Jefferson National Accelerator Facility, Newport News, VA 23606, USA}
\author{G. Cheng}
 \affiliation{Thomas Jefferson National Accelerator Facility, Newport News, VA 23606, USA}

\author{B. Coriton}
 \altaffiliation[Current address:  ]{ITER, 13067 St. Paul-lez-Durance, France}
 \affiliation{General Atomics, San Diego, CA 92121, USA}

\author{E. Daly}
 \affiliation{Thomas Jefferson National Accelerator Facility, Newport News, VA 23606, USA}
\author{P. Dhakal}
 \affiliation{Thomas Jefferson National Accelerator Facility, Newport News, VA 23606, USA}
\author{A. Gurevich}
 \affiliation{Center for Accelerator Science, Department of Physics, Old Dominion University, \\Norfolk, Virginia 23529, USA}

\author{F. Hannon}
 \altaffiliation[Current address:  ]{Phasespace Tech, 23734 Bj\"{a}rred, Sweden}
 \affiliation{Thomas Jefferson National Accelerator Facility, Newport News, VA 23606, USA}

\author{K. Harding}
 \affiliation{Thomas Jefferson National Accelerator Facility, Newport News, VA 23606, USA}
\author{L. Holland}
 \affiliation{General Atomics, San Diego, CA 92121, USA}

\author{F. Marhauser}
 \altaffiliation[Current address:  ]{SCK CEN,  2400 Mol, Belgium}
 \affiliation{Thomas Jefferson National Accelerator Facility, Newport News, VA 23606, USA}
 
\author{K. McLaughlin}
 \affiliation{General Atomics, San Diego, CA 92121, USA}
\author{D. Packard}
 \affiliation{General Atomics, San Diego, CA 92121, USA}
\author{T. Powers}
 \affiliation{Thomas Jefferson National Accelerator Facility, Newport News, VA 23606, USA}
\author{U. Pudasaini}
  \affiliation{Thomas Jefferson National Accelerator Facility, Newport News, VA 23606, USA}
\author{J. Rathke}
  \affiliation{TECHSOURCE, Inc., Los Alamos, NM 87544, USA}
\author{R. Rimmer}
  \affiliation{Thomas Jefferson National Accelerator Facility, Newport News, VA 23606, USA}
\author{T. Schultheiss}
  \affiliation{TJS Technologies LLC, Commack, NY 11725, USA}
\author{H. Vennekate}
   \affiliation{Thomas Jefferson National Accelerator Facility, Newport News, VA 23606, USA}
\author{D. Vollmer}
 \affiliation{General Atomics, San Diego, CA 92121, USA}

\date{3-22-2023}

\begin{abstract}

The higher efficiency of superconducting radio-frequency (SRF) cavities compared to normal-conducting ones enables the development of high-energy continuous-wave linear accelerators (linacs). Recent progress in the development of high-quality Nb$_3$Sn film coatings along with the availability of cryocoolers with high cooling capacity at 4~K makes it feasible to operate SRF cavities cooled by thermal conduction at relevant accelerating gradients for use in accelerators. A possible use of conduction-cooled SRF linacs is for environmental applications, requiring electron beams with energy of $1 - 10$~MeV and 1~MW of power. We have designed a 915~MHz SRF linac for such an application and developed a prototype single-cell cavity to prove the proposed design by operating it with cryocoolers at the accelerating gradient required for 1~MeV energy gain. The cavity has a $\sim 3$~$\mu$m thick Nb$_3$Sn film on the inner surface, deposited on a $\sim4$~mm thick bulk Nb substrate and a bulk $\sim7$~mm thick Cu outer shell with three Cu attachment tabs. The cavity was tested up to a peak surface magnetic field of 53~mT in liquid He at 4.3~K. A horizontal test cryostat was designed and built to test the cavity cooled with three Gifford-McMahon cryocoolers. The rf tests of the conduction-cooled cavity, performed at General Atomics, achieved a peak surface magnetic field of 50~mT and stable operation was possible with up to 18.5~W of rf heat load. The peak frequency shift due to microphonics was 23~Hz. These results represent the highest peak surface magnetic field achieved in a conduction-cooled SRF cavity to date and meet the requirements for a 1~MeV energy gain.
\end{abstract}

\maketitle


\section{\label{intro}Introduction}
The superconducting radio-frequency (SRF) technology is widely used in modern particle accelerator research facilities because of its higher efficiency compared to normal-conducting rf technology \cite{SRF1, SRF2}. SRF accelerators require liquid helium (LHe) cryoplants, often subcooled to 2~K, to cool the cavities well below the superconducting transition temperature. Whereas niobium has been the material of choice for SRF cavities, Nb$_3$Sn has emerged as a viable alternative in recent years \cite{Posen_2017}. Since the critical temperature of Nb$_3$Sn is nearly twice that of Nb, similar rf dissipation as that of Nb can be achieved at 4.3~K, instead of 2~K, at fields much smaller than the thermodynamic critical field, $B_c$, which reduces significantly the cost and complexity of the LHe cryoplant. The cooling power at 4~K from commercial closed-cycle refrigerators (CCR), also known as cryocoolers, has also been steadily increasing in the last few years, making it feasible to design SRF accelerators in which the cavities are cooled by thermal conduction. Proof of principle demonstration of conduction-cooled SRF cavities have been achieved in single-cell cavities at 650~MHz~\cite{Dhuley_2020}, 1.5~GHz~\cite{Ciovati_SUST} and 2.6~GHz~\cite{Stilin}, up to a peak surface magnetic field $B_p \sim 42$~mT, with different types of cryocoolers and methods to connect the cryocooler's cold stage (stage 2) to the cavity.

Conduction-cooling greatly simplifies the overall design and cost of an SRF accelerator, making the SRF technology feasible for industrial accelerators.
One among several possible applications of SRF industrial accelerators is environmental remediation, which requires continuous-wave electron beams with energies in the range 1 - 10~MeV and 1~MW beam power~\cite{Envacc_Report}. Conceptual designs for MW-class conduction-cooled SRF accelerators at 1~MeV and 10~MeV have been published~\cite{EnvAcc, Dhuley_PRAB}.
Given the high beam power requirement, the efficiency of the overall accelerator is dominated by the efficiency of the high-power rf source, particularly at low beam energy. 915~MHz magnetrons for industrial heating are the most efficient and cost effective commercial high-power rf sources, with power levels up to $\sim 100$~kW, making them an ideal choice for high-power industrial accelerators~\cite{Magnetrons}. R$\&$D efforts are ongoing to develop efficient power-combining schemes as well as methods to achieve the amplitude and phase control required to drive SRF cavities~\cite{Magn1, Magn2, Magn3, Magn4, Magn5, Magn6, Magn7, Magn8}.
In this contribution we present the conceptual design of a conduction-cooled 915~MHz, 1~MeV, 1~MW cw SRF electron linac along with the development of a prototype cavity aiming at demonstrating achieving the $B_p$-value which corresponds to the accelerating gradient required for a 1~MeV energy gain. The article is organized as follows: Sec.~\ref{sec:accel} presents the conceptual design of the 915~MHz accelerator, focused on the beam transport simulations and the SRF cavity design, Sec.~\ref{sec:protcav} presents the development of a multi-metallic 952.6~MHz single-cell prototype cavity and its performance in LHe at 4.3~K. Section~\ref{sec:tstrap} presents the results from cryogenic and mechanical tests of the Cu thermal strap used for the connection between the cavity and the cryocooler, Sec.~\ref{sec:htb} presents the design of a horizontal test cryostat (HTC) for the prototype cavity with three cryocoolers, Sec.~\ref{sec:cavHTC} presents the results from the rf tests of the prototype cavity cooled by conduction inside the HTC, as well as cavity microphonics and vibration measurements. The analysis and discussion of the experimental results are presented in Sec.~\ref{sec:disc} and a summary is given in Sec.~\ref{sec:conc}.

\section{\label{sec:accel}Design of a 915~MH\lowercase{z}, 1~M\lowercase{e}V, 1~MW cw electron linac}
The conceptual design for a 750~MHz, 1~MeV, 1~MW, cw, conduction-cooled SRF electron linear accelerator (linac) for environmental remediation was presented in Ref.~[\onlinecite{EnvAcc}]. Different options for the high-power rf source were discussed and it was estimated that $\sim60\%$ of the operational cost was due to electric power consumption, if either a klystron or multibeam inductive output tube were used. Magnetrons are the most efficient high-power rf sources and R$\&$D efforts are ongoing to make them capable of the amplitude and phase regulation required to drive SRF cavities as well as efficiently combine the power from multiple magnetrons up to 1~MW. Choosing the accelerator frequency to match that of high-power industrial magnetrons lowers the capital cost of the high-power rf source, compared to a frequency for which only a few devices are commercially available.

\subsection{\label{subsec:bt}Beam transport simulations}
\begin{figure}[htb]
\includegraphics[width=86mm]{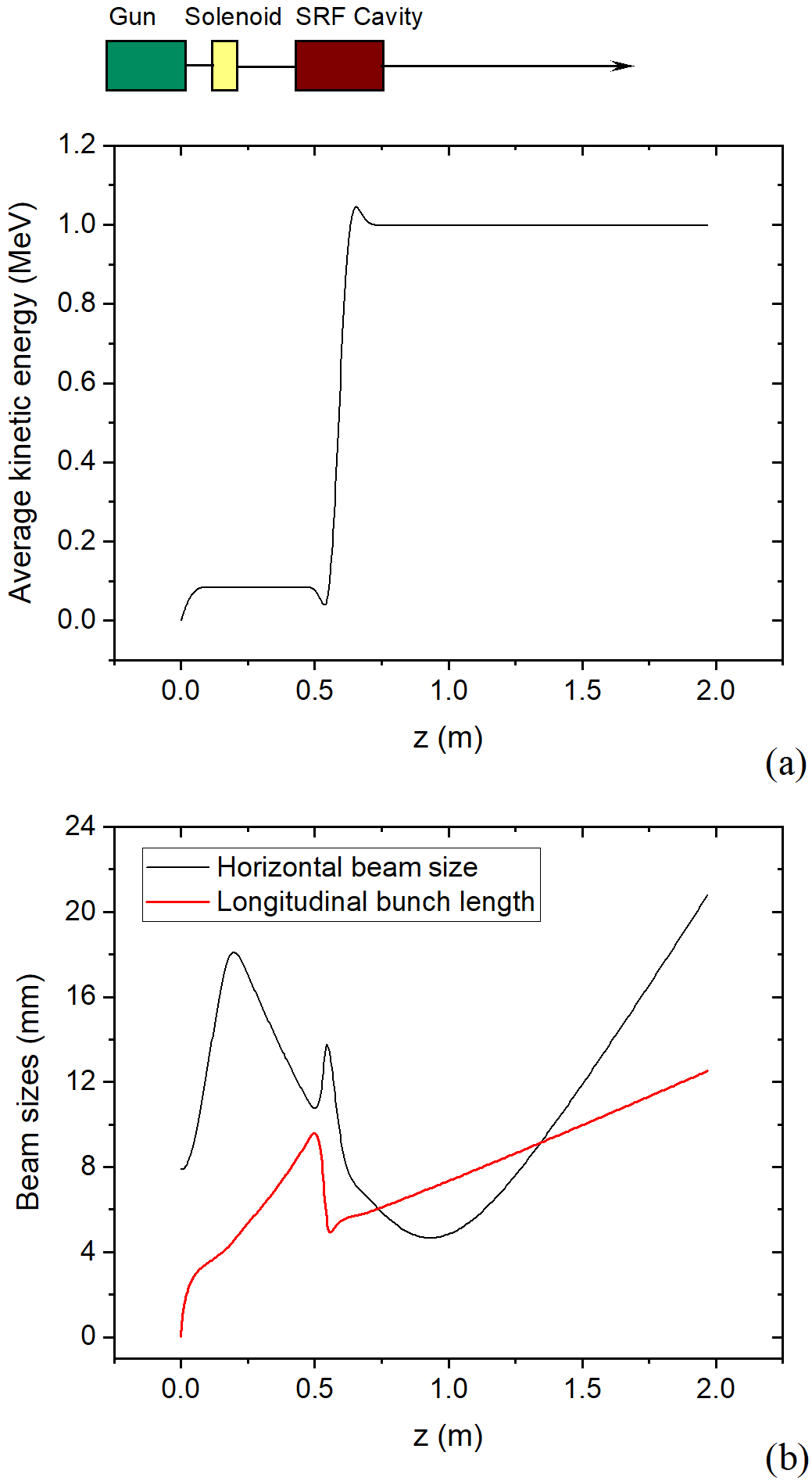}
\caption{\label{fig:bt} Average beam energy (a), transverse rms beam size and longitudinal rms bunch length (b) along the beam axis for a 915~MHz, 1~MeV, 1~A linac obtained from a simulation with GPT. The schematic layout on top is not to scale.}
\end{figure}

The accelerator design described in Ref.~[\onlinecite{EnvAcc}] has been re-evaluated at 915~MHz and it consists of the same main components: an rf gridded thermionic gun, a focusing solenoid, a cryomodule with a conduction-cooled single-cell SRF cavity, another focusing solenoid, raster magnets and a beam-exit window.
The beam transport simulations were carried out with with the particle tracking software General Particle Tracer (GPT)~\cite{GPT} using a simplified geometry which included the gun, the focusing solenoid and the SRF cavity. The cathode emitting area was assumed circular and planar, with a radius of 16~mm. The center positions of the solenoid and the cavity are 18~cm and 60~cm from the cathode, respectively. The main design parameters are listed in Table~\ref{table1}, resulting in a final kinetic energy of 1~MeV, with a root-mean-square (rms) energy spread of 100~keV at the exit of the cavity. The transverse rms beam radius is less than 14~mm inside the cavity, whereas the longitudinal rms bunch length is less than 10~mm, as shown in Fig.~\ref{fig:bt}. None of the electrons in the simulation hit the cavity inner surface.

\begin{table}
\caption{\label{table1}
Main design parameters for the 915~MHz, high-power cw electron linac and the SRF cavity.}
\begin{ruledtabular}
\begin{tabular}{cc}
\multicolumn{2}{c}{Accelerator parameters}\\
\hline
Gun voltage (kV) & -85\\
Solenoid peak magnetic field on axis (mT)& 18\\
SRF cavity off-crest phase & 0\degree \\
SRF cavity peak electric field on axis (MV/m) & 16\\
Beam energy (MeV)& 1.0\\
Beam current (A)& 1.0\\
\hline
\multicolumn{2}{c}{SRF cavity geometry}\\
\hline
Equator diameter (mm) & 145.2\\
Equator ellipse major axis (mm)& 58.3\\
Equator ellipse minor axis (mm)& 34.3\\
Wall angle (\degree)& 5\\
Radius of iris circle (mm) & 4.9\\
Iris radius (mm)& 43\\
Cell length (mm)& 86\\
\hline
\multicolumn{2}{c}{SRF cavity electromagnetic parameters}\\
\hline
$E_p/E_{acc}$ & 2.61 \\
$B_p/E_{acc}$ [mT/(MV/m)]& 3.90 \\
$G$ ($\Omega$) & 176.3\\
$R/Q$ ($\Omega$)& 26.2\\
\end{tabular}
\end{ruledtabular}
\end{table}

\subsection{\label{subsec:cavdes}Cavity design}

According to the GPT simulation, the electric field on-axis which results in a beam energy of 1~MeV can be achieved with a 915~MHz single-cell cavity with an iris-to-iris cell length of 86~mm, which corresponds to a geometric $\beta$ of 0.53, where $\beta$ is the speed of the electron relative to the speed of light. The cell shape was designed as a compromise between minimizing the peak surface fields, maximizing the cavity aperture, ease of surface treatments and mechanical stability. The parameterization of the cell shape followed the one described in Ref.~[\onlinecite{buildcav}] and the final value of the geometric parameters are listed in Table~\ref{table1}. The diameter of the beam tube was enlarged to 148~mm on one side of the cavity to facilitate the propagation of higher order modes (HOMs). The main electromagnetic parameters obtained from a simulation with the electromagnetic field solver Superfish~\cite{SF} are listed in Table~\ref{table1}. $E_p$ and $B_p$ are the peak surface electric and magnetic fields, respectively, $E_{acc}$ is the effective accelerating field, $G$ is the geometry factor and $R/Q$ is linac definition of the characteristic the shunt impedance. The value of the operational peak electric field on axis listed in Table~\ref{table1} corresponds to $E_{acc}=10.6$~MV/m, $B_p=41.5$~mT and $E_p = 27.8$~MV/m.
Multipacting (MP) simulations were carried out with the code FishPact~\cite{donoghue:srf05} and no stable electron trajectories with a final impact energy $> 25$~eV were found below $E_{acc}=12.2$~MV/m, which is above the operational $E_{acc}$-value.



Two identical coaxial fundamental power couplers (FPCs) are mounted to side ports, 180\degree apart, on the small-diameter beam tube of the cavity. Each FPC needs to provide 457~kW of rf power, which corresponds to an external-$Q$ value of $Q_{ext}=2.7 \times 10^4$ at the target $E_{acc}$. The relatively low $Q_{ext}$-value is due to the high beam loading and results in a bandwidth of the cavity resonance of $\sim34$~kHz. This relatively large bandwidth makes the cavity less susceptible to detuning due to microphonics. The distance to the cavity iris and the length of the inner conductor were set to achieve this $Q_{ext}$. The tip of the FPC's center conductor has a ``pringle" shape to maximize the coupling to the electric field while minimizing the penetration ($\sim 1$~mm) inside the beam tube. The FPC design was scaled to 915~MHz from a well established one for a high-current SRF injector~\cite{FPC}.

HOM analyses were carried out using the three-dimensional (3D) electromagnetic software package CST Studio Suite~\cite{CST} following the same methodology described in Refs.~[\onlinecite{EnvAcc, marhauser:pac09}]. Both wakefield and lossy Eigenmode simulations were carried out, in good agreement to each other, to assess the most parasitic beam-induced longitudinal and transverse HOM impedances. The simulations confirmed, by utilizing beamline absorbers placed outside the cryomodule, that the dipole HOM impedances are kept well below the corresponding threshold impedance of a single-pass beam breakup instability, while the dissipated power levels in the absorber, dominated by longitudinal HOMs, can be well handled by a conventional absorber.

\subsection{\label{subsec:thermal}Cavity thermal analysis}
The cavity is intended to be built from 4~mm thick high-purity Nb sheets. A few micrometers-thick Nb$_3$Sn thin film would be formed on the cavity inner surface by the vapor diffusion method~\cite{Posen_2017, Uttar_JVSTA}. The cavity flanges are made of bulk Nb and vacuum sealing with mating flanges is done by In wire.

The integrated cavity and cryomodule design described in Ref.~[\onlinecite{EnvAcc}] used a thick high-purity Cu layer deposited on the cavity outer surface to improve the thermal stabilization of the system. An experimental proof of concept was achieved in a 1.5~GHz single-cell cavity in which a high-purity Cu layer of thickness $\geq 5$~mm as well as an equatorial bolt-on ring were grown on the outer surface of a Nb/Nb$_3$Sn cavity by electroplating, with a thin interface layer deposited by cold-spray~\cite{Ciovati_SUST}. Growing the structures to connect the cavity to the cryocoolers' 2$^{\text{nd}}$ stage by electroplating can be very time consuming, therefore we propose a different approach consisting of a two-steps electroplating. In the first step, a Cu layer $\sim2$~mm thick is electroplated onto the thin cold-sprayed Cu layer. The layer is machined on a lathe to create attachment posts. An equator ring and a beam-tube plate machined from oxygen-free high-conductivity (OFHC) Cu (C10100 alloy) are mounted to the attachment posts on the cavity and electroplated in place during the second electroplating step, adding $\sim 5$~mm of Cu. A final lathe machining of the Cu structure is needed to achieve the final dimensions and surface finish of the cavity outer layer. The thickness of the equator ring and beam-tube plate tapers down towards the attachment posts, allowing the electroplated Cu to build-up in those areas. 
The thermal conductivity of the electroplated Cu was measured on samples and corresponds to that of Cu with a residual resistivity ratio of $\sim 300$~\cite{Ciovati_SRF21}.

\begin{figure}[htb]
\includegraphics[width=86mm]{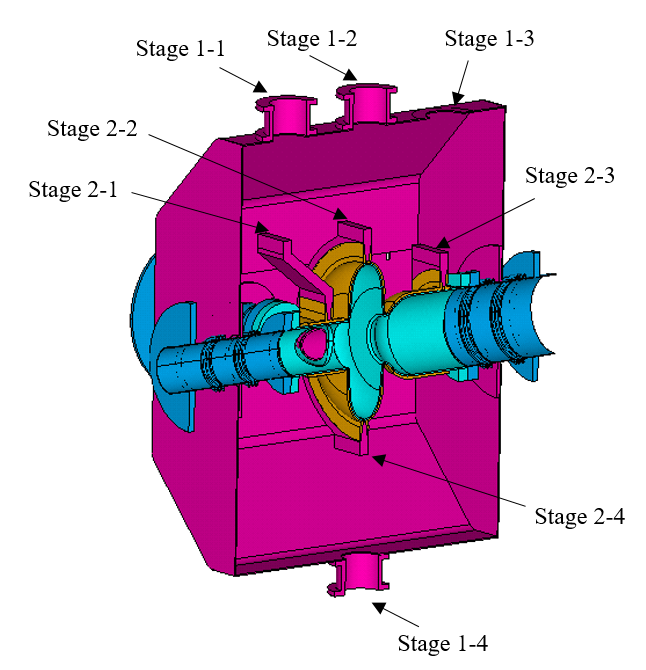}
\caption{\label{fig:3Dlayout} 3D model used for the thermal analysis. Each color represents a different material: cyan indicates Nb, blue indicates stainless steel, orange represents electroplated Cu, and fuchsia represents Cu with RRR=100. The nomenclature “Stage 1-3” means stage 1 of cryocooler number 3.}
\end{figure}

Figure~\ref{fig:3Dlayout} shows a 3D layout used for the thermal analysis of the cavity, FPCs, thermal shield and four cryocoolers with the finite element code Ansys~\cite{ansys}. The temperature-dependent thermal conductivities of stainless steel and niobium used in the thermal analysis were the same as those used in Ref.~[\onlinecite{EnvAcc}]. The thermal conductivity of Cu with $RRR=100$ from Ref.~[\onlinecite{Cu_NIST}] was used for OFHC Cu. The thermal conductivity of the electroplated Cu was taken from Ref.~[\onlinecite{Ciovati_SRF21}]. The Nb$_3$Sn thin film was not included in the model for simplification. Considering the thermal conductivity of Nb$_3$Sn from Ref.~[\onlinecite{Nb3Sn_thermcond}], the thermal conductance per unit area of 2~$\mu$m thick Nb$_3$Sn at 4~K is comparable to that of 4~mm thick Nb with residual resistivity ratio ($RRR$) $\sim 300$.

\begin{figure}[htb]
\includegraphics[width=86mm]{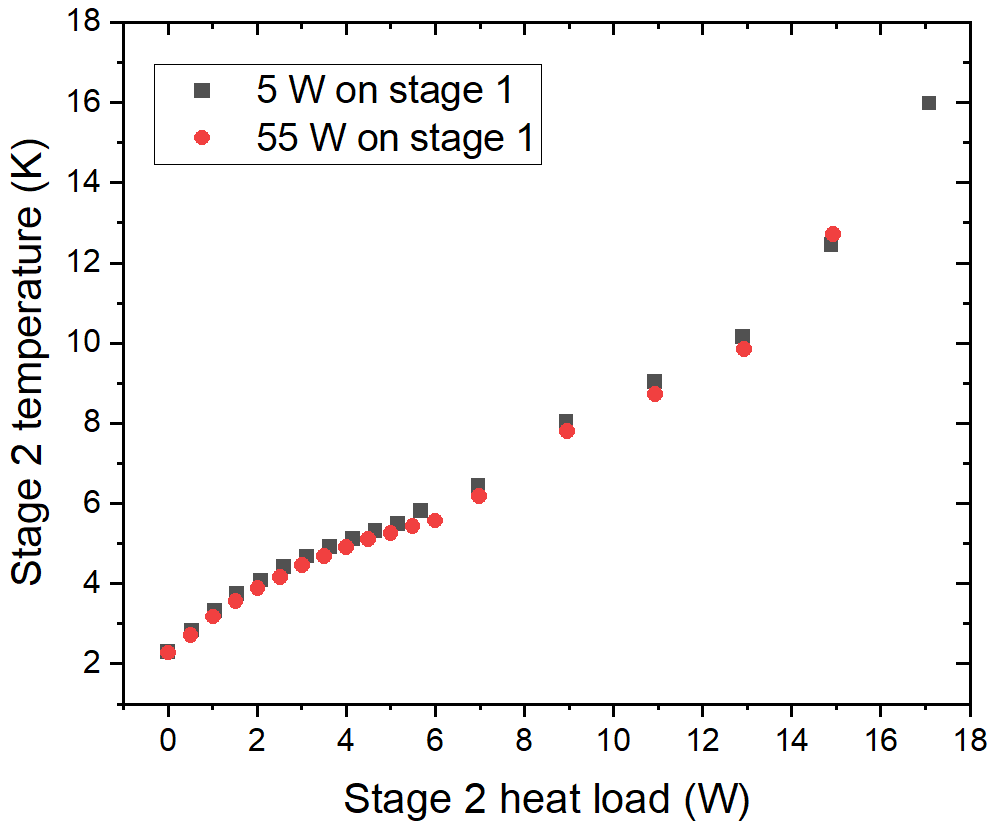}
\caption{\label{fig:CCRheatmap} Stage 2 temperature as a function of the stage 2 heat load for two different stage 1 heat load values for a GM cryocooler model RDE418-D4 measured at Jefferson Lab.}
\end{figure}

The cryocoolers are Gifford-McMahon (GM) type (model RDE-418D4, Sumitomo Cryogenics of America, Allentown, PA) with a nominal cooling power of 2~W at 4~K. The full cryocooler heat capacity map provided by the vendor was used in the thermal analysis and a conservative $2\%$ derating factor was applied to the cryocooler mounted upside-down, as suggested by the vendor. The stage 2 temperature as a function of the heat load applied to the same stage was measured at Jefferson Lab for the same cryocooler model and it is shown in Fig.~\ref{fig:CCRheatmap}. The temperature of the stage 2 does not change significantly for a heat load of up to $\sim80$~W applied to stage 1 and for a fixed heat load of up to $\sim10$~W applied to stage 2.

\begin{figure}[htb]
\includegraphics[width=86mm]{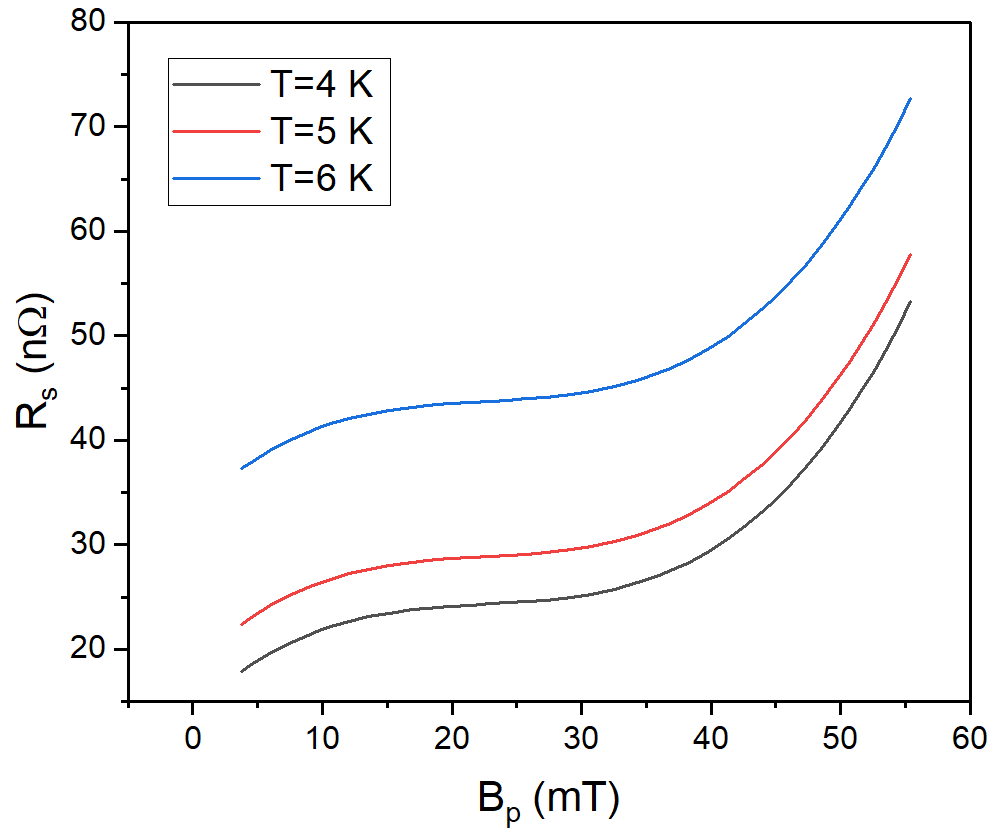}
\caption{\label{fig:Rs_empirical} Empirical field dependence of the surface resistance used in the thermal analysis calculated at three different temperatures.}
\end{figure}

Copper thermal straps were envisioned to connect stage 2 of the cryocoolers to the cavity. A temperature-independent surface contact conductance of 172.4~kW/(m$^2$ K), derived from Ref.~[\onlinecite{Dillon_2017}], was used in the thermal analysis. The temperatures of the outer surface of the beam tubes and FPCs were fixed at 300~K. A static heat load of 1~W was applied uniformly to the cavity outer surface. The rf power density applied to the cavity's inner surface is given by $\frac{1}{2}R_s \left( B/\mu_0 \right)^2$, where $B$ is the magnetic field over the cavity surface for the TM$_{010}$ accelerating mode and $R_s$ is the surface resistance of Nb$_3$Sn. Nb$_3$Sn cavities often display a significant field dependence of the surface resistance. For the thermal analysis, we considered $R_s(B_p, T)=a_0+a_1B_p+a_2B_p^2+a_3B_p^3+a_4\text{exp}(-a_5/T)$, where the 3$^{\text{rd}}$-order polynomial term is an empirical fit of the $R_s(B_p)$ measured at 4.3~K in the prototype Nb$_3$Sn/Nb cavity discussed in Sec.~\ref{subsec:nb3sn} and the last term is a least-squares fit of the Bardeen-Cooper-Schrieffer (BCS) surface resistance computed numerically with the code developed by Halbritter~\cite{BCS}. Figure~\ref{fig:Rs_empirical} shows the empirical $R_s(B_p)$ at three different temperatures used in the analysis. Finally, the rf heat load from the two FPCs, each with 474~kW of input power \footnote{The power into the FPC was slightly higher than the target value because the $Q_{ext}$ of the FPC in the 3D model was not exactly at the target value.}, is included in the model. A target $B_p$-value of 47~mT was chosen for the thermal analysis, corresponding to a $\sim13\%$ margin above the operational value. The thermal conductance of the thermal strap is used as a parameter to obtain convergence in the iterative simulation at the target $B_p=47$~mT.

\begin{figure}[htb]
\includegraphics[width=86mm]{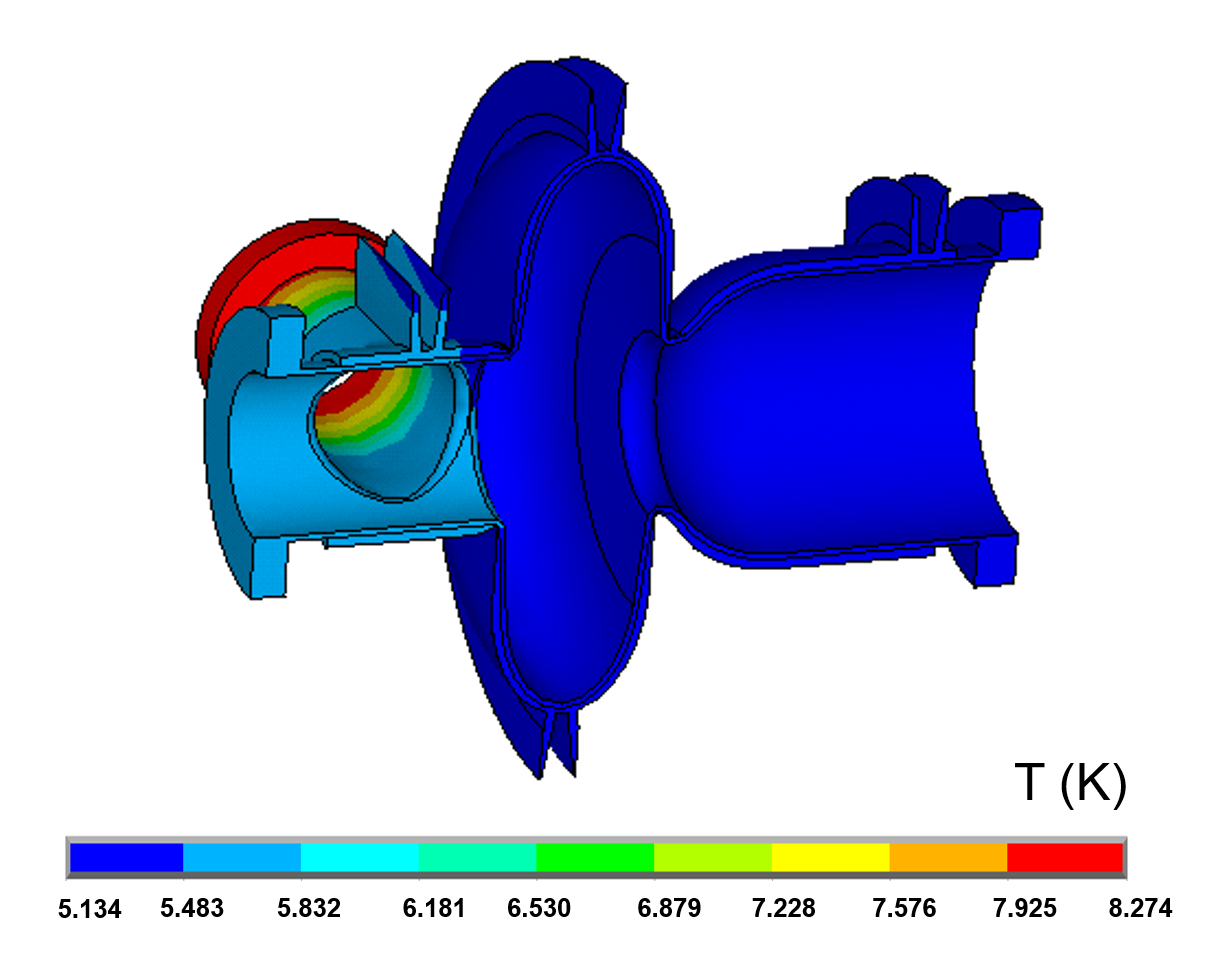}
\caption{\label{fig:tempdist} Temperature distribution for the cavity with $\sim1$~MW of rf input power and operating at $B_p=47$~mT.}
\end{figure}

The finite element solution was solved with a two loop iterative procedure.  The inner loop required setting of the CCR stage temperatures and iterating temperatures and rf heat loads determined by rf resistance and fields.  The outer loop consists of reading the CCRs' heat load reactions and looking up their temperatures from the CCR's capacity map for the next iteration. Small decreasing cryocooler temperature differences from one iteration to the next indicated a converging solution.  Increasing differences from one iteration to the next signified a diverging solution.  Fields were increased until we could not get a converged solution.
Figure~\ref{fig:tempdist} shows the temperature profiles the cavity, obtained from the thermal analysis for $B_p=47$~mT and with a thermal strap thermal conductance of 17.5~W/K at 4~K and 39.7~W/K at 10~K. The total heat load into stage 1 of the cryocoolers is 75~W.  The total heat load into stage 2 of the four cryocoolers is 13.2~W, which is the sum of 0.7~W of radiant heat from the FPCs' inner conductors and pringle tip, 3.3~W of rf heat load from the Nb$_3$Sn, 1~W of static heat load distributed on the cavity outer surface and 8.2~W of ambient conduction from the beam pipes and FPCs' outer conductors.
 
\section{\label{sec:protcav}952.6 MH\lowercase{z} prototype cavity}
The funding available for the project did not allow fabricating a prototype single-cell cavity of the same shape as the one discussed in Sec.~\ref{subsec:cavdes}, however a 952.6~MHz Nb single-cell cavity, which was designed and built as a prototype for the proposed Jefferson Lab Electron Ion Collider~\cite{JLEIC1} was available after it had successfully completed the surface processing and cryogenic rf test~\cite{JLEIC2}. The cavity frequency is sufficiently close to 915~MHz for the cavity to serve as a valid prototype to prove the concepts for conduction-cooled SRF cavities discussed in Sec.~\ref{sec:accel}. The cell length is 141~mm, the shape of the equator is circular with 54.8~mm radius, the shape of the iris is elliptical with 21.8~mm major axis and 15.9~mm minor axis, with a flat straight segment tangent to both the iris ellipse and equator circle. The iris and equator inner diameters are 110~mm and 276.2~mm, respectively. The cavity design followed the same rationale as that described in Ref.~[\onlinecite{Marhauser_FCC}]. The main electromagnetic parameters, computed with CST Studio Suite, are listed in Table~\ref{table:mech}.

\begin{figure}[htb]
\includegraphics[width=86mm]{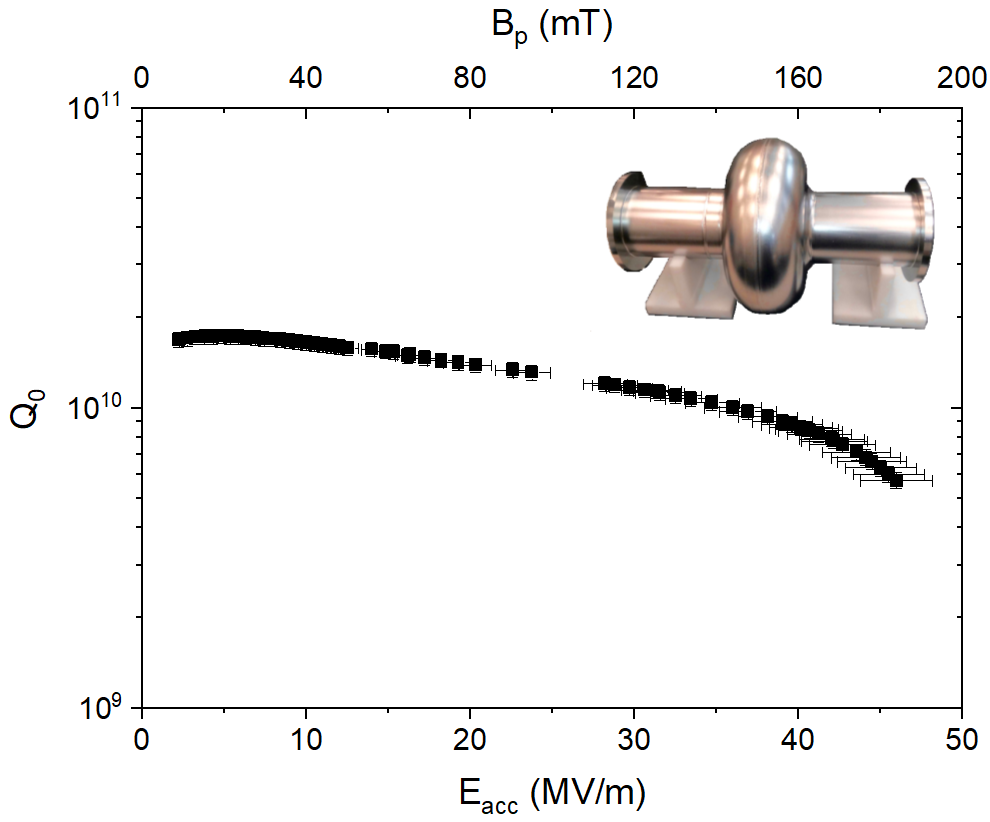}
\caption{\label{fig:QvsE_Nb} Results from the rf test in superfluid He at 2.0~K of the 952.6~MHz prototype single-cell Nb cavity. A picture of the cavity is shown in the inset.}
\end{figure}

The cavity was fabricated at Jefferson Lab with conventional cavity fabrication methods, starting from 4~mm thick, high-purity Nb sheets. The beamline flanges, with 165~mm outer diameter and 11~mm thickness were made from reactor-grade Nb. Indium wire (99.99$\%$ pure) 1.52~mm in diameter was used for the vacuum sealing of the cavity flanges to 316 stainless steel circular plates with rf feedthroughs for the input and transmitted power antennas and with a pump-out port. The best cavity performance was achieved after 30~$\mu$m removal from the inner surface by electropolishing (EP), degreasing in ultra-pure water with a detergent and ultrasonic agitation, high-pressure rinse (HPR) with ultra-pure water in an ISO 4 clean room, assembly of the test hardware, evacuation to $\sim10^{-8}$~mbar and baking at 120~\degree C for 12~h. The results from the cryogenic rf test in a vertical cryostat filled with superfluid He at 2.0~K are shown in Fig.~\ref{fig:QvsE_Nb}. The cavity was limited by quench at $B_p=184$~mT. MP was processed at $B_p\sim52$~mT and 96~mT. The highest x-ray dose rate, measured at the maximum rf field on the cryostat's top plate was 65~$\mu$Sv/h, which is a relatively low value.

\begin{figure}[htb]
\includegraphics[width=86mm]{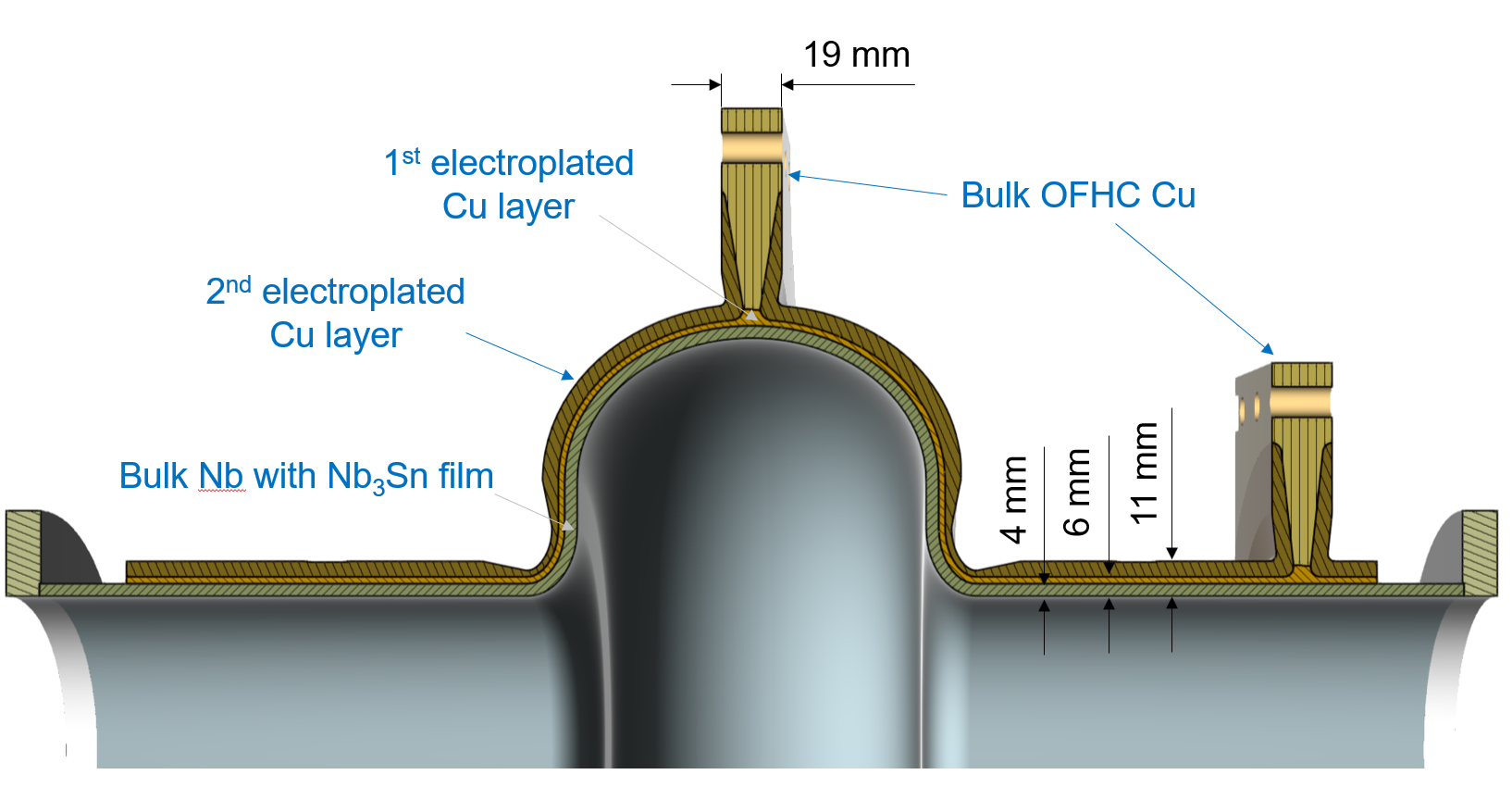}
\caption{\label{fig:Cushell} Cross-section layout of the 952.6 MHz prototype cavity for conduction-cooling demonstration.}
\end{figure}

In order to verify experimentally the conduction-cooling design concept discussed in Sec.~\ref{subsec:thermal}, the cavity will have to be coated with a Nb$_3$Sn film on the inner surface and a thick copper outer layer. The Nb$_3$Sn coating process and the results from the cryogenic rf tests following the coating are discussed in Sec.~\ref{subsec:nb3sn}. The Cu coating process follows the concepts discussed in Sec.~\ref{subsec:thermal}: deposition of a thin Cu layer by cold-spray, deposition of $\sim2$~mm thick Cu layer by electroplating, assembly of an equator ring and a beam-tube attachment plate, machined from OFHC Cu. A second electroplating step, growing a $\sim5$~mm thick Cu layer which bonds the equator ring and attachment plate to the cavity, completes the formation of the Cu outer shell of the cavity. The fabrication process of the Cu outer layer is discussed in details in Sec.~\ref{subsec:Cu}. Figure~\ref{fig:Cushell} shows a cross-section layout of the cavity. The choice of the thickness of the electroplated Cu layer was dictated by the need to deposit sufficient material to bond the attachment ring and plate to the cavity and to have some safety margin, in case the thermal conductivity of the Cu was not as good as that measured on the small sample. The Cu is expected to be thinner at the irises because of the reduced current density in those regions, leading to a lower plating rate. The purpose of the attachment plate on the beam tube was to mimic the one that would be required on the 915~MHz cavity to intercept the heat from the FPCs, as shown in Fig.~\ref{fig:3Dlayout}.

\subsection{\label{subsec:952eng}Engineering analysis}
A mechanical analysis of the cavity was done using Ansys to determine the cavity stiffness and tuning sensitivity as well as the stress distribution in the Nb resulting from the cool-down to 4~K, given the different coefficient of thermal expansion (CTE) between Cu and Nb. The stiffness analysis was done by fixing the axial position of the face of one flange, applying a 1~mm axial displacement to the face of the other flange and calculating the change in the cavity frequency and the force on the flange. The analysis was done for two cases, using material properties at 295~K and 4~K. The case at room temperature was analyzed for the cavity with and without the Cu outer structure. The material properties used for Nb were: Young's modulus $E=105$~GPa at 295~K, $E=118$~GPa at 4~K and Poisson's ratio $\nu=0.38$ at both temperatures. The material properties used for Cu were: Young's modulus $E=127.2$~GPa at 295~K, $E=151.8$~GPa at 4~K and Poisson's ratio $\nu=0.34$ at both temperatures. The results from the analysis at 295~K are listed in Table~\ref{table:mech}. The linear stiffness of the Nb/Cu cavity increases to $\sim109$~kN/mm at 4~K and, assuming a typical tuning range of 200~kHz at 4~K, a cold tuner would need to apply a force up to $\sim 16.6$~kN, which is well within the range of cold tuner designs already in use for SRF cavities~\cite{daly:pac05, pischalnikov:ipac18, Pischalnikov:IPAC2015, Padamsee}.
 
\begin{table}
\caption{\label{table:mech}
Electromagnetic parameters and mechanical properties at 295~K for the 952.6 MHz prototype cavity, with and without the Cu structure shown in Fig.~\ref{fig:Cushell}.}
\begin{ruledtabular}
\begin{tabular}{ccc}
\multicolumn{3}{c}{Electromagnetic parameters}\\
\hline
$E_p/E_{acc}$ & \multicolumn{2}{c}{1.90} \\
$B_p/E_{acc}$ & \multicolumn{2}{c}{4.00 mT/(MV/m)} \\
$G$ & \multicolumn{2}{c}{176.3 $\Omega$}\\
$R/Q(\beta=1)$ & \multicolumn{2}{c}{107.99 $\Omega$}\\
\hline
\multicolumn{3}{c}{Mechanical properties}\\
\hline
 & Nb & Nb/Cu\\
Tuning sensitivity & 1470 kHz/mm & 1313 kHz/mm\\
Stiffness & 13411 N/mm & 95089 N/mm \\
\end{tabular}
\end{ruledtabular}
\end{table}

To perform the analysis of the cool-down stress, a coefficient of linear thermal expansion $\alpha_L = 5$~ppm/K for Nb~\cite{Nb_CTE} and $\alpha_L = 11.4$~ppm/K for Cu ~\cite{Cu_NIST} were used. The same $E$ and $\nu$ values used for the stiffness analysis were assigned to Nb. Cu was modeled with a bi-linear stress versus strain curve based on the data from Ref.~[\cite{Cu_NIST}] for OFHC and electroformed Cu: yield strength $YS$(295~K)=52~MPa, $YS$(28~K)=61~MPa , tangent modulus $TM$(295~K)=1.7~GPa, $TM$(28~K)=1.5~GPa for OFHC Cu; $YS$(295~K)=105~MPa, $YS$(28~K)=145~MPa, $TM$=3.3~GPa for electroformed Cu. $E=114$~GPa and $\nu=0.34$ were assigned to both types of Cu, independent of temperature. Unlike both Cu and Nb, the Nb$_3$Sn softens considerably as the temperature decreases to 4~K. Temperature dependent $E$ and $\nu$ values from Ref.~[\onlinecite{Nb3Sn_elastic_constants}] were used in the model.

\begin{figure}[htb]
\includegraphics[width=86mm]{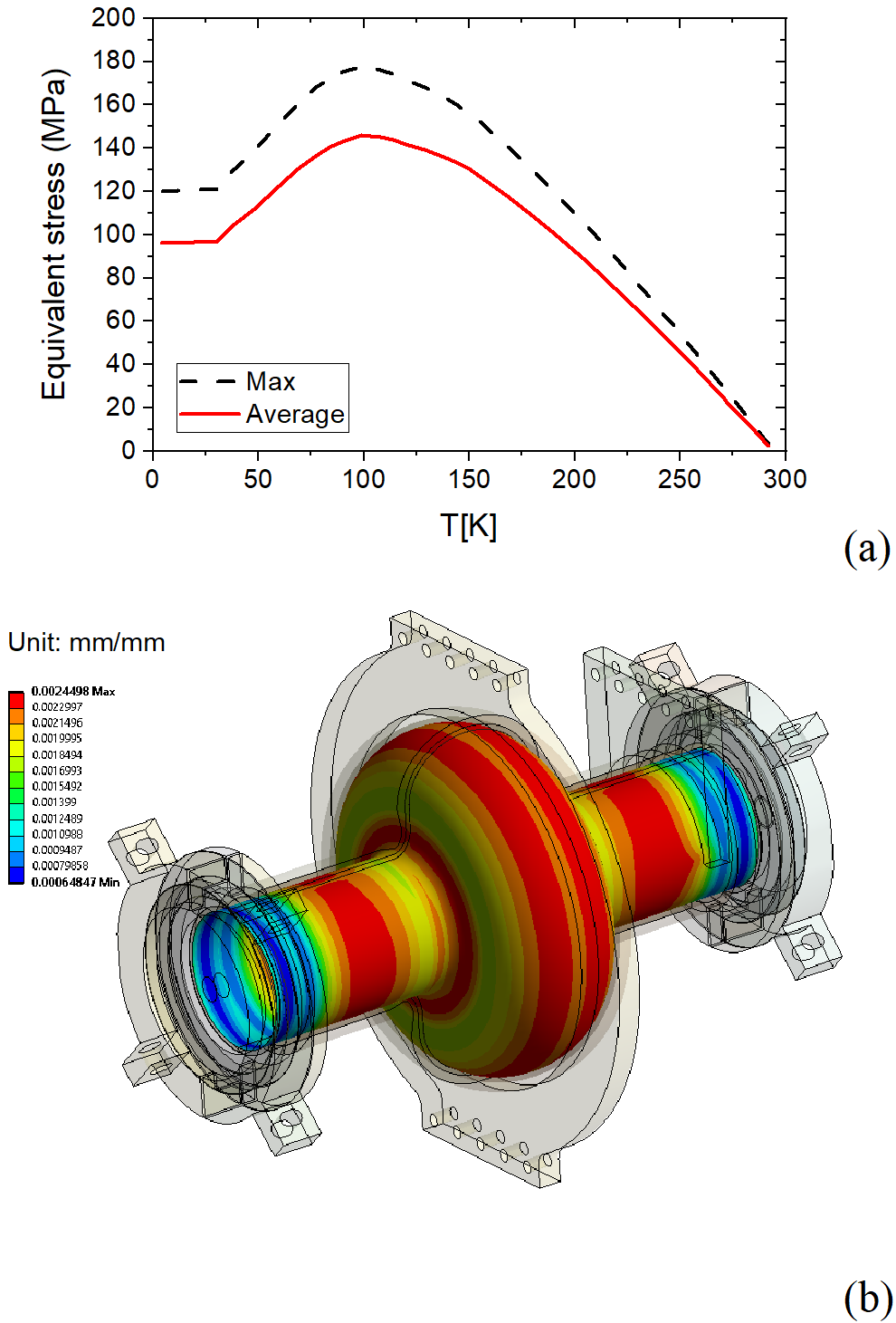}
\caption{\label{fig:Cu_mech}(a) Maximum and average equivalent (von Mises) stress as a function of temperature and (b) equivalent strain at 4~K in the Nb$_3$Sn layer for the prototype cavity with the Cu structure shown in Fig.~\ref{fig:Cushell} on the outer surface.}
\end{figure}

The results from the simulations indicate that the maximum stress in Nb remains well below the yield strength. Local yielding in the iris region is predicted in the Cu layer at 4~K. Figure~\ref{fig:Cu_mech} shows a plot of the average and maximum equivalent stress in the Nb$_3$Sn layer as a function of temperature as well as the equivalent strain at 4~K. The non-monotonic temperature dependence of the equivalent stress in Nb$_3$Sn could result from the decrease of $E$ by a factor of $\sim3$ between 300~K and 4~K. Tensile strain is predicted in the Nb$_3$Sn film at the equator.

A thermal analysis of the cavity cooled by three RDE-418D4 CCRs was done with Ansys in order to verify achieving the target $B_p=47$~mT. The same material properties, cryocooler heat map, interface thermal conductance and $R_s(B_p)$ used for the analysis of the 915~MHz cavity discussed in Sec.~\ref{subsec:thermal} were used for the simulation of the 952.6~MHz prototype cavity. A static heat load of 1~W was applied uniformly to the cavity outer surface and $98\%$ of the heat capacity was used for the CCR with upside-down orientation. The results showed that the cavity should be thermally stable at $B_p=47$~mT with up to 6~W of additional heat applied to the beam tube flange on the side with the CCR. Under these conditions, the cavity temperature would be $\sim 5.3$~K, the rf heat load would be 3.5~W and the total heat load to the CCRs' stage 2 would be 10.25~W. 

\subsection{\label{subsec:nb3sn}Nb$_3$Sn coating}
Following the rf test of the bare Nb cavity, the end-flanges were disassembled and the cavity was degreased, HPRed and Nb plates were mounted on the cavity flanges inside the clean room. The bottom plate had a crucible where 99.999$\%$ pure Sn granules and 99.99$\%$ pure SnCl$_2$ powder were placed. Another crucible with Sn and SnCl$_2$ was placed inside a Nb shelf welded to the top cover plate. The cavity was mounted on a vertical stand outside the clean room, which is inserted in the vacuum furnace~\cite{Eremeev_RSI}. The vapor diffusion process consisted of nucleation at 500~\degree C for 1~h followed by coating at 1200~\degree C for 6 h. A visual inspection of the cavity inner surface after the thin-film coating showed large areas with shiny appearance at the equator, which have been associated with ``patchy" regions with significantly thinner coating layer than non-shiny areas~\cite{Trenikhina_2018, pudasaini:srf2021}. After the coating process, the cavity was degreased, HPRed and assembled with the test hardware and evacuated on a vertical test stand. The standard instrumentation mounted to the cavity consisted of three single-axis cryogenic flux-gate magnetometers (FGMs) directed along the cavity axis and mounted at the equator, $\sim120$\degree apart, and Cernox temperature sensors mounted at the iris and equator.

The cavity was cooled with LHe to 4.3~K inside a vertical cryostat at Jefferson Lab, minimizing the temperature gradient across the iris near the Nb$_3$Sn transition temperature $T_c\sim18$~K. The $Q_0$ at $B_p=3.9$~mT was $1.3 \times 10^{10}$ and it decreased exponentially with increasing $B_p$ to $2.4 \times 10^9$ at 50.8~mT, where MP occurred and could not be processed. The cavity was kept under vacuum on the vertical test stand for about two months and it was cooled down again to 4.3~K for a re-test to try processing the MP. However, the $Q_0$ at 3.9~mT had degraded to $2.6 \times 10^{9}$, decreasing exponentially to $8.2 \times 10^{8}$ at 14.3~mT, in spite of a uniform cool-down across $T_c$. The frequency of the cavity was $\sim3.2$~MHz lower than after the first test. After warm-up to room temperature, the cavity shape was inspected with a 3D laser coordinate measuring machine, while still under vacuum on the vertical test stand. It was found that the cavity shape had become ``re-entrant" as the irises were pushed inward by $\sim0.7$~mm on each side as a result of the 1~atm differential pressure and material creep due the softening of the Nb by the annealing at 1200~\degree C.

\begin{figure}[htb]
\includegraphics[width=86mm]{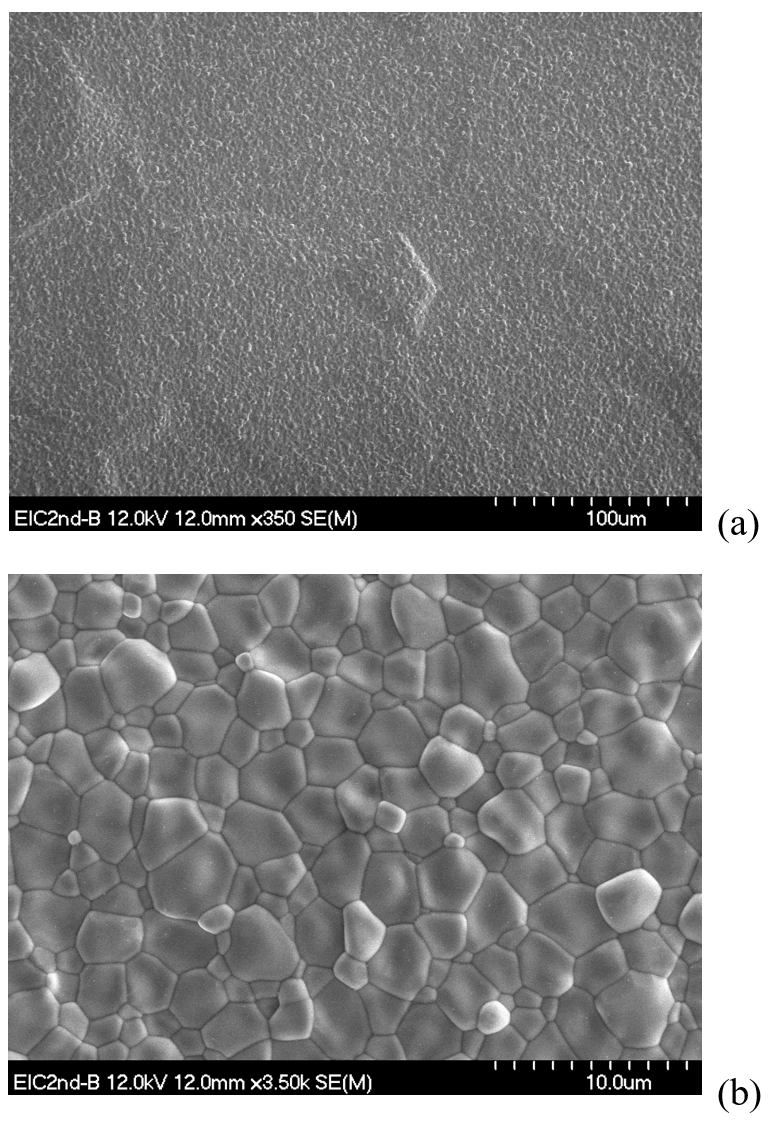}
\caption{\label{fig:Nb3Sn} (a) low (x350) and (b) high magnification (x3.5k) SEM images of the Nb$_3$Sn film on a Nb witness sample after the second coating.}
\end{figure}

After disassembly, the cavity was mechanically tuned back to the initial frequency, then $\sim20\;\mu$m were removed from the inner surface by EP, followed by degreasing, HPR and assembly with Nb end-plates. In order to achieve a more uniform coating, the required amount of Sn and SnCl$_2$ granules were distributed in three crucibles: one on the top plate, one at the bottom plate and one at at approximately mid-height, hanging from the top plate. Nb witness samples were placed on the top and bottom plates. The cavity was coated with the same temperature profiles as for the first coating. A visual inspection of the inner cavity surface showed a uniform, ``matte" finish. Figure~\ref{fig:Nb3Sn} shows an image of the Nb$_3$Sn thin film formed on the witness sample at the top plate, measured with a secondary electron microscope (SEM). The thin-film was uniform, with a Sn concentration of $(24.75 \pm 0.30)$~at.$\%$, close to that of stoichiometric Nb$_3$Sn, measured by energy dispersive x-ray spectroscopy. However, nanometer-size Sn residues were found to be distributed over the entire surface. Similar results were observed on the witness sample placed on the bottom plate. The film thickness was measured from the witness samples using SEM images taken from the cross-section prepared with a focused ion beam and it was $(2.18 \pm 0.48)$~$\mu$m and $(2.87 \pm 0.29)$~$\mu$m for the 1$^{\text{st}}$ and 2$^{\text{nd}}$ coating, respectively.

\begin{figure}[htb]
\includegraphics[width=86mm]{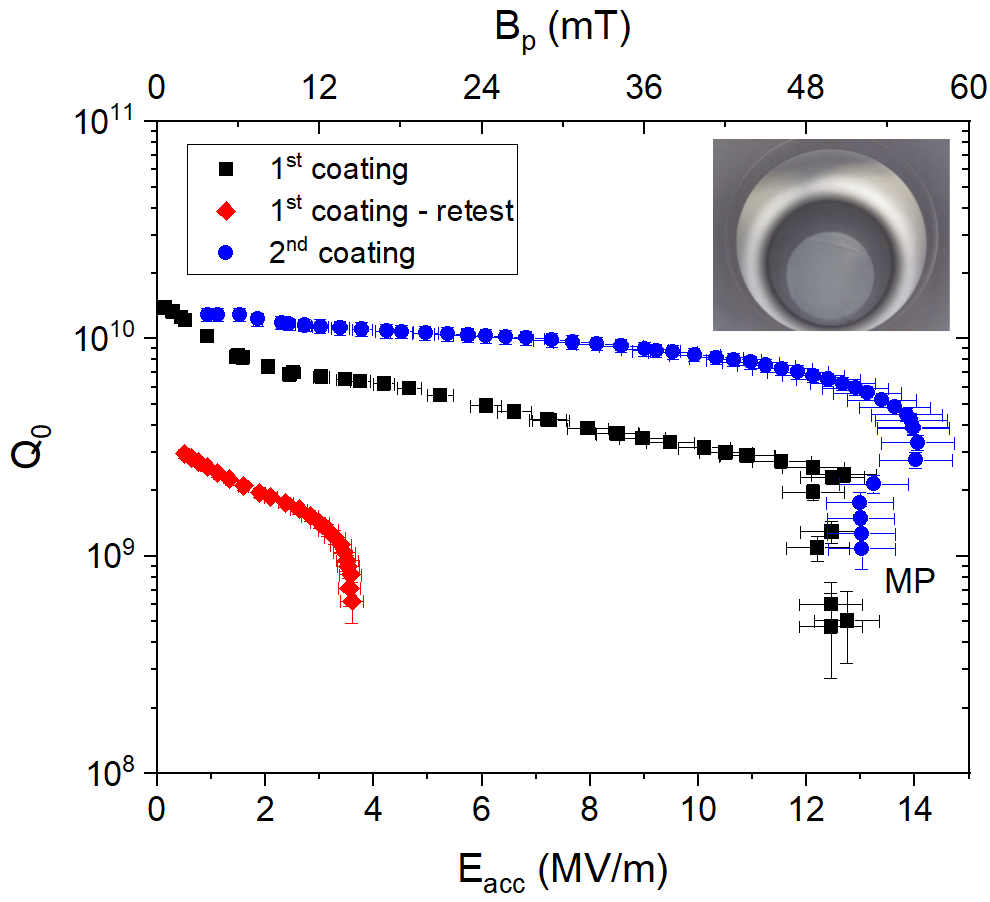}
\caption{\label{fig:QvsE_Nb3Sn} Results from the rf tests in LHe at 4.3~K of the 952.6~MHz prototype single-cell cavity after coating the inner surface with Nb$_3$Sn. The inset shows a picture of the inner surface after forming the Nb$_3$Sn film.}
\end{figure}

The cavity was degreased, HPRed, assembled with the test hardware, mounted on a vertical test stand with restraining plates to prevent deformation due to differential pressure after evacuation, evacuated to $\sim10^{-8}$~mbar and cooled in LHe at 4.3~K. The cavity was tested three times, each with different cool-down conditions in terms of residual dc magnetic field at the cavity and temperature gradient across the cell.  Figure~\ref{fig:QvsE_Nb3Sn} shows a summary of the cavity performance in LHe at 4.3~K after forming the Nb$_3$Sn film on the inner surface. The iris-to-iris axial temperature gradient, $dT/dz$, and the residual magnetic field, $B_a$, at $T_c$ for the test after the 2$^{\text{nd}}$ coating shown in Fig.~\ref{fig:QvsE_Nb3Sn} were 0.64~K/m and 2.5~$\mu$T, respectively. The cavity reached up to 56.4~mT and was limited by strong MP at 52~mT. The x-ray dose rate at the maximum rf field was 21~$\mu$Sv/h.
The residual resistance, $R_{res}$, increased linearly with increasing axial temperature gradient and residual magnetic field at the rates $dR_{res}/d(dT/dz)=(9 \pm 4)$~n$\Omega$/(K/m) and $dR_{res}/dB_a = (7 \pm 5)$~n$\Omega /\mu$T, respectively. These values are consistent with those mentioned in Ref.~[\onlinecite{Hall}] for a 1.3~GHz cavity.

The resonant frequency and $Q_0$ were measured as a function of the cavity temperature with a vector network analyzer (VNA) during warm-up from 4.3~K to obtain the temperature dependence of the change in rf penetration depth, $\Delta \lambda$~\cite{Halb_JAP}, and of the surface resistance averaged over the inner surface of the cavity, $R_s=G/Q_0$. The data between 15~K and $T_c$ were fitted using the two-fluid model for $\Delta \lambda (T)$ and Halbritter's code for $R_s(T)$. The London penetration depth, $\lambda_L=80$~nm and the BCS coherence length, $\xi_0=6$~nm were considered material constants~\cite{Hein}. The values of the parameters obtained from the least-square fits were: $T_c = (17.79 \pm 0.05)$~K, energy gap $\Delta /k_B = (41.6 \pm 0.2)$~K, normal electrons' mean free path $\ell = (3.78 \pm 0.08)$~nm and rf penetration depth at 0~K $\lambda_0 = (129 \pm 2)$~nm. The uncertainty on the $T_c$-value includes both statistical and systematic uncertainties.

\subsection{\label{subsec:Cu}Copper coating}
Following the rf tests after the 2$^{nd}$ coating, the cavity was disassembled, degreased, dried and aluminum end-caps with an O-ring seal were mounted to the cavity flanges. The cavity was shipped to Concurrent Technologies Corporation (CTC), Johnstown, PA to deposit a thin layer of copper on the cavity outer surface by the cold-spray method, which provides good adhesion of Cu to Nb. $99.9\%$ pure Cu powder sized at -325 mesh was fed into a de Laval nozzle at a rate of 150 standard liter per minute, along with He gas at 4.13~MPa and 400~\degree C to achieve a high impact velocity of the Cu powder onto the Nb. The cavity temperature only increased by a few degrees Celsius above room temperature during deposition. The layer thickness was obtained from the difference between cavity dimensions before and after coating, measured with calipers, and it was between $178 - 330$~$\mu$m. The deposition process took about $\sim20$~min.

\begin{figure}[htb]
\includegraphics[width=86mm]{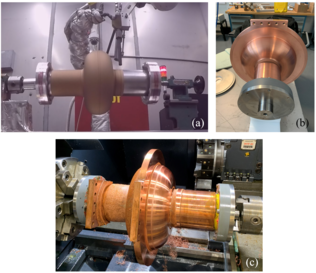}
\caption{\label{fig:Cu_coating} Pictures taken during the fabrication of the Cu structure on the outer cavity surface: during cold-spray (a), during fitting of the equator ring on the Cu layer after the 1$^{\text{st}}$ electroplating and machining (b) and during the final machining (c).}
\end{figure}

\begin{figure}[htb]
\includegraphics[width=86mm]{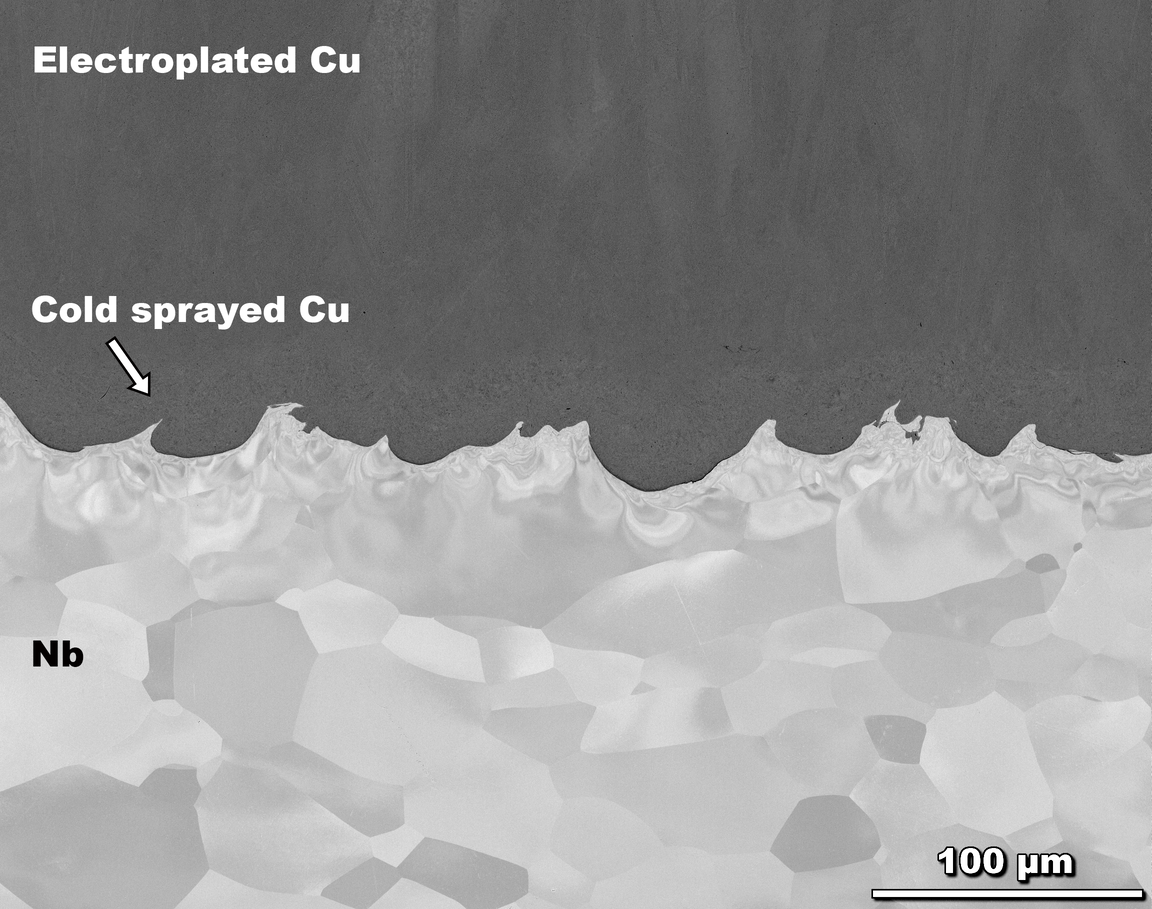}
\caption{\label{fig:NbCu_SEM} SEM image of the cross-section of a Nb/Cu sample with the interface Cu layer between Nb and electroplated Cu being deposited by cold-spray.}
\end{figure}

Electroplating of the 1$^{\text{st}}$ and 2$^{\text{nd}}$ Cu layers was done at AJ Tuck Co., Brookfield, CT, using a proprietary recipe resulting in $> 99.99\%$ pure Cu. Machining of the cavity outer surface to the required dimensions and surface finish after each electroplating step was done at Jefferson Lab using both computerized numerical control lathe and milling machines. The areas with the bolt-holes were machined to a 0.4~$\mu$m average surface roughness. The beam-pipe flanges were kept sealed with end-caps throughout all of the steps required to produce the Cu structure on the cavity. Figure~\ref{fig:Cu_coating} shows images of the cavity through different steps of the Cu deposition. Figure~\ref{fig:NbCu_SEM} shows an SEM image of the cross-section of a Nb/Cu sample, in which the Cu was deposited in a similar way as for the cavity. Post-processing analysis of the SEM images, acquired over a sample length of $\sim 5.9$~mm, showed a total interface length of $\sim 48$~mm and the debonded length was only $2\%$.

\begin{figure}[htb]
\includegraphics[width=86mm]{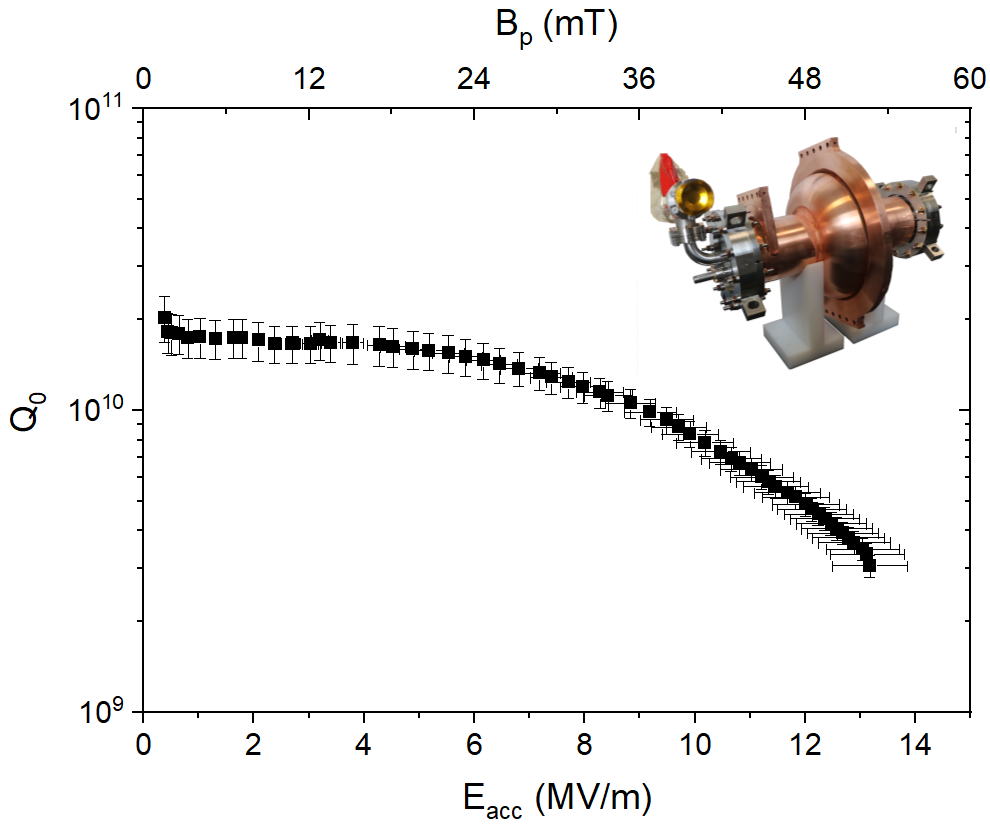}
\caption{\label{fig:QvsE_CuNbNb3Sn} Results from the rf test in LHe at 4.3~K of the 952.6~MHz prototype single-cell cavity after coating the outer surface with Cu. The inset shows a picture of the fully assembled cavity.}
\end{figure}

After the final machining of the outer surface, the cavity was degreased, HPRed, assembled with the test hardware. An all-metal valve and a burst disk were added to the pump-out port on the stainless steel end-cap on one side of the cavity. The cavity was evacuated on a vertical test stand to $\sim10^{-8}$~mbar and inserted in a vertical cryostat at Jefferson Lab. The cavity was cooled to 4.3~K at a rate of $< 2$~K/min in the range $30-300$~K. Two temperature cycles above $T_c$ were done to minimize the temperature gradient across the cavity. The cavity performance in LHe at 4.3~K is shown in Fig.~\ref{fig:QvsE_CuNbNb3Sn}, after cool-down with $dT/dz=1.4$~K/m and $B_a=0.6\;\mu$T at $T_c$, limited by MP at $B_p=53$~mT. No x-rays were detected above background during the test. The pressure sensitivity was $df/dP = -24$~Hz/mbar, compared to $-233$~Hz/mbar measured for the cavity without the Cu outer structure. The ambient magnetic field at the cavity equator was measured during the rf test by three FGMs distributed at $\sim120$\degree along the circumference. The magnetic field measured by one sensor increased linearly with the power dissipated in the cavity at a rate of $\sim85\; \mu$T/mW. The magnetic field measured by another sensor jumped by $\sim0.6\; \mu$T during a MP-induced quench.

\begin{figure}[htb]
\includegraphics[width=86mm]{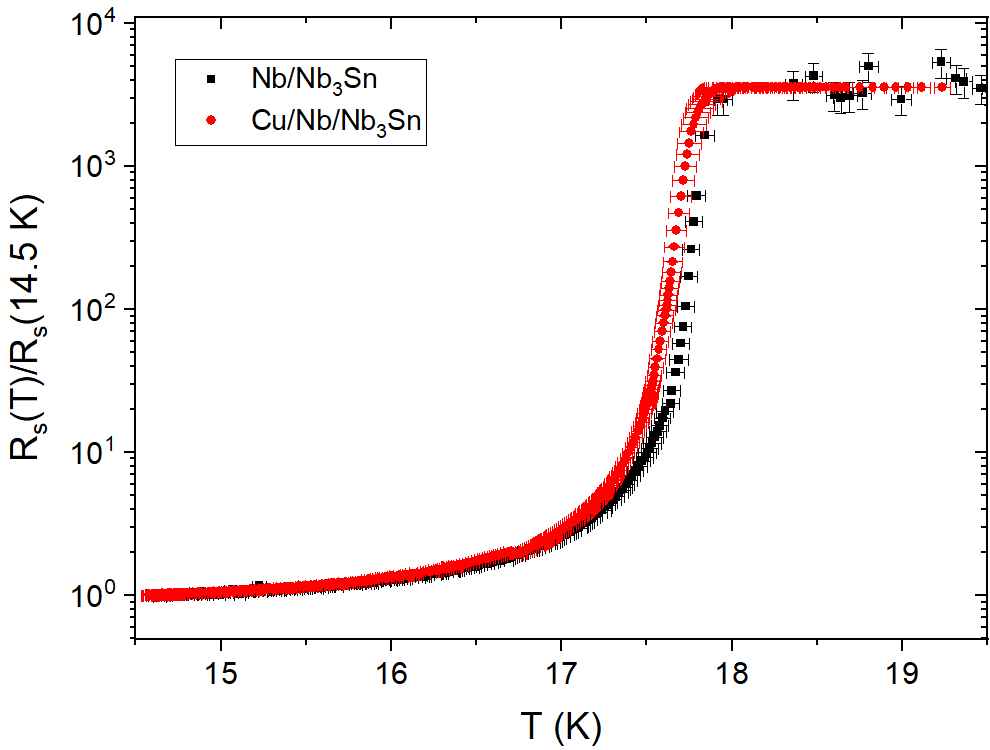}
\caption{\label{fig:Rs_T} $R_s(T)/R_s(15~\text{K})$ between $14.5 - 19.5$~K measured before and after growing the Cu structure on the cavity outer surface.}
\end{figure}

The resonant frequency and $Q_0$ were measured with a VNA and rf amplifiers during warm-up between 15~K and 295~K. The change in rf penetration depth as a function of temperature below $T_c$ was fitted with the two-fluid model, resulting in $\lambda_0=(138 \pm 1)$~nm, $\ell = (3.1 \pm 0.1)$~nm and $T_c = (17.73 \pm 0.08)$~K. The $Q_0(T)$ was measured at $B_p=1.2$~mT between $4.3-8.4$~K with the same self-excited loop rf system used for the high-power test and $R_s(T)=G/Q_0(T)$ was fitted with $R_s(T)=R_{BCS}(T) + R_{res}$, where $R_{BCS}$ was calculated numerically with Halbritter's code with fixed $T_c = 17.73$~K, $\ell = 3.1$~nm, $\lambda_L=80$~nm, $\xi_0=6$~nm. The fit parameters were $\Delta /k_B = (39.0 \pm 0.1)$~K and $R_{res} = (12.7 \pm 0.7)$~n$\Omega$. 
Figure~\ref{fig:Rs_T} shows a plot of $R_s(T)/R_s(14.5~\text{K})$ between $14.5 - 19.5$~K measured before and after growth of the Cu structure on the cavity outer surface showing a possible broadening of the transition by $\sim0.1$~K for the cavity after Cu deposition.

\begin{figure}[htb]
\includegraphics[width=86mm]{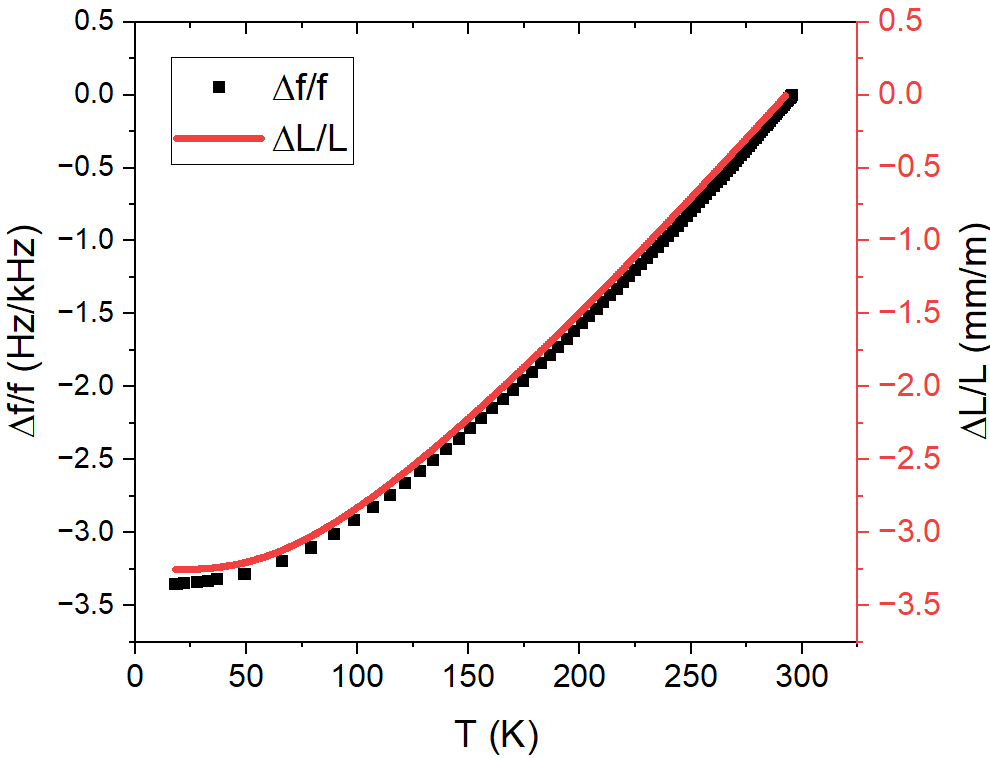}
\caption{\label{fig:df_dL} Normalized shift in resonance frequency of the Cu/Nb/Nb$_3$Sn cavity along with the normalized thermal expansion of Cu as a function of temperature.}
\end{figure}

The frequency shift of the cavity under vacuum between 295~K and 4.3~K was $\sim 3.23$~MHz, compared to $\sim 1.40$~MHz before Cu deposition. This difference is due to the different CTE between Cu and Nb: Fig.~\ref{fig:df_dL} shows the measured $\Delta f/f = \left[ f(295~\text{K}) - f(T) \right]/f(295~\text{K})$ for the Cu/Nb/Nb$_3$Sn cavity, showing good agreement with $\Delta L/L = \left[L(T) - L(295~\text{K})\right]/L(295~\text{K})$ for Cu from Ref.~[\onlinecite{Cu_NIST}]. 

\begin{figure}[htb]
\includegraphics[width=86mm]{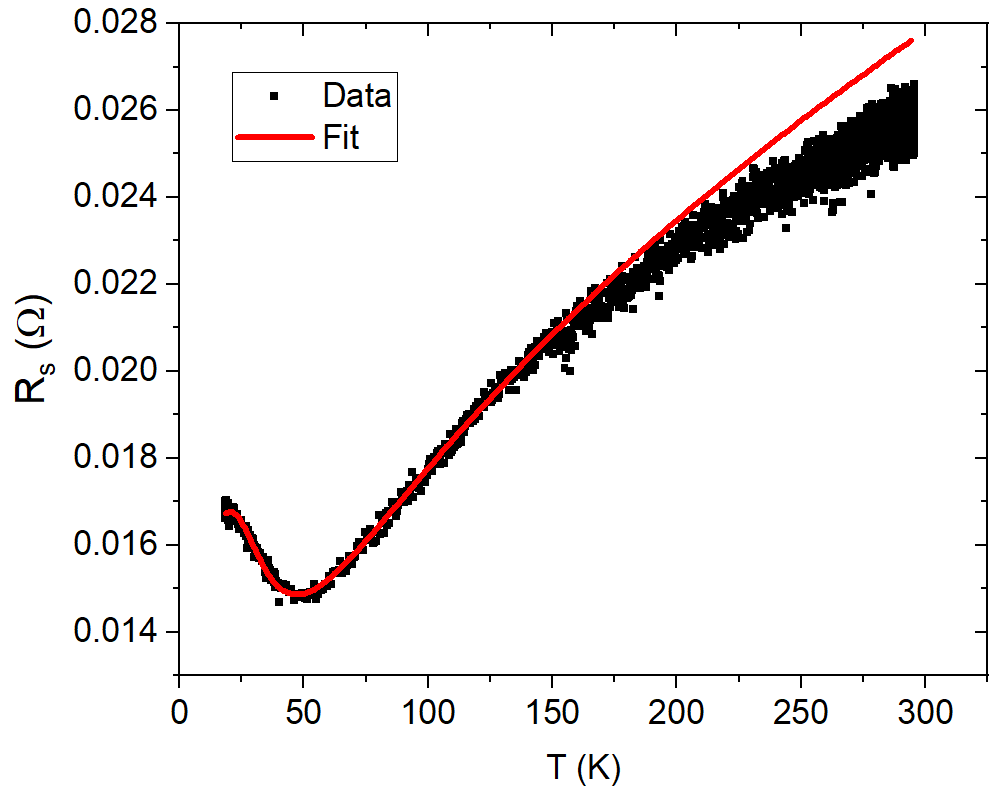}
\caption{\label{fig:Rs_nc} $R_s(T)$ in the normal state measured for the Cu/Nb/Nb$_3$Sn cavity. The solid line is a least-square fit with the real part of Eq.~(\ref{Eq1}).}
\end{figure}

The surface resistance as a function of temperature was also measured in the normal state and showed a minimum at $\sim 50$~K. To understand the mechanism of this minimum, consider the surface impedance $Z=-i\omega E(0) \mu_0/B$ of a thin film of thickness $d$ and conductivity $\sigma_1(T)$ deposited on a substrate with conductivity $\sigma_2(T)$, exposed to a parallel rf field:

\begin{equation}
Z=(1-i) \sqrt{\frac {\omega \mu_0}{2\sigma_1}} \frac {\beta + \text{tanh}[(1-i)q]} {\beta \, \text{tanh}[(1-i)q]+1},
\label{Eq1}
\end{equation}
where $\beta = \sqrt{\sigma_2/\sigma_1}$ and $q=d \sqrt{\mu_0\sigma_1\omega/2}$ is the ratio of the film thickness divided by the skin depth in the film. Figure~\ref{fig:Rs_nc} shows a least-square fit of the real part of Eq.~(\ref{Eq1}) to the experimental data using the temperature-dependent dc conductivities measured on a bulk Nb sample and a thin film Nb$_3$Sn sample on sapphire. The parameters obtained from the curve fitting were: $d=(5.1 \pm 0.2)$~$\mu$m, the $RRR$ of Nb$_3$Sn, $RRR_1=(21 \pm 63)$ and that of the Nb layer near the interface with the thin film, $RRR_2=(100 \pm 56)$ and the room temperature resistivity of Nb$_3$Sn, $\rho_1(\text{295 K})=(129 \pm 12)$~$\mu\Omega \, \text{cm}$. As one can see, Eq.~(\ref{Eq1}) captures the observed behavior of $R_s(T)$ and shows that the minimum in $R_s(T)$ at $\sim 50$~K in Fig.~\ref{fig:Rs_nc} reflects a crossover from $R_s^{Nb_3Sn}(T)$ at low $T$, for which the skin depth in Nb$_3$Sn is smaller than $d$, to $R_s^{Nb}(T)$ at high $T$, for which the skin depth in Nb$_3$Sn becomes much larger than the layer thickness and the contribution of Nb$_3$Sn to $E_s$ is negligible.

\section{\label{sec:tstrap}Thermal link assembly}
\begin{figure}[htb]
\includegraphics[width=65mm]{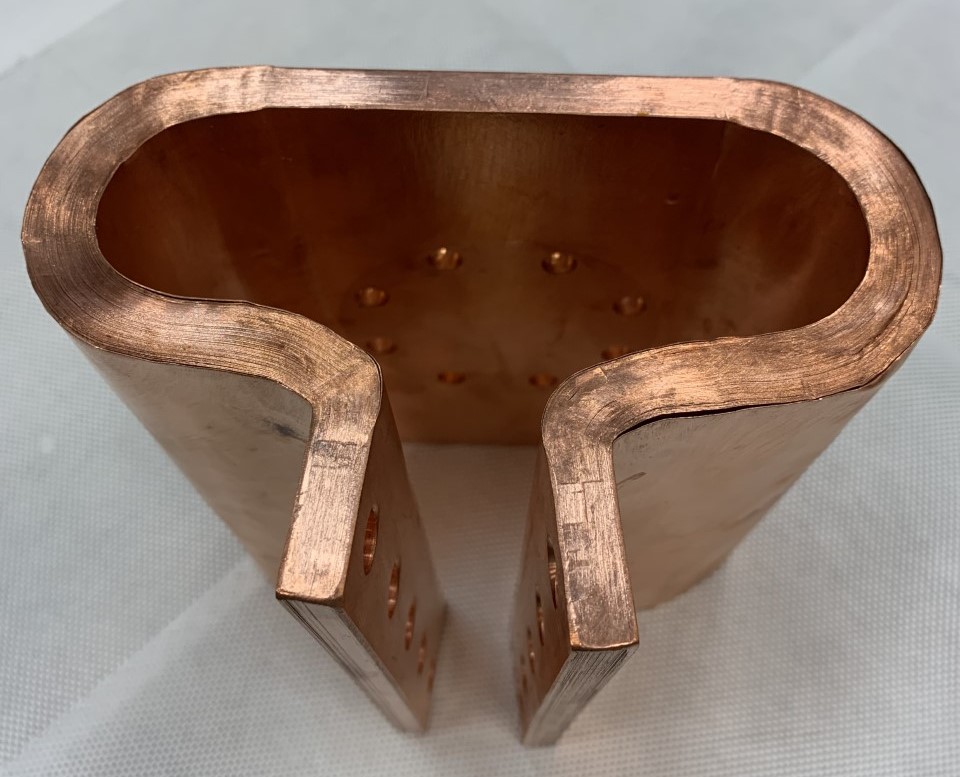}
\caption{\label{fig:TLA} Picture of the TLA used to connect the cavity to stage 2 of the CCR.}
\end{figure}

\begin{figure}[htb]
\includegraphics[width=86mm]{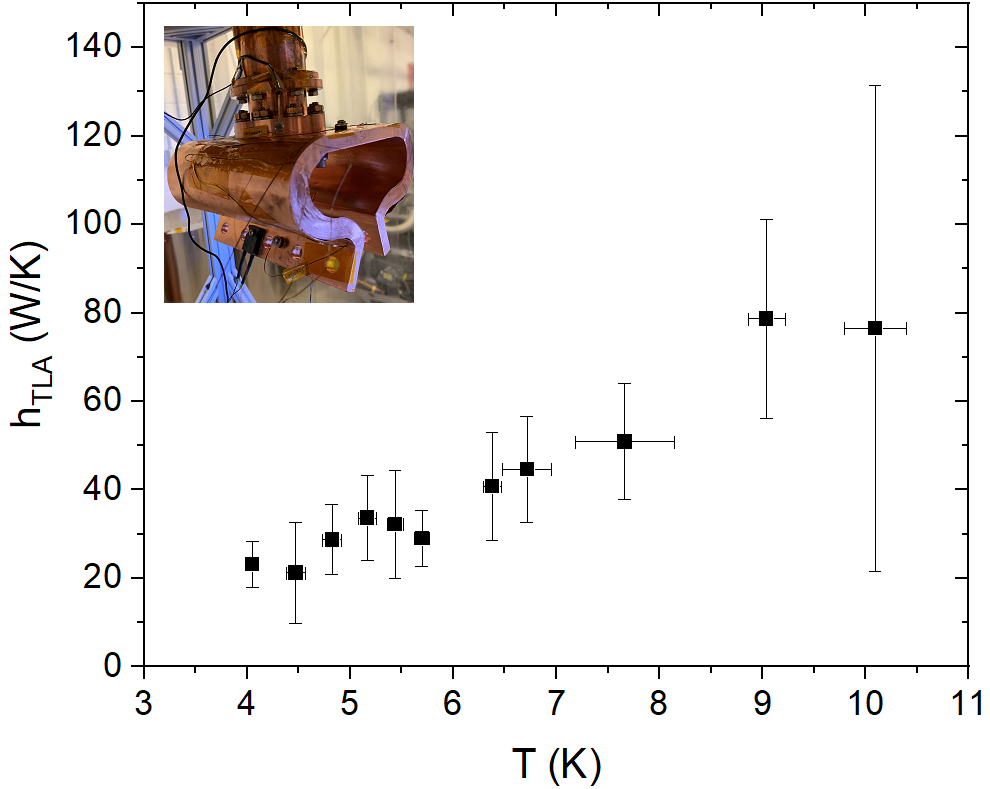}
\caption{\label{fig:h_TLA} Thermal conductance of the Cu TLA shown in Fig.~\ref{fig:TLA} as a function of temperature. The inset shows a picture of the TLA mounted to stage 2 of the CCR at Jefferson Lab.}
\end{figure}

An essential component of a conduction-cooled SRF cryomodule is the thermal link assembly (TLA) connecting stage 2 of a CCR to the cavity. Besides maximizing the thermal conductance between the two, the TLA needs to have sufficient flexibility to allow for misalignments between the cavity and stage 2 of the CCR, as well as providing compliance for thermal contraction during cool-down. The specifications in terms of the required thermal conductance, $h_{TLA}$, were determined from the thermal analysis described in Sec.~\ref{subsec:thermal}: 17.5~W/K at 4~K and 39.7~W/K at 10~K. The TLAs were manufactured by Absolut System, Seyssinet-Pariset, France, by stacking $\sim85$, 100~$\mu$m thick Cu foils, C10300 alloy, of $\sim 113$~mm length and 152~mm width. The foil stack was bent into a "double S" shape and $\sim25$~mm long sections at the ends were press-welded in an inert gas atmosphere. A 74~mm wide central section was also press-welded and bolt-hole patterns were machined in the press-welded regions. Press-welding is a cost-effective method to fabricate TLAs with no internal thermal interfaces, which must be avoided in order to achieved the required thermal conductance. The TLAs were cleaned and baked at the vendor, as discussed in Ref.~[\onlinecite{Trollier}]. Figure~\ref{fig:TLA} shows a picture of the TLA. One TLA was mounted to stage 2 of a Sumitomo RDE418-D4 CCR through an adapter OFHC Cu cylinder and calibrated Cernox resistance-temperature devices (RTDs) were mounted to the following locations: one sensor on stage 2, one sensor on the cylinder adapter, two sensors on the wide press-welded area, one sensor each on the press-welded end sections. Two heaters were also mounted to the end sections. The setup was inserted in a vacuum vessel and cooled to $\sim2.4$~K with the CCR. The difference between the temperature near the heater and that on the cold side area of the TLA, $\Delta T$, was measured as a function of the heater power, $P_h$, which was the same on each end of TLA. The TLA thermal conductance was calculated as $h_{TLA}=2P_h/(\Delta T_1 + \Delta T_2)$, where $\Delta T_1$ and $\Delta T_2$ are the $\Delta T$ from each side of the TLA. Figure~\ref{fig:h_TLA} shows $h_{TLA}$ as a function of the average temperature. Each data point is the average of 3 measurement cycles. The measured thermal conductance was 23~W/K at 4~K and 77~W/K at 10~K, exceeding the specifications.

The stiffness of the TLA was measured at room temperature by measuring the force as a function of displacement for three orthogonal directions. The force per unit length was found to be 18~N/mm in the direction normal to the plane with the circular bolt-hole pattern, 22~N/mm in the direction perpendicular to the plane with the linear bolt-hole pattern, and 14~N/mm in the direction along the plane with the linear bolt-hole pattern.

\section{\label{sec:htb}Horizontal test cryostat design and assembly}
A horizontal test cryostat (HTC), which allows rf testing of the prototype cavity cooled by 3 CCRs, was designed and procured. It has a cylindrical grade 304 stainless steel vacuum vessel with two dome-shaped, O-ring sealed end caps. The overall dimensions are 150~cm in length and 1.1~m outer diameter. Two closely-spaced CCRs are mounted at the top, the third one being mounted upside-down at the bottom, using dedicated insertion fixtures. The vacuum flanges of one of the CCRs at the top and the one at the bottom were mounted to a bellows assembly, bolted to weldments on the vacuum vessel, allowing adjusting the position of their stage 2 inside the vessel. The cavity is held inside the vessel by 4 pairs of nitronic rods and it is thermally isolated from them by G10 washers. TLAs connect stage 2 of each CCR to the cavity. A286 stainless steel bolts with hard-tempered silicon-bronze nuts were used to make the connections, along with stainless steel blocks on both sides of each joint to distribute the pressure uniformly over the contact area. The Cu contact surfaces were machined to an average roughness of 0.4~$\mu$m and a thin layer of Apiezon N grease was applied over the contact areas. The clamping pressure estimated for the torque on the bolts is $\sim 53$~MPa.

\begin{figure}[htb]
\includegraphics[width=86mm]{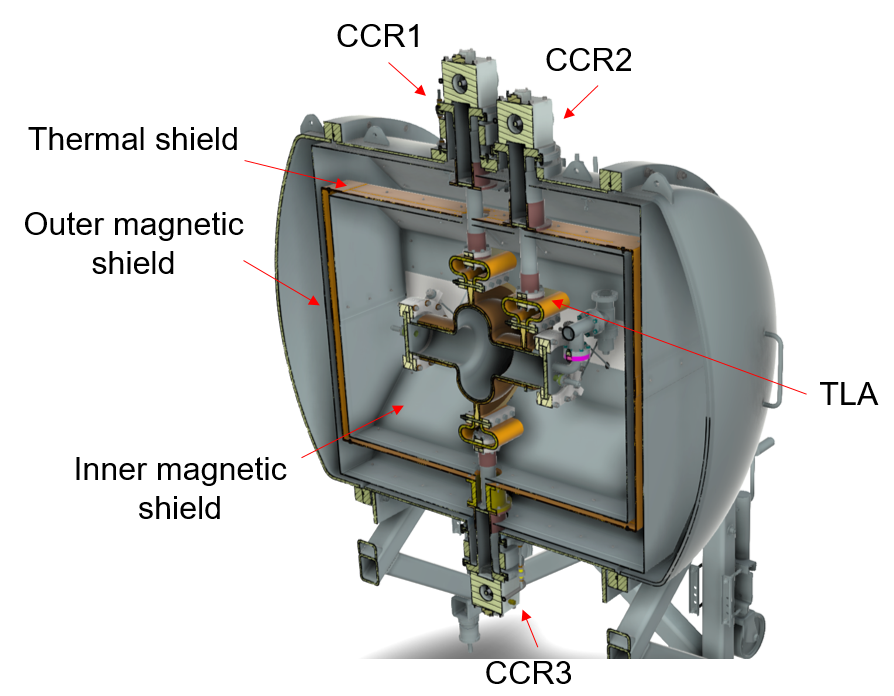}
\caption{\label{fig:HTC} Cross-section view of the 3D layout of the HTC on a support cart.}
\end{figure}

An outer magnetic shield, a thermal shield and an inner magnetic shield are installed between the vacuum vessel shell and the cavity. The inner and outer magnetic shields were made by Ad-vance Magnetics, Inc., Rochester, IN, USA from 1~mm thick Cryoperm 10 and Ad-Mu-80, respectively. The thermal shield was made out of 2~mm thick OFHC copper, and it is attached to the stage 1 flanges of one of the top and the bottom CCRs. Both the vacuum vessel and the thermal shield were manufactured by Master Machine $\&$ Tool, Inc., Newport News, VA, USA.
The parts for each of the shields consisted of 6 side panels and 4 end caps, all fastened together during the assembly, split between an upper and a lower sub-assembly. The design of the magnetic shields was validated by a magnetostatic analysis using Ansys, predicting an ambient magnetic field of less than 0.5~$\mu$T anywhere at the cavity surface.
The biggest challenge with the mechanical design of the HTC was to ensure that the mechanical loads on both stages of the cryocoolers were below the manufacturer's specifications: 49~N, 294~N and 981~N for horizontal, tensile and compressive loads on stage 2, respectively and 98~N, 294~N, 1961~N for horizontal, tensile and compressive loads on stage 1. The top sub-assembly of the thermal shield is hung from the top sub-assembly of the outer magnetic shield by stainless steel wires. The bottom sub-assembly of the thermal shield is mounted to a Cu spool-piece, attached to stage 1 of the bottom cryocooler. The top and bottom thermal shield sub-assemblies are allowed to slide over each other during cool-down.
A thermal and mechanical analysis of the HTC was done using Ansys. The TLAs were modeled as an orthotropic material and the material properties were adjusted to match the measured stiffness and thermal conductance. The nitronic rods's pretension was set to achieve null displacement at the cavity flanges and interfaces with the TLAs at 300~K. The temperature distributions from the thermal analysis were used to determine the stresses at the cryocoolers' heat stations. The results validated the design as all mechanical loads to the cryocoolers after cool-down were within the specifications. Figure~\ref{fig:HTC} shows a cross-section of the 3D model of the HTC.

\begin{figure}[htb]
\includegraphics[width=86mm]{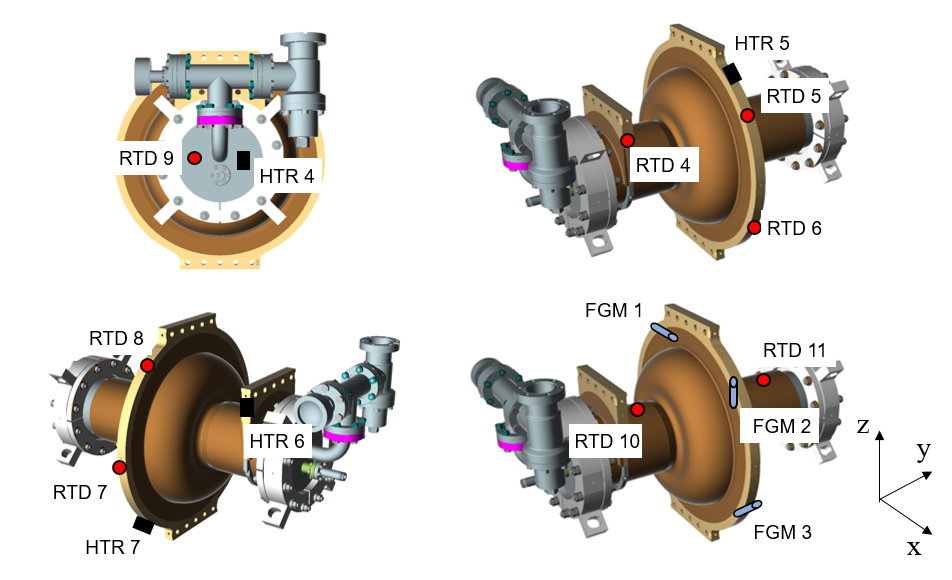}
\caption{\label{fig:inst} Approximate locations of heaters (black rectangles), temperature sensors (red circles) and FGMs (blue cylinders) installed on the prototype cavity inside the HTC.}
\end{figure}

The vacuum vessel has three side-ports: one for the rf and instrumentation panel, one for evacuation and one for a vacuum gauge and a burst disk. A flexible rf input cable (TFlex-405, Times Microwave Systems, Inc., Wallingford, CT, USA) and a semi-rigid rf pick-up cable (ULT-05, Keycom, Tokyo, Japan) are connected between the vacuum-side of an rf feedthrough with two Subminiature A (SMA) connectors on the instrumentation panel and the N-type rf feedthroughs mounted on the cavity. These cables were selected for their low thermal conductivity and both cables have a thermal intercept mounted to the thermal shield~\cite{cheng:srf2019}. The instrumentation consisted of 16 Cernox CX-1050 RTDs and 3 cryogenic FGMs. The RTDs were distributed as follows: one on stage 2 of each CCR, one on each TLA, one each on the top and bottom thermal shield sub-assemblies and eight on the cavity assembly at the approximate locations shown in Fig.~\ref{fig:inst}. A section of the leads of the RTDs attached to the cavity was wrapped around copper bobbins mounted to a Cu plate, bolted to stage 1 of CCR 1 to intercept the heat from the warm end of the wires. The 3 FGMs were attached with Kapton tape to the Cu equator ring at the approximate locations shown in Fig.~\ref{fig:inst}, each aligned to three orthogonal components of the ambient field.

\begin{figure}[htb]
\includegraphics[angle=-90, width=60mm]{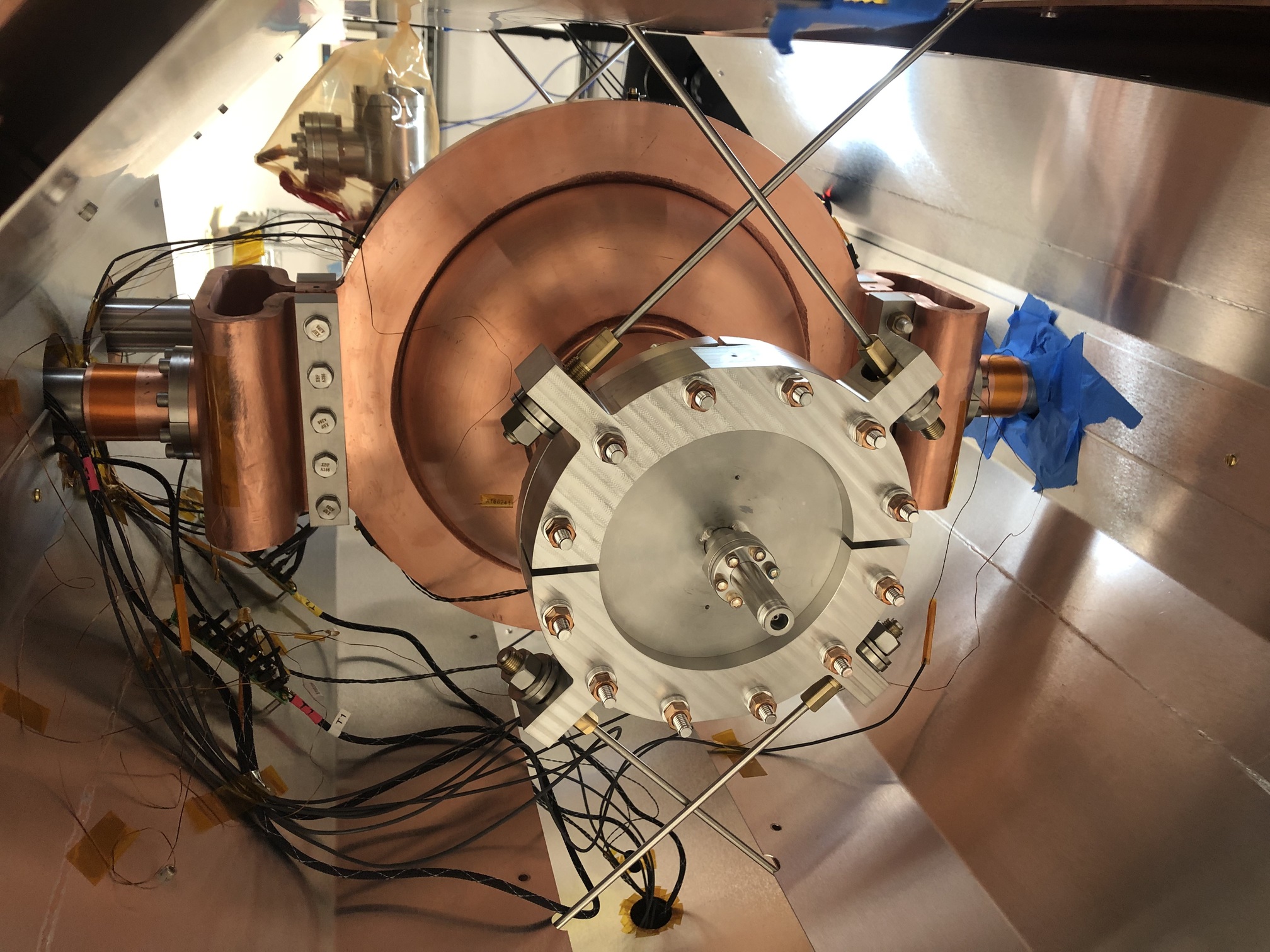}
\caption{\label{fig:cavGA} Picture of the prototype cavity being assembled in the HTC at General Atomics.}
\end{figure}

In addition, 7 low-inductance ceramic heaters (HTRs) were distributed as follows: one each on the top and bottom thermal shield sub-assemblies, one on the Cu bobbins mount plate and 4 on the cavity at the approximate locations shown in Fig.~\ref{fig:inst}. The heaters wires consisted of Cu twisted pairs and a section of the wires of the heaters attached to the cavity assembly was also epoxied to Cu bobbins mounted to the same Cu plate which has the bobbins for the RTDs' leads. Figure~\ref{fig:cavGA} shows a picture of the prototype cavity being assembled inside the HTC at General Atomics. The cavity and stage 2 of the CCRs were wrapped with 15 layers of multi-layered insulation (MLI). No MLI was wrapped around the thermal shield.

\section{\label{sec:cavHTC}Results from conduction-cooled prototype cavity}
After the last test in LHe at Jefferson Lab, the cavity was shipped under static vacuum to General Atomics where a $\sim 50$~m$^2$ laboratory space was setup to assemble the cavity into the HTC and to perform the cool-down, data acquisition and rf testing. Both the data acquisition programs, written in LabVIEW, and the low-level rf (LLRF) system were adapted from the ones used at Jefferson Lab for SRF cavity cool-down and testing~\cite{powers:srf2019}. The room with the HTC was also equipped with 5 x-ray area monitors (Model 375, Ludlum Measurements, Inc., Sweetwater, TX, USA). After the assembly of the HTC was completed, it was evacuated to $1.6\times 10^{-3}$~Pa prior to cool-down.

\subsection{\label{subsec:cooldown}Initial cooldown}
\begin{figure}[htb]
\includegraphics[width=86mm]{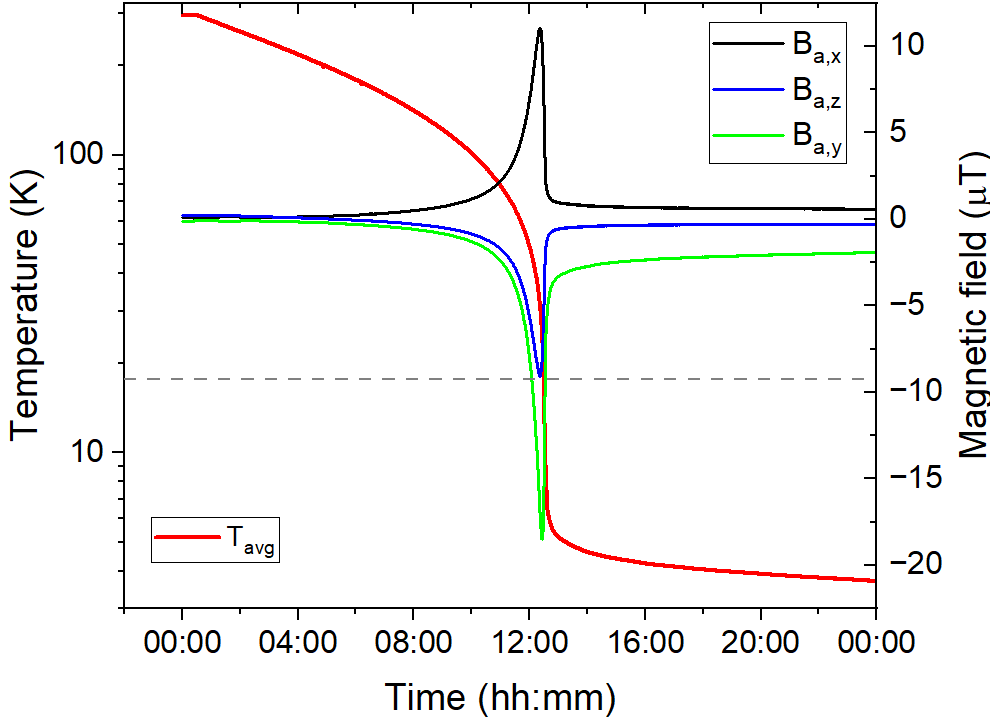}
\caption{\label{fig:cooldown1} Average temperature and ambient magnetic field at the cavity equator as a function of time, starting at the beginning of the cool-down, during the first cool-down of the cavity in the HTC. The dashed line is at $T_c \sim 17.7$~K.}
\end{figure}

The ambient magnetic field at the cavity equator prior to cool-down was $B_{a,x}=0.09$~$\mu$T, $B_{a,y}=-0.13$~$\mu$T and $B_{a,z}=-0.22$~$\mu$T. Figure~\ref{fig:cooldown1} shows the average temperature of the equator ring, $T_{avg}$ and the ambient magnetic field at the equator during the cool-down. The temperature measured by RTD4 (shown in Fig.~\ref{fig:inst}) and the RTD on stage 2 of CCR 1 was significantly higher than nearby sensors, likely due to improper mounting of those two sensors. $T_{avg}$ reached $\sim 3.7$~K after one day and it took three more days for $T_{avg}$ to reach 2.76~K. The temperature of the cavity beam tubes was within 0.1~K of $T_{avg}$.
Based on the CCRs capacity map, it is estimated that the total static heat load to stage 2 of the CCRs is $\sim1$~W, and that the static heat loads on stage 1 of CCR 2 and CCR 3 are $\sim40$~W and $\sim65$~W, respectively.
A significant change in the ambient magnetic field at the cavity equator ring occurred during cool-down, likely due to thermoelectric currents for bimetallic surfaces, which can result in spontaneous magnetic flux upon cooling down below critical temperatures of 18~K for Nb$_3$Sn and 9~K for Nb~\cite{VANHARLINGEN1982, Thermoelectric_SC}. During the cool-down, the magnitude of $B_a$ reached a maximum of $\sim 22$~$\mu$T at $T_{avg}=29$~K.

\subsection{\label{subsec:rf}Rf test results}
\begin{figure}[htb]
\includegraphics[width=86mm]{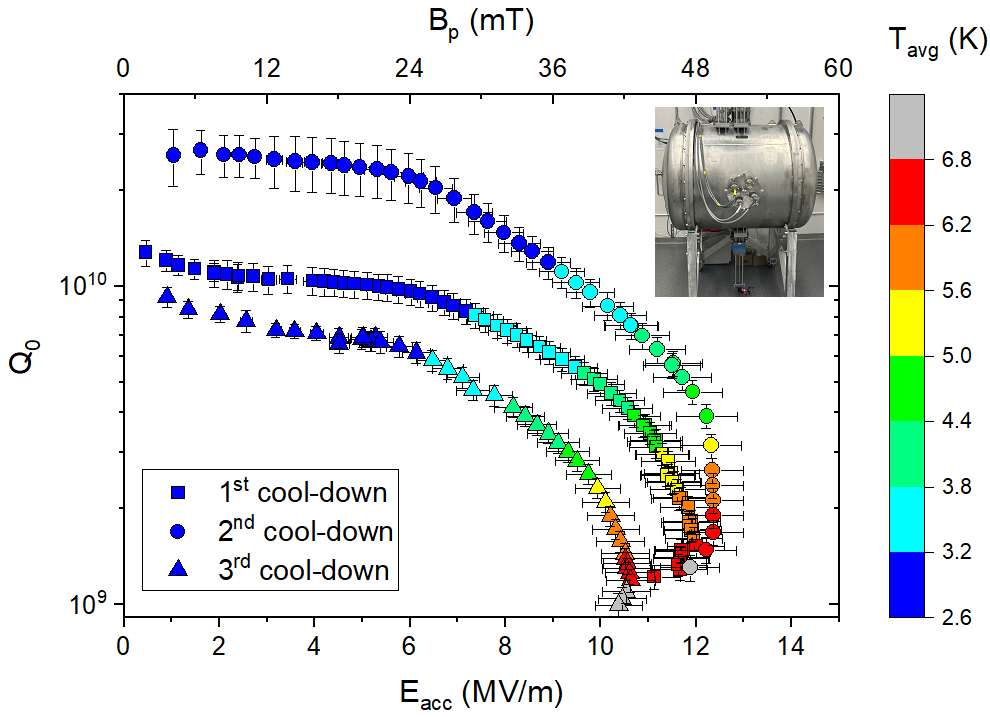}
\caption{\label{fig:QvsE} $Q_0$ as a function of the rf field for the conduction-cooled prototype cavity measured at General Atomics for different cool-down conditions summarized in Table~\ref{table:cooldown}. All the data points were measured in steady-state conditions and the symbols' colors correspond to the average equator ring temperature. The cavity performance was limited by thermal instability beyond the lowest measured $Q_0$-value. The inset shows a photo of the HTC.}
\end{figure}

Figure~\ref{fig:QvsE} shows a plot of $Q_0$ measured as a function of the rf field after three different cool-downs in the HTC. The cool-down conditions are summarized in Table~\ref{table:cooldown} and described in details in what follows. Each data point in Fig.~\ref{fig:QvsE} is taken in steady-state conditions, achieved $2-4$~min after each step increasing the rf field. Unlike when cooling with a LHe bath, the cavity temperature increases significantly with increasing rf power dissipation, $P_{rf}$, as the CCRs' cooling power increases with temperature. No x-ray dose was detected by the area monitors above their sensitivity limit of 1~$\mu$Sv/h. Additional Geiger-M\"{u}ller tubes were placed close to the HTC during one of the rf tests and no radiation was detected.

\begin{figure}[htb]
\includegraphics[width=86mm]{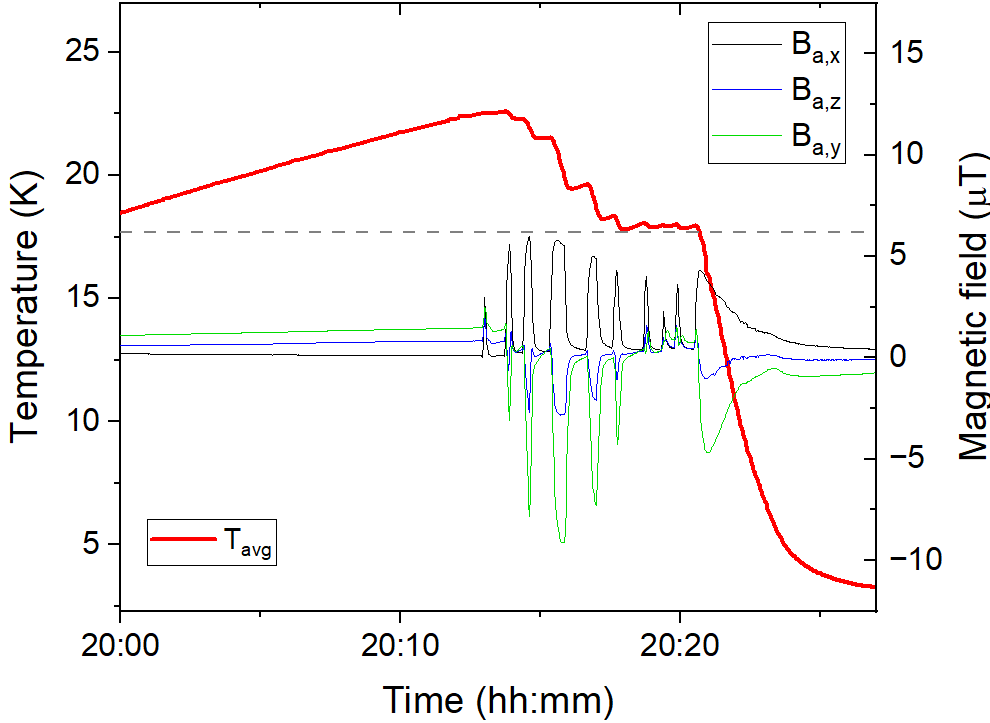}
\caption{\label{fig:2ndcooldown} $T_{avg}$ and $B_a$ along three orthogonal directions at the cavity equator during the second cool-down for which the CCRs were power cycled on and off. The dashed line is at $T_c \sim 17.7$~K.}
\end{figure}

After the rf test, a temperature cycle through $T_c$ was performed by turning off the CCRs until $T_{avg}$ reached about 23~K. Then the CCRs were power-cycled on and off to slowly lower the cavity temperature to 17.5~K, below which the CCRs were kept on, cooling the cavity to a base temperature $T_{avg}=2.65$~K. Figure~\ref{fig:2ndcooldown} shows $T_{avg}$ and the ambient magnetic field at the cavity equator ring during this second cool-down. The slower cool-down across $T_c$, compared to the $1^{\text{st}}$ cool-down, resulted in lower $B_a$. The results from the rf test after the $2^{\text{nd}}$ cool-down are shown in Fig.~\ref{fig:QvsE}. Lower $B_a$ resulted in a higher $Q_0$ at low rf field, compared to the $1^{\text{st}}$ test, because of a lower $R_{res}$. The maximum $B_p$ was 50~mT, corresponding to $E_{acc}=12.4$~MV/m. Increasing further the rf input power resulted in a sharp increase in $P_{rf}$ and both $Q_0$ and $B_p$ dropped as the cavity becomes undercoupled. Thermal instability was encountered above $P_{rf} \sim 20$~W. The thermal instability manifested as a sudden drop in both $E_{acc}$ and $T_{avg}$ and jumps in $B_a$, at which point the rf input power was quickly reduced to prevent further thermal runaway. There was no hysteresis in the $Q_0(B_p)$ curve upon reducing the rf field. The maximum temperature difference around the cavity equator ring was $< (70 \pm 10)$~mK. The maximum difference between the temperature of the beam tube on the side with the CCR, $T_{BT2}$, minus $T_{avg}$ was $< (-50 \pm 30)$~mK and the maximum difference between the temperature measured on the beam tube on the side without CCR, $T_{BT1}$, minus $T_{avg}$ was $< (170 \pm 30)$~mK.
To better determine the maximum stable rf heat load for long-term operation of the cavity, it was attempted to hold the cavity rf heat load for 1~h, starting at $P_{rf} \sim 20$~W but thermal instability occurred after 20 min. The test was repeated decreasing $P_{rf}$ in 0.5~W steps until it was successfully completed at $P_{rf} \sim 18.5$~W. All temperature readings were at steady state during those conditions, $T_{avg}$ was between $6.45 - 6.50$~K throughout the 1~h long test.

\begin{figure}[htb]
\includegraphics[width=86mm]{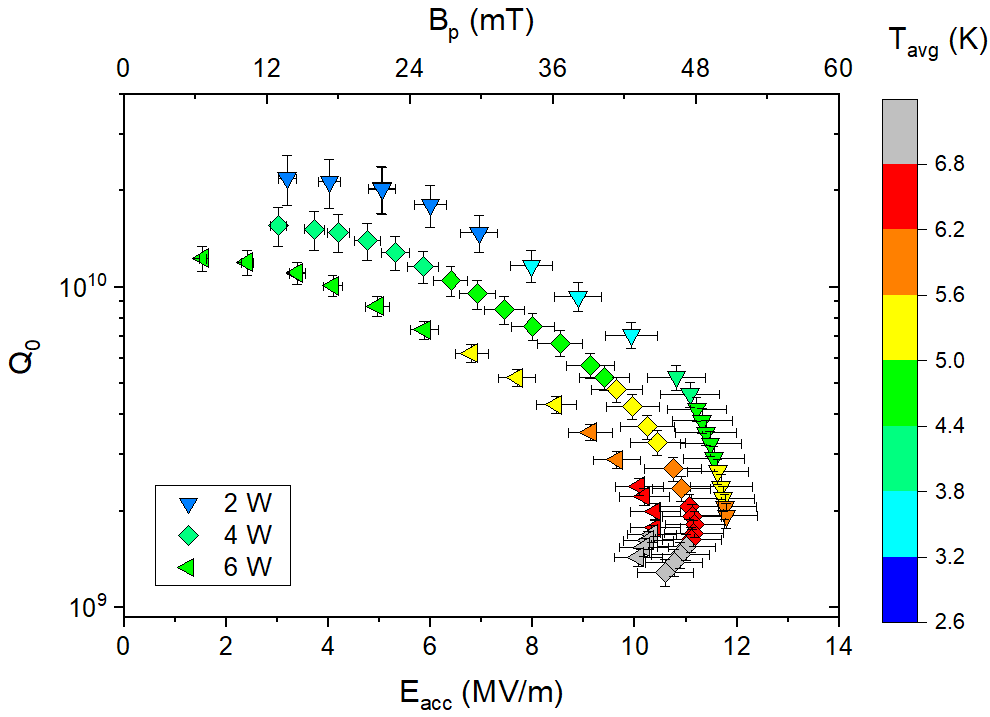}
\caption{\label{fig:QvsE_HTR6} $Q_0$ as a function of the rf field for the conduction-cooled prototype cavity measured at General Atomics after the second cool-down for different values of dc heat applied by HTR 6. All the data points were measured in steady-state conditions and the symbols' colors correspond to the average equator ring temperature. The cavity performance was limited by thermal instability beyond the lowest measured $Q_0$-value, except for $P_{dc}=2$~W for which the cavity test was stopped before reaching thermal instability.}
\end{figure}

\begin{figure}[htb]
\includegraphics[width=86mm]{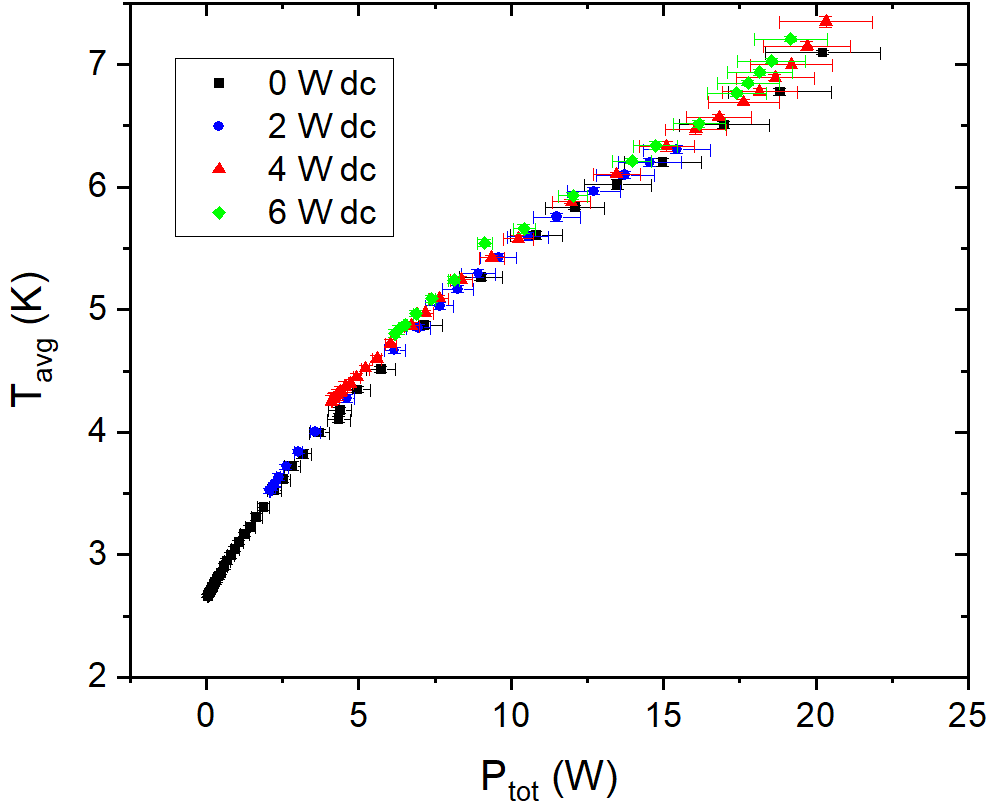}
\caption{\label{fig:Tavg_vs_Pt} $T_{avg}$ as a function of the total heat load, $P_{tot} = P_{rf} + P_{dc}$, for different values of $P_{dc}$. For each value of $P_{dc}$, $P_{rf}$ was increased up to the thermal stability limit, except for $P_{dc}=2$~W.}
\end{figure}

Additional $Q_0(B_p)$ curves were measured while applying $P_{dc}=2-6$~W to HTR 6 and the results are shown in Fig.~\ref{fig:QvsE_HTR6}. Further rf tests were done while applying $2-6$~W to HTR 4 and the results were similar to those shown in Fig.~\ref{fig:QvsE_HTR6}, for the same heater power. Figure~\ref{fig:Tavg_vs_Pt} shows a plot of the $T_{avg}$ as a function of the sum of the cavity rf heat load plus any dc heat from HTR 6, $P_{tot}$, showing that adding dc heat to the Cu layer shifts $T_{avg}$ at the lowest $P_{rf}$ upward and that the total maximum stable heat load remains constant.

\begin{figure}[htb]
\includegraphics[width=86mm]{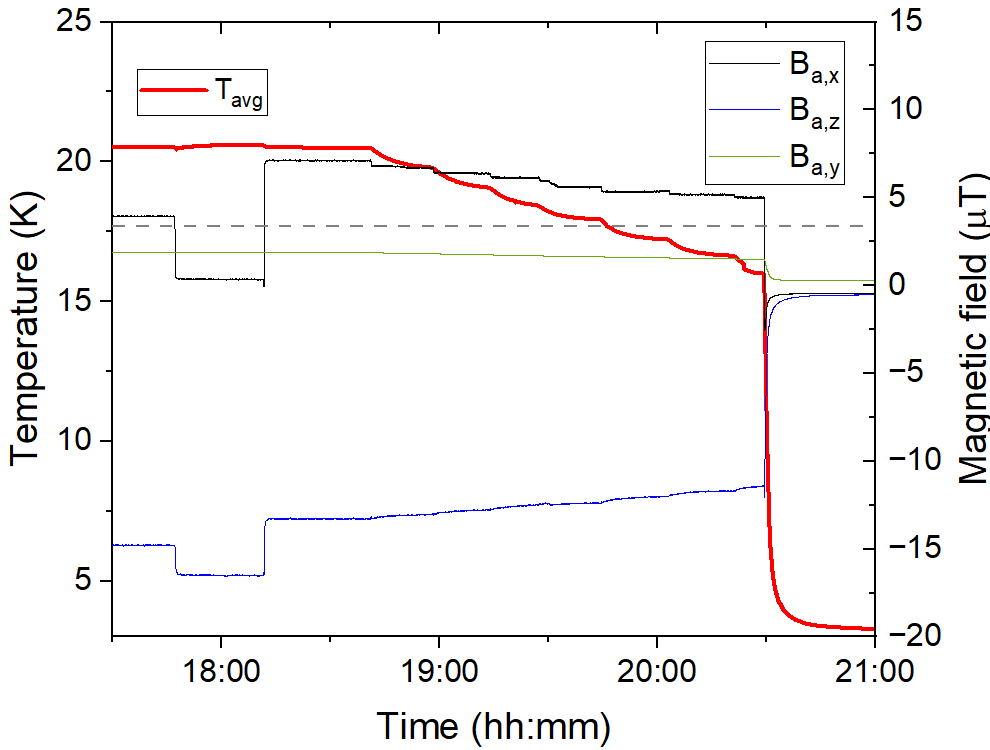}
\caption{\label{fig:3rdcooldown} $T_{avg}$ and $B_a$ along three orthogonal directions at the cavity equator during the third cool-down for which the cavity temperature was controlled by adjusting the power to heaters 5-7. The heaters location is shown in Fig.~\ref{fig:inst}. The dashed line is at $T_c \sim 17.7$~K.}
\end{figure}

Another temperature cycle through $T_c$ was performed by applying 11.5~W to heaters 5 and 7 and 31~W to heater 6 to reach $T_{avg}\sim20$~K. The three heaters are mounted on the cavity close to the TLAs as shown in Fig.~\ref{fig:inst}, and the heaters' power was chosen to minimize the magnitude of $B_a$. The heaters' power was then reduced in small steps until $T_{avg}$ was 16~K, below which the heaters were turned off. $T_{avg}$ and $B_a$ during the 3$^{\text{rd}}$ cool-down across $T_c$ are shown in Fig.~\ref{fig:3rdcooldown}. The cavity was cooled to a base temperature $T_{avg}=2.65$~K and the high-power rf test was repeated, the results being shown in Fig.~\ref{fig:QvsE}. The ambient magnetic field during the 3$^{\text{rd}}$ cool-down was significantly higher than during the 2$^{\text{nd}}$ cool-down, as listed in Table~\ref{table:cooldown}, resulting in a higher $R_{res}$ and lower $Q_0$. It was verified that thermo-currents rather than the heaters themselves or their leads were responsible for producing the high ambient magnetic field during cool-down, since the change in $B_a$ was less than $\sim 0.1$~$\mu$T when applying up to 20~W of power to each heater with the cavity at room temperature.
Table~\ref{table:cooldown} gives a summary of $B_a$ at the cavity equator ring and the temperature gradient along the cavity axis between the beam tube on the side without CCR and the equator ring, $dT/dy|_1$, and between the equator ring and the beam tube on the side with CCR, $dT/dy|_2$, when $T_{avg} \sim 17.7$~K, during the cool-down prior to the high-power rf tests with results shown in Fig.~\ref{fig:QvsE}. Overall, the cool-down data show a correlation between high $B_a$ and a low $Q_0$, because of a higher $R_{res}$. The cool-down data also suggest a complex distribution of the magnetic field and temperature gradients at the cavity, preventing simple quantitative estimates of the $R_{res}$ from such data. For example, it is known that the trapped flux sensitivity depends on the orientation of $B_a$ with respect to the cavity axis~\cite{kugeler_PRAB2020, Miyazaki_PRAB2021}.

\begin{figure}[htb]
\includegraphics[width=86mm]{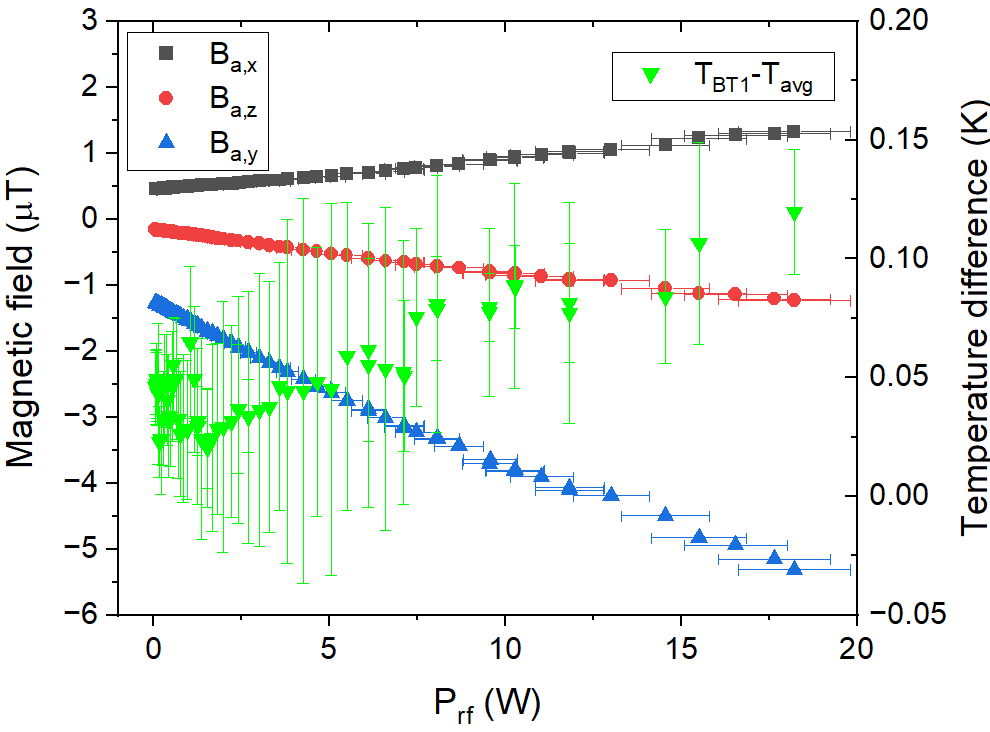}
\caption{\label{fig:Ba_vs_Prf} $B_a$ at different locations of the cavity equator ring, along three orthogonal directions, and temperature difference $T_{BT1}-T_{avg}$ as a function of $P_{rf}$, measured during the high-power rf test after the $1^{\text{st}}$ cool-down.}
\end{figure}

It was noticed during the rf tests that the magnitude of the ambient field at the cavity equator ring increased nearly linearly with increasing $P_{rf}$ as shown, as an example, in Fig.~\ref{fig:Ba_vs_Prf} for the rf test after the $1^{\text{st}}$ cool-down. For the same value of $P_{rf}$, the magnitude of $B_a$ increased linearly with increasing $P_{dc}$ also. Figure~\ref{fig:Ba_vs_Prf} shows a correlation between the difference $T_{BT1} - T_{avg}$ and $B_a$.

\begin{figure}[htb]
\includegraphics[width=86mm]{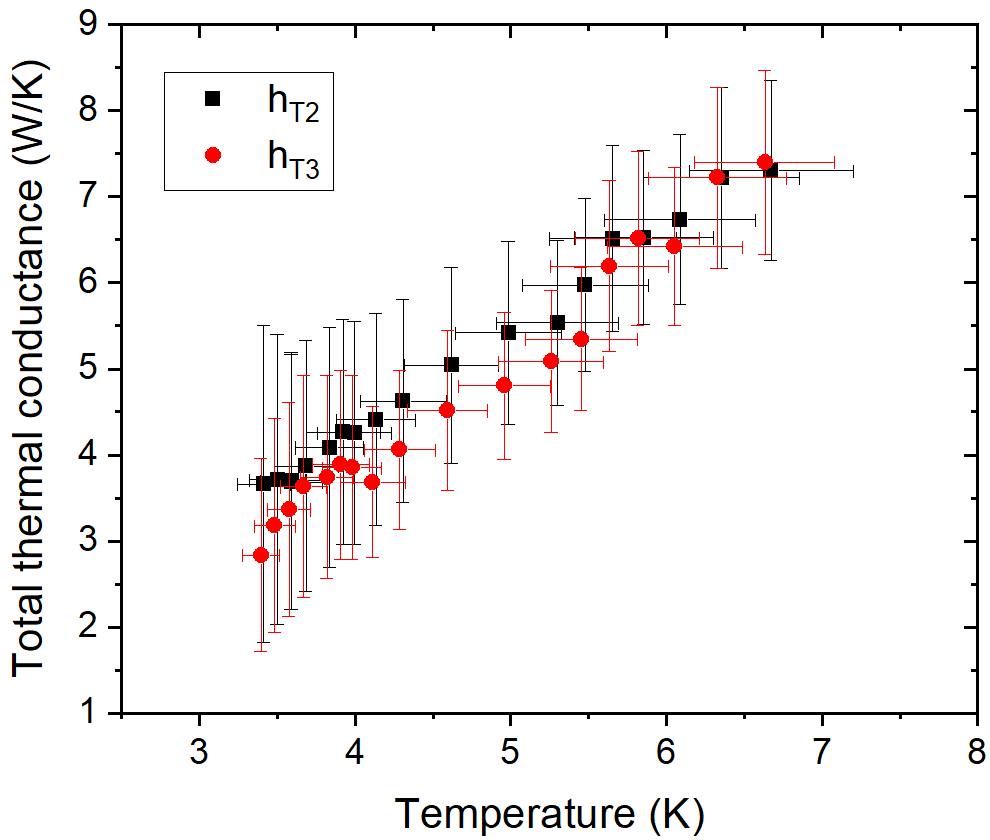}
\caption{\label{fig:thermalcond} Total thermal conductance between the cavity and CCRs 2 and 3 as a function of temperature obtained from the data measured during the high-power rf test after the $2^{\text{nd}}$ cool-down.}
\end{figure}

The rf heat load and temperatures measured during the high-power rf test after the $2^{\text{nd}}$ cool-down allowed calculating the total thermal conductance, $h_T$, of the system comprising of the interface between stage 2 of the CCR and the TLA, the TLA and the interface between the TLA and the cavity, assuming that each of the 3 CCRs absorbs 1/3 of $P_{rf}$, which is acceptable given the uniformity of the temperature in the Cu layer. Figure~\ref{fig:thermalcond} shows a plot of $h_{T3} = P_{rf}/3(T_{avg}-T_{CCR3})$ and $h_{T2} = P_{rf}/3(T_{BT2}-T_{CCR2})$ as a function of  $T_{avg3}=(T_{avg}+T_{CCR3}+T_{TLA3})/3$ and $T_{avg2}=(T_{BT2}+T_{CCR2}+T_{TLA2})/3$, respectively, where $T_{CCR}$ is the temperature of stage 2 of the CCR and $T_{TLA}$ is the temperature of the TLA.

\begin{table}
\caption{\label{table:cooldown}
Ambient magnetic field at the cavity equator ring and temperature gradient at either side of the equator ring when $T_{avg} = 17.7-17.8$~K for each of the three cool-downs. The $x$, $y$ and $z$ directions are shown in Fig.~\ref{fig:inst}.}
\begin{ruledtabular}
\begin{tabular}{>{\centering\arraybackslash}p{1.4cm}>{\centering\arraybackslash}p{1cm}>{\centering\arraybackslash}p{1cm}>{\centering\arraybackslash}p{0.8cm}>{\centering\arraybackslash}p{0.8cm}>{\centering\arraybackslash}p{0.8cm}}
\textrm{Cool-down No.}&
\textrm{$dT/dy|_1$ (K/m)}&
\textrm{$dT/dy|_2$ (K/m)}&
\textrm{$B_{a,x}$ ($\mu$T)}&
\textrm{$B_{a,y}$ ($\mu$T)}&
\textrm{$B_{a,z}$ ($\mu$T})\\
\hline
1 & -5.1 & 8.1 & 6.2 & -15 & -4.6\\
2 & -5.1 & 6.3 & 4.3 & -1.7 & -0.2\\
3 & 3.0 & -3.9 & 5.4 & 1.6 & -12.3\\
\end{tabular}
\end{ruledtabular}
\end{table}

\subsection{\label{subsec:microphonics}Microphonics}
The digital LLRF system used to measure the cavity $Q_0(B_p)$ provides analog outputs of the in-phase, $I(t)$, and quadrature, $Q(t)$, amplitudes of the transmitted voltage from the cavity. Because we are using a frequency tracking rf system, the change in frequency relative to the center frequency of the cavity resonance, $\delta f(t)$, can be calculated as~\cite{Powers_microphonics}:

\begin{equation}
\delta f(t) = \frac {Q(t) \frac{dI}{dt} - I(t) \frac{dQ}{dt} }{2 \pi \left(I^2 + Q^2 \right)}.
\label{Eq2}
\end{equation}

\begin{figure}[htb]
\includegraphics[width=86mm]{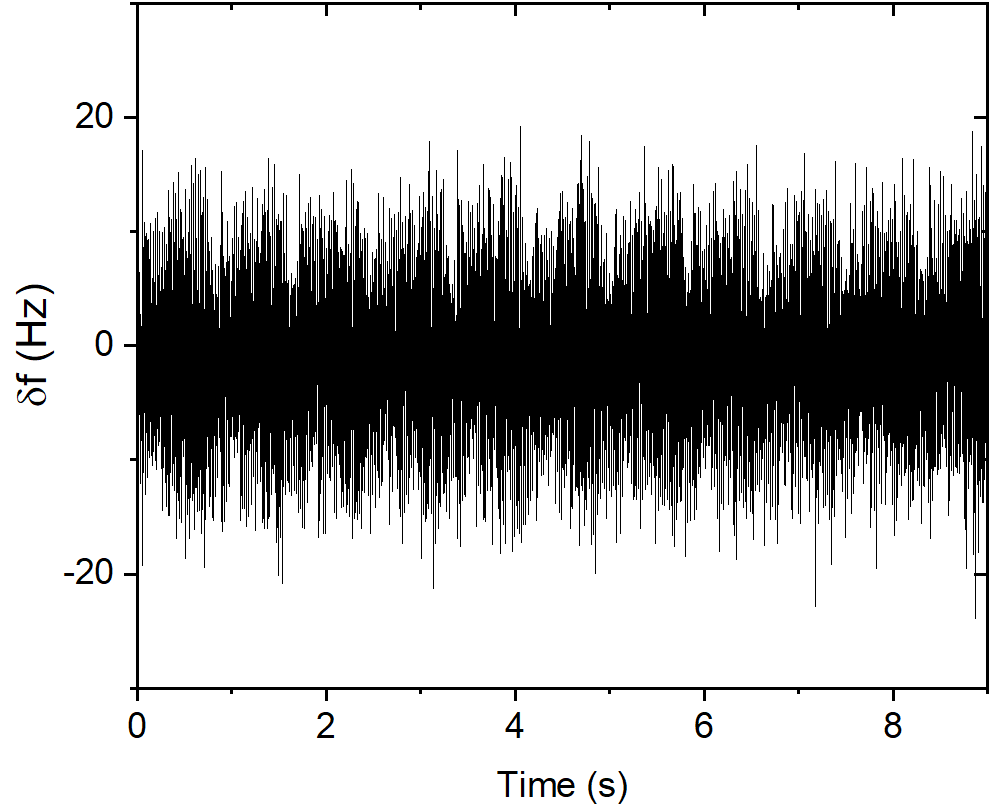}
\caption{\label{fig:deltaf} Snapshot of the shift in the resonant frequency of the prototype cavity inside the HTC, due to microphonics.}
\end{figure}

\begin{figure}[htb]
\includegraphics[width=86mm]{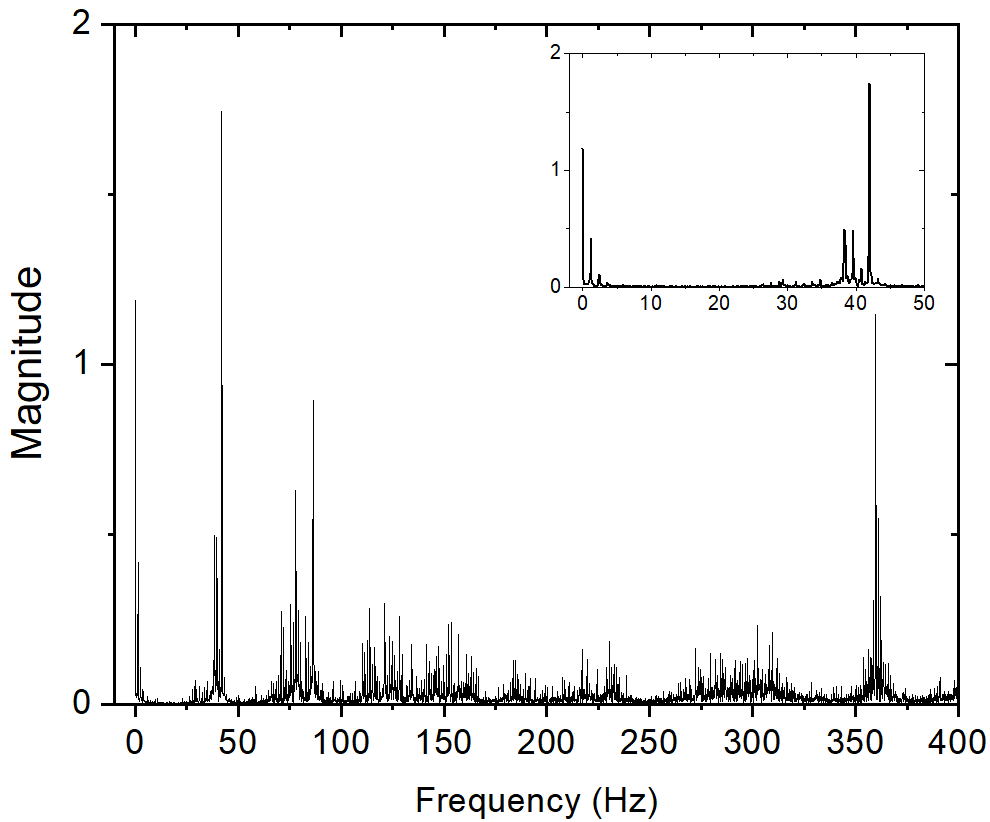}
\caption{\label{fig:FFT} Magnitude of the FFT of $\delta f(t)$ shown in Fig.~\ref{fig:deltaf}. The inset shows an expanded view of the peaks below 50 Hz.}
\end{figure}

The $I(t)$ and $Q(t)$ signals are digitized by a sound and vibration device (model USB-4431, National Instruments, Austin, TX, USA) connected to a computer, acquiring data through a LabVIEW program. Multiple 3 min long data sets were acquired at $B_p=14$~mT with a rate of 20~kS/s. The data was filtered through a 1~kHz low-pass filter. Figure~\ref{fig:deltaf} shows a snapshot of $\delta f(t)$ acquired with the 3 CCR and the turbo-pump cart pumping on the vacuum vessel being on. The average peak excursion of $|\delta f(t)|$ from 3 sets of 3 min-long data was $\delta f_{pk}=(23.0 \pm 0.9)$~Hz.
Figure~\ref{fig:FFT} shows a fast-Fourier transform (FFT) of $\delta f(t)$. The frequencies of the three highest peaks are 42~Hz, 360~Hz and 1.2~Hz. 
Additional data were taken after turning off CCR2 and both CCR2 and CCR3, in which case $\delta f_{pk}$ decreased to $(20.8 \pm 0.4)$~Hz and $(18.7 \pm 1.3)$~Hz, respectively.  No significant changes were observed when turning off the turbo-pump cart. 

\begin{figure}[htb]
\includegraphics[width=86mm]{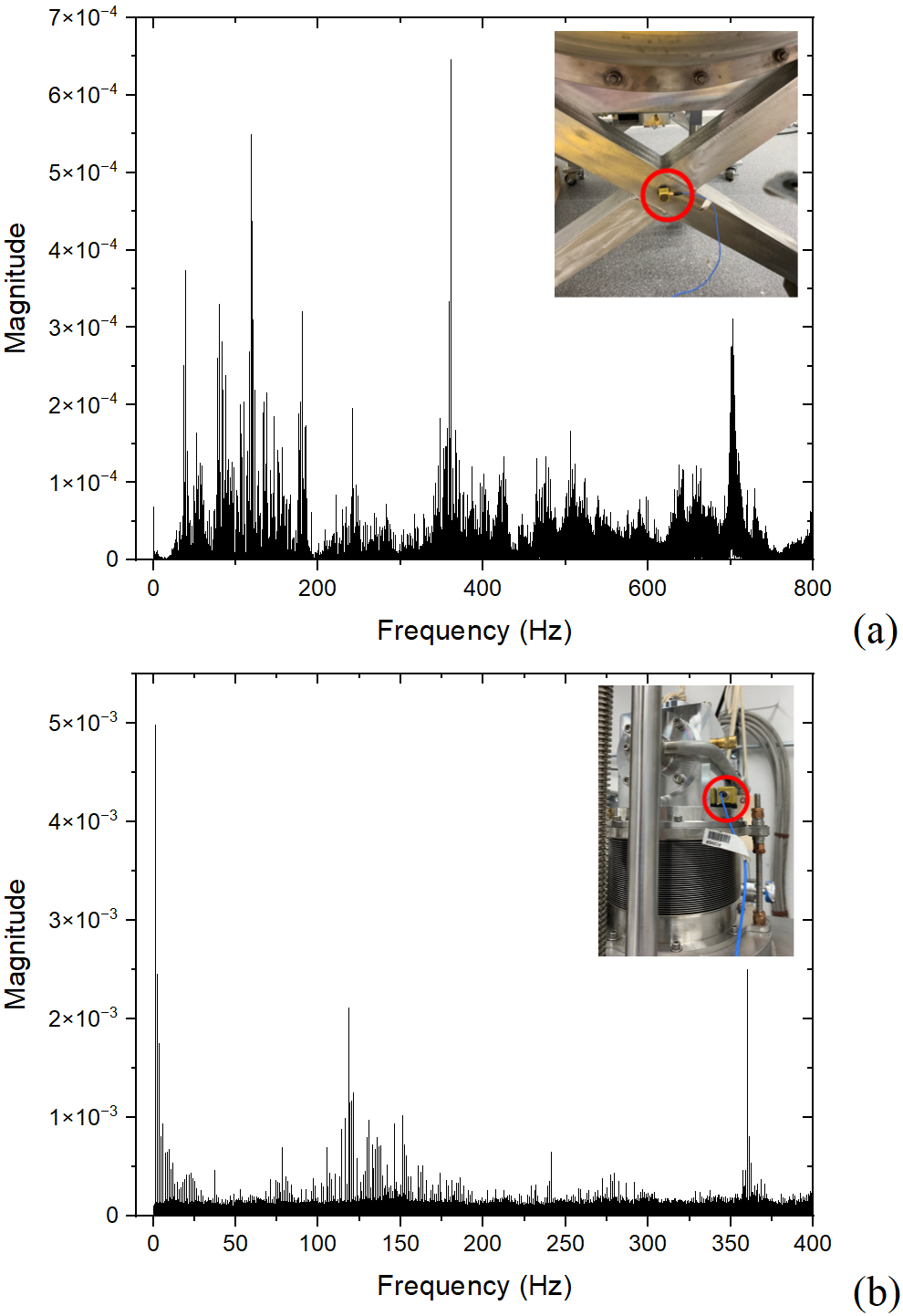}
\caption{\label{fig:accel} Magnitude of the FFT of the vertical acceleration measured on the HTC support cart (a) and on top of CCR1. The accelerometer is circled in red in the inset photos.}
\end{figure}

A tri-axial accelerometer (model 356B18, PCB Piezotronics, Inc., Depew, NY, USA) was used to measure the acceleration due to mechanical vibrations at different locations on the HTC. The data acquisition system consisted of the same setup used to acquire the transmitted voltage signals from the LLRF. 3 min-long data sets with a rate of 2~kS/s were measured at each location. Figure~\ref{fig:accel} shows the magnitude of the FFT of the acceleration in the vertical direction at two representative locations: one on the HTC support cart and one on top of CCR1. The peak acceleration amplitude measured on the support cart was $0.06g$, where $g$ is the acceleration of gravity and the spectrum matches that of the acceleration measured on the floor. The top three frequencies with the highest magnitude in Fig.~\ref{fig:accel}(a) are at 361.1~Hz, 118.8~Hz and 39.6~Hz.
The peak acceleration amplitude measured on top of CCR1 was $1.7g$ and the top three frequencies with the highest magnitude in Fig.~\ref{fig:accel}(b) are at 360~Hz, 1.2~Hz and 2.4~Hz.

The peak at 1.2~Hz in the spectra shown in Figs.~\ref{fig:FFT} and \ref{fig:accel} corresponds to the  frequency of the displacer moving inside the CCRs' cold stage. The peak at $\sim40$~Hz and its higher harmonics were caused by the CCRs' He compressors, which were located in the same room as the HTC, $\sim 2$~m away.

\section{\label{sec:disc}Data Analysis and Discussion}
\subsection{\label{subsec:1} Surface resistance field-dependence}
\begin{figure}[htb]
\includegraphics[width=86mm]{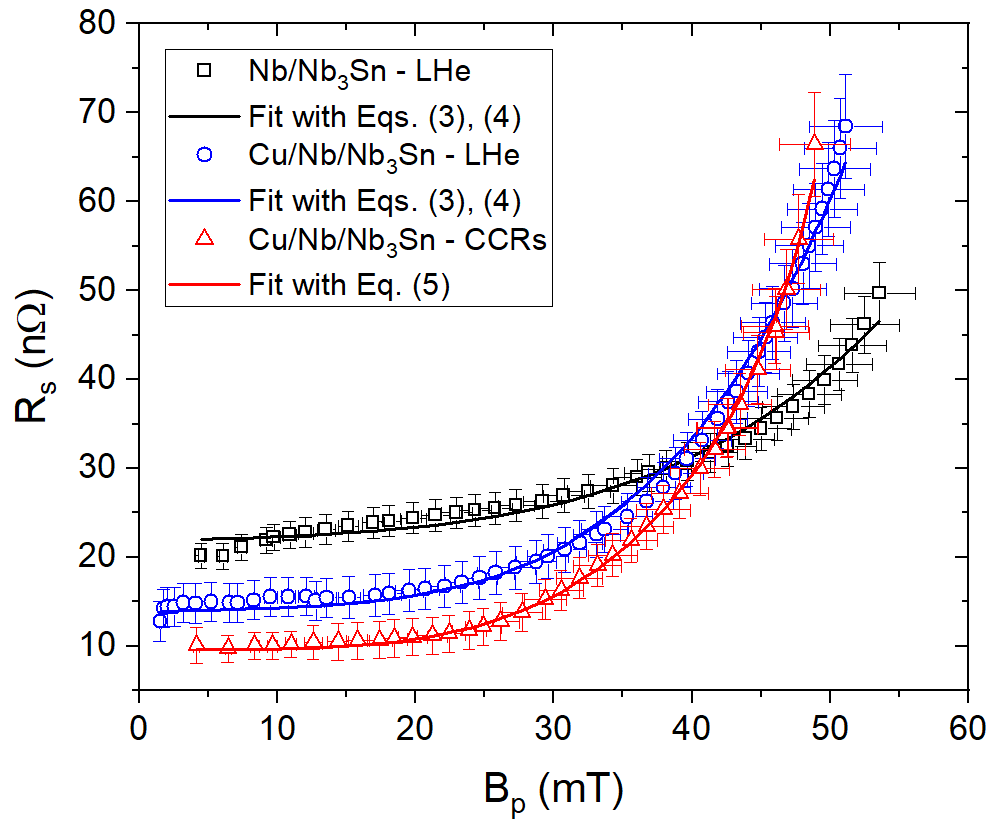}
\caption{\label{fig:TFBM} $R_s(B_p)$ data for the Nb prototype cavity cavity with Nb$_3$Sn film on the inner surface cooled by LHe and with Cu outer layer tested both in LHe and with cryocoolers. Solid lines are least-squares fit to each data set.}
\end{figure}

The data show a stronger increase of the surface resistance with increasing $B_p$ after the Cu layer was deposited onto the outer cavity surface. The data have been analyzed with a one-dimensional thermal feedback model~\cite{Bauer_PhysicaC, Gurevich_RAST} for three cases: (1) Nb/Nb$_3$Sn cooled by LHe, (2) Cu/Nb/Nb$_3$Sn cooled by LHe and (3) Cu/Nb/Nb$_3$Sn cooled by CCRs. 
The following equations were solved numerically to find the temperature of the inner ($T_m$) and outer ($T_s$) surfaces of the cavity as a function of $B_p$ for case 1:

\begin{equation}
\frac{1}{2}R_s(T_m, B_p) \left( \frac{B_p}{\mu_0} \right)^2 = h_{cav}(T_m-T_s)
\label{Eq3}
\end{equation}
\begin{equation}
h_{cav}(T_m-T_s) = h_1(T_s - T_0)^{2.5},
\label{Eq4}
\end{equation}
where $T_0=4.3$~K is the LHe bath temperature and $h_1=1$~kW/(m$^2$ K$^{2.5}$) is the nucleate boiling heat transfer coefficient in He I~\cite{VanSciver1986}. $\frac{1}{h_{cav}}=\frac{d_{Nb}}{\kappa_{Nb}}+\frac{d_{Nb_3Sn}}{\kappa_{Nb_3Sn}}$ is the thermal resistance through the cavity wall, $d_{Nb}=3.8$~mm is the thickness of the Nb substrate, $d_{Nb_3Sn}=3$~$\mu$m is the thickness of the Nb$_3$Sn film, $\kappa_{Nb}=88$~W/(m~K) is the thermal conductivity of Nb at 4.3~K and $\kappa_{Nb_3Sn}=49$~mW/(m~K) is the thermal conductivity of Nb$_3$Sn at 4.3~K~\cite{Nb3Sn_thermcond}.
For case 2 the term $d_{Cu}/\kappa_{Cu}$ is added to the cavity thermal resistance, where $d_{Cu}=7$~mm is the thickness of the Cu layer and $\kappa_{Cu}=2365$~W/(m K) is the thermal conductivity of the electroplated Cu at 4.3~K.
For case 3 the following equation is solved numerically:
\begin{equation}
\frac{1}{2}R_s(T_m, B_p) \left( \frac{B_p}{\mu_0} \right)^2 = h_{cav}(T_{avg})[T_m-T_{avg}(B_p)],
\label{Eq5}
\end{equation}
reflecting the fact that the temperature of the outer surface is no longer a constant. The thermal conductivities of Cu, Nb and Nb$_3$Sn at $T_{avg}$ are used to calculate $h_{cav}(T_{avg})$.
The following phenomenological field dependence of $R_s$, based on a series expansion of $R_{res}(B_p^2)$ up to the second order, was used in Eqs.~(\ref{Eq3}-\ref{Eq5}):

\begin{multline}
R_s(T_m,B_p)=R_{BCS}(T_m) + \\
R_{res}\left[1+\gamma \left( \frac{B_p}{B_c} \right)^2 + \delta \left( \frac{B_p}{B_c} \right)^4 \right],
\label{Eq6}
\end{multline}
where $B_c=500$~mT is the thermodynamic critical field of Nb$_3$Sn at 0~K~\cite{Posen_PhysRevSTAB.17.112001}. The following simplified expression, valid for $T < T_c/2$, was considered for $R_{BCS}(T_m)$~\cite{Gurevich_RAST}:

\begin{equation}
R_{BCS}(T_m) = \frac{A}{T_m}e^{-\Delta/k_BT_m},
\label{Eq7}
\end{equation}
with $A=1.7\times 10^{-4}$~$\Omega$K and $\Delta /k_B = 41.6$~K for case 1, $\Delta /k_B = 39$~K for cases 2 and 3. A least-squares fit of the data for the 2$^{\text{nd}}$ coating shown in Fig.~\ref{fig:QvsE_Nb3Sn} for case 1 and of the data shown in Fig.~\ref{fig:QvsE_CuNbNb3Sn} for case 2 with Eqs.~(\ref{Eq3}) and (\ref{Eq4}) was done to determine the value of the fit parameters $R_{res}$, $\gamma$ and $\delta$. For case 1 we obtained $R_{res}=20.5$~n$\Omega$, $\gamma=24$, $\delta = 6.6 \times 10^3$, whereas $R_{res}=8.9$~n$\Omega$, $\gamma=3\times10^{-6}$, $\delta = 4.8\times 10^4$ for case 2. For case 3, the data set $R_s(B_p)$ and $T_{avg}(B_p)$ from Fig.~\ref{fig:QvsE} for the 2$^{\text{nd}}$ cool-down was used for a least-squares fit with Eq.~(\ref{Eq5}). Since the cavity was kept under vacuum between the rf tests considered for cases 2 and 3, the values of $\gamma$ and $\delta$ were kept constant and $R_{res}=9.6$~n$\Omega$ was the only fit parameter for case 3. Figure~\ref{fig:TFBM} shows $R_s(B_p)$ for the three cases considered, along with the results from the least-squares fits, which are in good agreement with the data. The higher $R_{res}$ for case 1 compared to cases 2 and 3 could be due to the higher $B_a$ along the cavity axis for case 1. 

Besides the field dependence of $R_s$ described by Eq.~(\ref{Eq6}), another possibility was considered multiplying $R_{BCS}(T_m)$ by a factor $\left[1+ \theta \left( \frac{B_p}{B_c} \right)^2 \right]$, where the $B_p^2$-term represents pair-breaking by the rf current, and neglecting the second-order term in the series of $R_{res}(B_p^2)$. However, the quality of the fits was poorer and the field dependence of $R_s$ could not be reproduced keeping the same values of $\theta$ and $\gamma$ between cases 2 and 3. Adding a fixed thermal boundary resistance term to the cavity thermal resistance, because of the Nb/Cu interface, as an additional fit parameter for cases 2 and 3 did not improve significantly the quality of the fit of the case 3 data. This suggests that the main source of the field dependence of $R_s$ at $B_p \ll B_c$ comes from the residual resistance. The field dependence of $R_{res}$ at $B_p \ll B_c$ may  result from weakly-coupled grain boundaries in Nb$_3$Sn~\cite{Nb3Sn_GB, Nb3Sn_GB2, Nb3Sn_GB3, Nb3Sn_GB4} or trapped vortices~\cite{GB_diss_1, GB_diss_2, GB_diss_3}.

The analysis of $R_s(B_p)$ data with Eqs. (\ref{Eq3}-\ref{Eq7}) suggests that the quality of the Nb$_3$Sn film degraded after depositing the thick Cu outer layer, given the increase of $\delta$ by a factor of $\sim7$ between cases 1 and 2. Similar results were measured in a 1.5~GHz Cu/Nb/Nb$_3$Sn single-cell cavity~\cite{Ciovati_SUST}. The quadratic term provides a much smaller contribution to the observed increase of $R_s$ with increasing rf field.

The superconducting properties of Nb$_3$Sn are very sensitive to strain~\cite{Godeke_2006, DeMarzi_2013, Ding_2021} and additional stress in the film occurs after depositing the thick Cu outer layer, because of the differential thermal contraction between Cu and the Nb substrate. Since the residual stresses in the film after formation on the Nb substrate are unknown, it is difficult to estimate the absolute strain value in the film after deposition of the Cu layer and cool-down to 4~K. If one neglects any initial residual stress, the maximum strain in the film from Fig.~\ref{fig:Cu_mech}(b) can be estimated to be of the order of $0.2\%$. Such strain value would result in a $\sim1\%$ decrease of $T_c$, consistent with the results from the cavity rf measurements at low field, and $\sim25\%$ decrease of $J_c$. Compositional and morphological changes across grain boundaries in Nb$_3$Sn makes them prime suspects for producing residual rf losses~\cite{Nb3Sn_GB, Gurevich_2017, Junki_Nb3Sn}. The additional strain from the cool-down stresses could result in weaker coupling among grain boundaries, producing an upturn of the field dependent residual rf losses.

\subsection{\label{subsec:2} Cu outer layer}
\begin{figure}[htb]
\includegraphics[width=86mm]{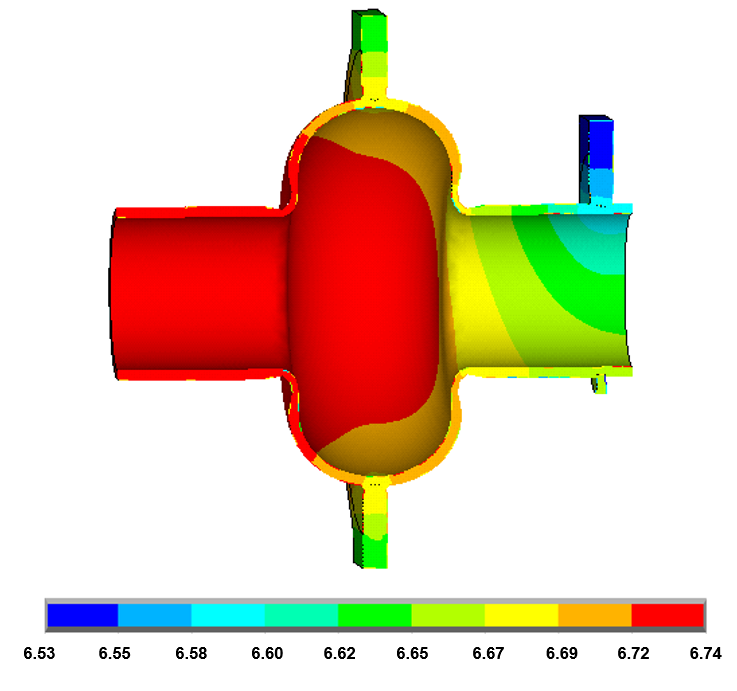}
\caption{\label{fig:ansys_thermal} Temperature distribution for the 952.6~MHz prototype cavity cooled by 3 CCRs, calculated with Ansys for $B_p = 47$~mT at the maximum $P_{rf} = 16.9$~W which resulted in a stable solution. The temperature color map scale is in units of Kelvin.}
\end{figure}

The thermal analysis of the prototype cavity discussed in Sec.~\ref{subsec:952eng} was repeated using Ansys in order to validate the simulation with the test results. The main difference compared to the initial analysis was the use of the overall thermal conductance values from Fig.~\ref{fig:thermalcond}, $h_T = 4.3$~W/K at 4~K and $h_T = 11.1$~W/K at 10~K, instead of $h_T = 16.8$~W/K at 4~K and $h_T = 36.5$~W/K at 10~K calculated from the specified TLA thermal conductance and the interface heat transfer coefficient assumed for the initial analysis. In addition, the stage 2 temperature versus cooling power curve shown in Fig.~\ref{fig:CCRheatmap}, which is $\sim10\%$ better than the nominal data from the vendor, was used in the revised analysis. The process to determine the highest rf heat load at which the finite-element analysis converged also had to be modified as follows: an iterative solving process was applied to all of the CCRs' stage 2 and a very small damping value of $\sim 0.05$, multiplying the difference in the stage 2 temperatures from the solution of one iteration step to the next, needed to be used when determining the stage 2 temperatures set for the subsequent iteration step. Figure~\ref{fig:ansys_thermal} shows the cavity temperature distribution at $B_p=47$~mT at the highest rf heat load of 16.9~W, which resulted in a converging solution with Ansys. A multiplying factor of 5.2 was applied to the field-dependent part of the empirical $R_s(B_p, T)$ used in the initial analysis, in order to match the measured rf heat load at 47~mT. This factor reflects the steeper increase of $R_s(B_p)$ than that from the empirical formula, which is based on data prior to the Cu coating. The maximum stable cavity temperature and rf heat load calculated with Ansys are consistent with the experimental results, given the uncertainty in the measurement of the overall thermal conductance. The maximum stable cavity temperature $T_{max}\sim 6.7$~K from the Ansys simulation is consistent with the temperature for which $dP_{rf}/dT$ exceeds $dh_T/dT$, meaning that above $T_{max}$ the rf heat can no longer be conducted effectively to the CCRs.

In order to better understand the role of the thick Cu outer layer in the maximum stable rf heat load that can be achieved, the Ansys thermal analysis was repeated with the following modifications: (i) the Cu layer was removed, (ii) the Cu equator ring and beam-tube attachment were replaced by 4~mm thick high-$RRR$ Nb equator ring and beam-tube attachment and (iii) the same empirical $R_s(B_p, T)$ used for the initial analyis (Fig.~\ref{fig:Rs_empirical}) was used in the simulation. The Nb equator ring and beam-tube attachment could be electron-beam welded to the 4~mm thick Nb cavity prior to Nb$_3$Sn coating. Using the same procedure mentioned above, the simulation resulted in a maximum stable cavity temperature $T_{max}\sim 5.5$~K at $B_p = 44$~mT and $P_{rf}=4.8$~W.

The results of the Ansys analysis show that the presence of a thick Cu outer layer significantly enhances the maximum stable rf heat load that could be achieved in a conduction cooled SRF cavity. However, the increase of the Nb$_3$Sn surface resistance with increasing $B_p$ after the deposition of the Cu layer did not allow to fully exploit the higher thermal stability towards achieving higher accelerating gradients. The presence of an outer layer with high thermal conductivity could prove to be particularly beneficial when considering the additional heat loads from the FPCs and from the beam pipes extending between the cavity flanges and the outside of a cryomodule to be used in an accelerator.
Another benefit of the Cu outer layer is the increase in the mechanical stability of the cavity, making it less sensitive to microphonics and less susceptive to cracking of the Nb$_3$Sn film. The occurrence of cracks in the film because of high stresses due to, for example, differential pressure, transportation or handling, would severely degrade $R_s$~\cite{Pudasaini_2020, eremeev:srf2019-mop015}.

Whereas the presence of the Cu outer layer helps achieving a uniform cool-down of the cavity across $T_c$ when cooled with LHe, higher temperature gradients occur when cooling with CCRs. The presence of thermoelectric currents between the Cu layer and the Nb, both during the cool-down and during the high-power rf test, is manifested by changes in the local magnetic field at the cavity surface. This phenomenon makes controlling the uniformity of the cavity temperature and therefore the local residual field by external heaters somewhat more challenging than in absence of the Cu layer. Modeling the thermoelectric currents and residual magnetic field for the prototype cavity in the HTC configuration will be part of future work.

A drawback of having the Cu outer layer on the Nb/Nb$_3$Sn cavity is that if any kind of damage to the Nb$_3$Sn thin film was to occur, a substantial rework of the cavity to remove the Cu layer would be necessary, given the high temperature required to form a new Nb$_3$Sn layer. The Cu ring and attachment plate could be cut by wire electro-discharge machining and the remainder of the Cu could be etched away by immersing the cavity in nitric acid. The Cu shell would need to be re-formed on the cavity outer surface after re-coating the inner surface with Nb$_3$Sn. Given the large margin in terms of the maximum rf  dissipation that was demonstrated with our setup, it should be possible to limit the amount of Cu to perhaps only the equator region and the beam-tube region where a high-power FPC would be connected to the cavity. Further engineering analysis to optimize the amount of Cu needed should allow finding a good compromise between handling the required heat loads and mitigating the drawbacks of the Cu layer mentioned above. 

\subsection{\label{subsec:3} Type of cryocooler}
Pulse-tube (PT) cryocoolers with the same nominal cooling capacity as that of the GM-type we have used for this project are also being considered for conduction cooling of SRF cavities. The main advantage of the PT-type CCR, compared to the GM-type, is the absence of a mechanical displacer, therefore reducing the amplitude of mechanical vibrations of stage 2~\cite{CCR_vibration}. However, the MW-class beam-power envisioned for conduction-cooled SRF accelerators sets the loaded-$Q$ of the cavity to the order of $\sim1\times 10^5$, corresponding to a bandwidth of $\sim 9$~kHz. A frequency shift of less than $\sim 1/10$ of the cavity bandwidth due to microphonics would not require any additional overhead in the high-power rf source. The $\delta f_{pk}$ we have measured on a cavity with 3 GM-type CCRs is similar to that measured on cavities cooled by superfluid He inside cryomodules and with loaded-$Q$ values of the order of $\sim1\times 10^7$~\cite{Bachimanchi:IPAC2015-THXB1}.

A major drawback of PT CCRs is that the wall-plug efficiency is significantly lower than that of GM-type, for the same cooling capacity. For example, the PT CCR model PT420 (Cryomech, Syracuse, NY, USA) requires a wall-plug power of 12.5~kW at 60~Hz~\cite{PT}, compared to 7.6~kW at 60~Hz for the GM-type model RDE-418D4~\cite{GM}, both providing a cooling power of 2~W at 4.2~K. In addition, the cost of a PT CCR is significantly higher than that of a GM CCR with the same nominal cooling capacity at 4.2~K.

\subsection{\label{subsec:4} Thermal link and interface materials}
High-purity (99.999$\%$) Al is an ideal material for a TLA because its thermal conductivity is higher than that of OFHC Cu at 4~K~\cite{Thermal_cond_Al}. This material is being used to fabricate thermal links for conduction-cooled SRF cavity application at Fermilab~\cite{Dhuley_TLA, Dhuley_thermalRes, Dhuley_2020}. We used TLAs made of oxygen-free Cu due to its lower cost, greater availability and to avoid a contact potential between the TLA and the Cu cavity and stage 2, which could produce thermo-currents. In addition, press-welding of Cu foils is a low-cost solution to fabricate a flexible TLA with no thermal resistance between the foils. A more complex and expensive method, such as electron-beam welding, would be needed to achieve a similar outcome if high-purity Al foils were to be used~\cite{Trollier}. Finally, given the proximity between the CCR stage 2 and the cavity, the thermal conductivity of the TLA material plays less of a role compared to the contact thermal resistance of the connections between the cavity and stage 2 of the CCR. This is evident by comparing Figs.~\ref{fig:h_TLA} and \ref{fig:TLA}.

Regarding the type of material to be used as part of the thermal interface between stage 2 of the CCR, the TLA and the cavity, both thin In foils and Apiezon N grease have been considered. No definitive answer was found in the literature about which of the two provides the highest thermal conductance across the surface of different metals at the temperatures of interest~\cite{Dillon_2017, JointTR1, JointTR2, JointTR3, JointTR4}. However, the disassembly of a joint in which an In foil is used between the mating surfaces is more complex than that in which grease is used. Using In as part of a joint often requires the use of custom tools to break the joint apart, to remove the In from the surfaces and, in the case of In being removed from Nb, soaking the Nb in nitric acid to remove staining. The chances of leaving scratches onto the surface during In removal is also higher, which would require additional mechanical polishing and/or machining.

\section{\label{sec:conc}Summary and outlook}
We have scaled the design of a 1~MW, 1~MeV conduction-cooled SRF linac to 915~MHz to take advantage of the availability of low-cost and efficient high-power, commercial rf magnetrons. The key component to be developed for such an accelerator is a single-cell SRF cavity cooled by conduction using CCRs and operating at $B_p=41.5$~mT. We have re-purposed a 952.6~MHz single-cell Nb cavity to be the prototype to develop the technologies needed to demonstrate the operation of the cavity, cooled by CCRs, at these fields. The cavity was coated with a Nb$_3$Sn film on the inner surface. A method was developed to produce a thick, high-purity Cu layer with three CCR attachment locations onto the cavity outer surface. The cavity performance was measured at Jefferson Lab in a vertical cryostat filled with LHe before and after the Nb$_3$Sn coating and after the formation of the Cu outer layer. The performance of the Nb$_3$Sn-coated cavity was limited by MP at $B_p\sim53$~mT.

A horizontal test cryostat was designed and procured to be able to measure the cavity performance when cooled by three GM-type CCRs. As part of the cryostat development, a thermal link made of high-purity Cu was successfully designed, procured and tested, meeting the requirements in terms of both mechanical flexibility and thermal conductance.
The cavity was assembled inside the HTC at General Atomics and its performance was measured for different cool-down conditions and for different values of additional heat applied by local heaters mounted onto the cavity. The cavity achieved $B_p=50$~mT, which is the highest peak surface magnetic field reached in a conduction-cooled cavity to date and exceeding the performance requirement with a $\sim20\%$ margin. The cavity was operated stably, in a steady state, for 1~h at a maximum rf heat load of 18.5~W. These results validated all the technical design choices that were made and represent an important stepping-stone towards the development of conduction-cooled SRF cavities for industrial linac applications.

Whereas the addition of the Cu outer layer significantly extends the thermal limit and mechanical stability of the cavity, a stronger increase of $R_s(B_p)$ was found after depositing such layer. This could be due to strain in the film resulting from the differential CTE between Cu and Nb. A better understanding of the effect of strain on the field-dependent surface resistance of Nb$_3$Sn, along with the search for alternative materials with high thermal conductivity at 4~K and a CTE close to that of Nb, as well as further engineering design of the cavity outer layer would all be important paths for future R$\&$D. Given the ongoing efforts towards forming high-performance Nb$_3$Sn films directly onto a Cu substrate~\cite{Withanage_2021, Ilyina_2019, sun:srf2021-weotev03, ge:srf2019-tufub8}, the approach described in this article could be a proven solution to utilize Cu/Nb$_3$Sn cavities in a conduction-cooled SRF cryomodule for future accelerators.

Future work aims at the design of a 10~MeV, 1~MW electron linac for environmental remediation using a multi-cell, conduction-cooled SRF cavity and to validate some of the design aspects on a multi-cell prototype cavity to be tested in the HTC at General Atomics.

\section*{Acknowledgments}
The authors would like to acknowledge B. Golesich at CTC for the Cu cold-spray, A. Tuck at AJ Tuck Co. for the Cu electroplating and D. Combs of JLab's machine shop for the machining of the Cu outer layer. We would like to thank our colleagues from the SRF Cavity Production Group at Jefferson Lab for helping with the cavity cleaning, assembly and cool-down in the vertical cryostat. We would like to thank A. Cuffe, R. Bachimanchi, C. Wilson and P. Owen of Jefferson Lab for help preparing the LLRF modules for off-site use and software development. We would also like to thank M. Dale of Sumitomo Cryogenics of America for many useful discussions. Finally, we would like to acknowledge C. Bott and M. Pearce of Hampton Roads Sanitation District for useful discussions on the potential use of SRF accelerators for wastewater treatment.
This work is authored by Jefferson Science Associates, LLC and it was supported by the U.S. Department of Energy, Office of Science, Office of Accelerator Research $\&$ Development and Production, under contract No. DE-AC05-06OR23177. SB's microscopy work at the National High Magnetic Field Laboratory was partly supported by the U.S. Department of Energy, Office of Science, Office of High Energy Physics under Award No. DE-SC0009960.

\bibliography{PRAB_Envac2_bib}

\begin{thebibliography}{89}%
\makeatletter
\providecommand \@ifxundefined [1]{%
 \@ifx{#1\undefined}
}%
\providecommand \@ifnum [1]{%
 \ifnum #1\expandafter \@firstoftwo
 \else \expandafter \@secondoftwo
 \fi
}%
\providecommand \@ifx [1]{%
 \ifx #1\expandafter \@firstoftwo
 \else \expandafter \@secondoftwo
 \fi
}%
\providecommand \natexlab [1]{#1}%
\providecommand \enquote  [1]{``#1''}%
\providecommand \bibnamefont  [1]{#1}%
\providecommand \bibfnamefont [1]{#1}%
\providecommand \citenamefont [1]{#1}%
\providecommand \href@noop [0]{\@secondoftwo}%
\providecommand \href [0]{\begingroup \@sanitize@url \@href}%
\providecommand \@href[1]{\@@startlink{#1}\@@href}%
\providecommand \@@href[1]{\endgroup#1\@@endlink}%
\providecommand \@sanitize@url [0]{\catcode `\\12\catcode `\$12\catcode
  `\&12\catcode `\#12\catcode `\^12\catcode `\_12\catcode `\%12\relax}%
\providecommand \@@startlink[1]{}%
\providecommand \@@endlink[0]{}%
\providecommand \url  [0]{\begingroup\@sanitize@url \@url }%
\providecommand \@url [1]{\endgroup\@href {#1}{\urlprefix }}%
\providecommand \urlprefix  [0]{URL }%
\providecommand \Eprint [0]{\href }%
\providecommand \doibase [0]{https://doi.org/}%
\providecommand \selectlanguage [0]{\@gobble}%
\providecommand \bibinfo  [0]{\@secondoftwo}%
\providecommand \bibfield  [0]{\@secondoftwo}%
\providecommand \translation [1]{[#1]}%
\providecommand \BibitemOpen [0]{}%
\providecommand \bibitemStop [0]{}%
\providecommand \bibitemNoStop [0]{.\EOS\space}%
\providecommand \EOS [0]{\spacefactor3000\relax}%
\providecommand \BibitemShut  [1]{\csname bibitem#1\endcsname}%
\let\auto@bib@innerbib\@empty
\bibitem [{\citenamefont {Belomestnykh}(2013)}]{SRF1}%
  \BibitemOpen
  \bibfield  {author} {\bibinfo {author} {\bibfnamefont {S.}~\bibnamefont
  {Belomestnykh}},\ }\bibfield  {title} {\bibinfo {title} {Superconducting
  radio-frequency systems for high-$\beta$ particle accelerators},\ }in\ \href
  {https://doi.org/10.1142/S1793626812300006X} {\emph {\bibinfo {booktitle}
  {Reviews of Accelerator Science and Technology}}},\ Vol.~\bibinfo {volume}
  {5},\ \bibinfo {editor} {edited by\ \bibinfo {editor} {\bibfnamefont {A.~W.}\
  \bibnamefont {Chao}}\ and\ \bibinfo {editor} {\bibfnamefont {W.}~\bibnamefont
  {Chou}}}\ (\bibinfo  {publisher} {WORLD SCIENTIFIC},\ \bibinfo {address}
  {Singapore},\ \bibinfo {year} {2013})\ pp.\ \bibinfo {pages}
  {147--184}\BibitemShut {NoStop}%
\bibitem [{\citenamefont {Kelly}(2013)}]{SRF2}%
  \BibitemOpen
  \bibfield  {author} {\bibinfo {author} {\bibfnamefont {M.}~\bibnamefont
  {Kelly}},\ }\bibfield  {title} {\bibinfo {title} {Superconducting
  radio-frequency systems for low-beta particle accelerators},\ }in\ \href
  {https://doi.org/10.1142/S1793626812300071} {\emph {\bibinfo {booktitle}
  {Reviews of Accelerator Science and Technology}}},\ Vol.~\bibinfo {volume}
  {5},\ \bibinfo {editor} {edited by\ \bibinfo {editor} {\bibfnamefont {A.~W.}\
  \bibnamefont {Chao}}\ and\ \bibinfo {editor} {\bibfnamefont {W.}~\bibnamefont
  {Chou}}}\ (\bibinfo  {publisher} {WORLD SCIENTIFIC},\ \bibinfo {address}
  {Singapore},\ \bibinfo {year} {2013})\ pp.\ \bibinfo {pages}
  {185--203}\BibitemShut {NoStop}%
\bibitem [{\citenamefont {Posen}\ and\ \citenamefont
  {Hall}(2017)}]{Posen_2017}%
  \BibitemOpen
  \bibfield  {author} {\bibinfo {author} {\bibfnamefont {S.}~\bibnamefont
  {Posen}}\ and\ \bibinfo {author} {\bibfnamefont {D.~L.}\ \bibnamefont
  {Hall}},\ }\bibfield  {title} {\bibinfo {title} {Nb$_3${S}n superconducting
  radiofrequency cavities: fabrication, results, properties, and prospects},\
  }\href {https://doi.org/10.1088/1361-6668/30/3/033004} {\bibfield  {journal}
  {\bibinfo  {journal} {Superconductor Science and Technology}\ }\textbf
  {\bibinfo {volume} {30}},\ \bibinfo {pages} {033004} (\bibinfo {year}
  {2017})}\BibitemShut {NoStop}%
\bibitem [{\citenamefont {Dhuley}\ \emph {et~al.}(2020)\citenamefont {Dhuley},
  \citenamefont {Posen}, \citenamefont {Geelhoed}, \citenamefont {Prokofiev},\
  and\ \citenamefont {Thangaraj}}]{Dhuley_2020}%
  \BibitemOpen
  \bibfield  {author} {\bibinfo {author} {\bibfnamefont {R.~C.}\ \bibnamefont
  {Dhuley}}, \bibinfo {author} {\bibfnamefont {S.}~\bibnamefont {Posen}},
  \bibinfo {author} {\bibfnamefont {M.~I.}\ \bibnamefont {Geelhoed}}, \bibinfo
  {author} {\bibfnamefont {O.}~\bibnamefont {Prokofiev}},\ and\ \bibinfo
  {author} {\bibfnamefont {J.~C.~T.}\ \bibnamefont {Thangaraj}},\ }\bibfield
  {title} {\bibinfo {title} {First demonstration of a cryocooler conduction
  cooled superconducting radiofrequency cavity operating at practical cw
  accelerating gradients},\ }\href {https://doi.org/10.1088/1361-6668/ab82f0}
  {\bibfield  {journal} {\bibinfo  {journal} {Superconductor Science and
  Technology}\ }\textbf {\bibinfo {volume} {33}},\ \bibinfo {pages} {06LT01}
  (\bibinfo {year} {2020})}\BibitemShut {NoStop}%
\bibitem [{\citenamefont {Ciovati}\ \emph {et~al.}(2020)\citenamefont
  {Ciovati}, \citenamefont {Cheng}, \citenamefont {Pudasaini},\ and\
  \citenamefont {Rimmer}}]{Ciovati_SUST}%
  \BibitemOpen
  \bibfield  {author} {\bibinfo {author} {\bibfnamefont {G.}~\bibnamefont
  {Ciovati}}, \bibinfo {author} {\bibfnamefont {G.}~\bibnamefont {Cheng}},
  \bibinfo {author} {\bibfnamefont {U.}~\bibnamefont {Pudasaini}},\ and\
  \bibinfo {author} {\bibfnamefont {R.~A.}\ \bibnamefont {Rimmer}},\ }\bibfield
   {title} {\bibinfo {title} {Multi-metallic conduction cooled superconducting
  radio-frequency cavity with high thermal stability},\ }\href
  {https://doi.org/10.1088/1361-6668/ab8d98} {\bibfield  {journal} {\bibinfo
  {journal} {Superconductor Science and Technology}\ }\textbf {\bibinfo
  {volume} {33}},\ \bibinfo {pages} {07LT01} (\bibinfo {year}
  {2020})}\BibitemShut {NoStop}%
\bibitem [{\citenamefont {Stilin}\ \emph {et~al.}()\citenamefont {Stilin},
  \citenamefont {Holic}, \citenamefont {Liepe}, \citenamefont {Porter},
  \citenamefont {Sears},\ and\ \citenamefont {Sun}}]{Stilin}%
  \BibitemOpen
  \bibfield  {author} {\bibinfo {author} {\bibfnamefont {N.~A.}\ \bibnamefont
  {Stilin}}, \bibinfo {author} {\bibfnamefont {A.~T.}\ \bibnamefont {Holic}},
  \bibinfo {author} {\bibfnamefont {M.}~\bibnamefont {Liepe}}, \bibinfo
  {author} {\bibfnamefont {R.~D.}\ \bibnamefont {Porter}}, \bibinfo {author}
  {\bibfnamefont {J.}~\bibnamefont {Sears}},\ and\ \bibinfo {author}
  {\bibfnamefont {Z.}~\bibnamefont {Sun}},\ }\bibfield  {title} {\bibinfo
  {title} {{Conduction Cooling Methods for {Nb3Sn} {SRF} Cavities and
  Cryomodules}},\ }in\ \href {https://doi.org/10.18429/JACoW-IPAC2021-MOPAB391}
  {\emph {\bibinfo {booktitle} {Proc. IPAC'21}}}\ (\bibinfo  {publisher} {JACoW
  Publishing, Geneva, Switzerland})\ pp.\ \bibinfo {pages}
  {1192--1195}\BibitemShut {NoStop}%
\bibitem [{\citenamefont {Henderson}\ and\ \citenamefont
  {Waite}(2015)}]{Envacc_Report}%
  \BibitemOpen
  \bibfield  {author} {\bibinfo {author} {\bibfnamefont {S.}~\bibnamefont
  {Henderson}}\ and\ \bibinfo {author} {\bibfnamefont {T.}~\bibnamefont
  {Waite}},\ }\href@noop {} {\emph {\bibinfo {title} {Workshop on Energy and
  Environmental Applications of Accelerators, Argonne, IL}}},\ \bibinfo {type}
  {Tech. Rep.}\ (\bibinfo  {institution} {{U.S.} {D}epartment of {E}nergy},\
  \bibinfo {year} {2015})\BibitemShut {NoStop}%
\bibitem [{\citenamefont {Ciovati}\ \emph {et~al.}(2018)\citenamefont
  {Ciovati}, \citenamefont {Anderson}, \citenamefont {Coriton}, \citenamefont
  {Guo}, \citenamefont {Hannon}, \citenamefont {Holland}, \citenamefont
  {LeSher}, \citenamefont {Marhauser}, \citenamefont {Rathke}, \citenamefont
  {Rimmer}, \citenamefont {Schultheiss},\ and\ \citenamefont {Vylet}}]{EnvAcc}%
  \BibitemOpen
  \bibfield  {author} {\bibinfo {author} {\bibfnamefont {G.}~\bibnamefont
  {Ciovati}}, \bibinfo {author} {\bibfnamefont {J.}~\bibnamefont {Anderson}},
  \bibinfo {author} {\bibfnamefont {B.}~\bibnamefont {Coriton}}, \bibinfo
  {author} {\bibfnamefont {J.}~\bibnamefont {Guo}}, \bibinfo {author}
  {\bibfnamefont {F.}~\bibnamefont {Hannon}}, \bibinfo {author} {\bibfnamefont
  {L.}~\bibnamefont {Holland}}, \bibinfo {author} {\bibfnamefont
  {M.}~\bibnamefont {LeSher}}, \bibinfo {author} {\bibfnamefont
  {F.}~\bibnamefont {Marhauser}}, \bibinfo {author} {\bibfnamefont
  {J.}~\bibnamefont {Rathke}}, \bibinfo {author} {\bibfnamefont
  {R.}~\bibnamefont {Rimmer}}, \bibinfo {author} {\bibfnamefont
  {T.}~\bibnamefont {Schultheiss}},\ and\ \bibinfo {author} {\bibfnamefont
  {V.}~\bibnamefont {Vylet}},\ }\bibfield  {title} {\bibinfo {title} {Design of
  a cw, low-energy, high-power superconducting linac for environmental
  applications},\ }\href {https://doi.org/10.1103/PhysRevAccelBeams.21.091601}
  {\bibfield  {journal} {\bibinfo  {journal} {Phys. Rev. Accel. Beams}\
  }\textbf {\bibinfo {volume} {21}},\ \bibinfo {pages} {091601} (\bibinfo
  {year} {2018})}\BibitemShut {NoStop}%
\bibitem [{\citenamefont {Dhuley}\ \emph {et~al.}(2022)\citenamefont {Dhuley},
  \citenamefont {Gonin}, \citenamefont {Kazakov}, \citenamefont
  {Khabiboulline}, \citenamefont {Sukhanov}, \citenamefont {Yakovlev},
  \citenamefont {Saini}, \citenamefont {Solyak}, \citenamefont {Sauers},
  \citenamefont {Thangaraj}, \citenamefont {Zeller}, \citenamefont {Coriton},\
  and\ \citenamefont {Kostin}}]{Dhuley_PRAB}%
  \BibitemOpen
  \bibfield  {author} {\bibinfo {author} {\bibfnamefont {R.~C.}\ \bibnamefont
  {Dhuley}}, \bibinfo {author} {\bibfnamefont {I.}~\bibnamefont {Gonin}},
  \bibinfo {author} {\bibfnamefont {S.}~\bibnamefont {Kazakov}}, \bibinfo
  {author} {\bibfnamefont {T.}~\bibnamefont {Khabiboulline}}, \bibinfo {author}
  {\bibfnamefont {A.}~\bibnamefont {Sukhanov}}, \bibinfo {author}
  {\bibfnamefont {V.}~\bibnamefont {Yakovlev}}, \bibinfo {author}
  {\bibfnamefont {A.}~\bibnamefont {Saini}}, \bibinfo {author} {\bibfnamefont
  {N.}~\bibnamefont {Solyak}}, \bibinfo {author} {\bibfnamefont
  {A.}~\bibnamefont {Sauers}}, \bibinfo {author} {\bibfnamefont {J.~C.~T.}\
  \bibnamefont {Thangaraj}}, \bibinfo {author} {\bibfnamefont {K.}~\bibnamefont
  {Zeller}}, \bibinfo {author} {\bibfnamefont {B.}~\bibnamefont {Coriton}},\
  and\ \bibinfo {author} {\bibfnamefont {R.}~\bibnamefont {Kostin}},\
  }\bibfield  {title} {\bibinfo {title} {Design of a 10 {MeV}, 1000 {kW}
  average power electron-beam accelerator for wastewater treatment
  applications},\ }\href {https://doi.org/10.1103/PhysRevAccelBeams.25.041601}
  {\bibfield  {journal} {\bibinfo  {journal} {Phys. Rev. Accel. Beams}\
  }\textbf {\bibinfo {volume} {25}},\ \bibinfo {pages} {041601} (\bibinfo
  {year} {2022})}\BibitemShut {NoStop}%
\bibitem [{\citenamefont {Vyas}\ \emph {et~al.}(2016)\citenamefont {Vyas},
  \citenamefont {Verma}, \citenamefont {Maurya},\ and\ \citenamefont
  {Singh}}]{Magnetrons}%
  \BibitemOpen
  \bibfield  {author} {\bibinfo {author} {\bibfnamefont {S.~K.}\ \bibnamefont
  {Vyas}}, \bibinfo {author} {\bibfnamefont {R.~K.}\ \bibnamefont {Verma}},
  \bibinfo {author} {\bibfnamefont {S.}~\bibnamefont {Maurya}},\ and\ \bibinfo
  {author} {\bibfnamefont {V.}~\bibnamefont {Singh}},\ }\bibfield  {title}
  {\bibinfo {title} {Review of magnetron developments},\ }\href
  {https://doi.org/doi:10.1515/freq-2015-0196} {\bibfield  {journal} {\bibinfo
  {journal} {Frequenz}\ }\textbf {\bibinfo {volume} {70}},\ \bibinfo {pages}
  {455} (\bibinfo {year} {2016})}\BibitemShut {NoStop}%
\bibitem [{\citenamefont {Dexter}\ \emph {et~al.}(2011)\citenamefont {Dexter},
  \citenamefont {Burt}, \citenamefont {Carter}, \citenamefont {Tahir},
  \citenamefont {Wang}, \citenamefont {Davis},\ and\ \citenamefont
  {Rimmer}}]{Magn1}%
  \BibitemOpen
  \bibfield  {author} {\bibinfo {author} {\bibfnamefont {A.~C.}\ \bibnamefont
  {Dexter}}, \bibinfo {author} {\bibfnamefont {G.}~\bibnamefont {Burt}},
  \bibinfo {author} {\bibfnamefont {R.~G.}\ \bibnamefont {Carter}}, \bibinfo
  {author} {\bibfnamefont {I.}~\bibnamefont {Tahir}}, \bibinfo {author}
  {\bibfnamefont {H.}~\bibnamefont {Wang}}, \bibinfo {author} {\bibfnamefont
  {K.}~\bibnamefont {Davis}},\ and\ \bibinfo {author} {\bibfnamefont
  {R.}~\bibnamefont {Rimmer}},\ }\bibfield  {title} {\bibinfo {title} {First
  demonstration and performance of an injection locked continuous wave
  magnetron to phase control a superconducting cavity},\ }\href
  {https://doi.org/10.1103/PhysRevSTAB.14.032001} {\bibfield  {journal}
  {\bibinfo  {journal} {Phys. Rev. ST Accel. Beams}\ }\textbf {\bibinfo
  {volume} {14}},\ \bibinfo {pages} {032001} (\bibinfo {year}
  {2011})}\BibitemShut {NoStop}%
\bibitem [{\citenamefont {Kazakevich}\ \emph {et~al.}(2014)\citenamefont
  {Kazakevich}, \citenamefont {Johnson}, \citenamefont {Flanagan},
  \citenamefont {Marhauser}, \citenamefont {Yakovlev}, \citenamefont {Chase},
  \citenamefont {Lebedev}, \citenamefont {Nagaitsev}, \citenamefont
  {Pasquinelli}, \citenamefont {Solyak}, \citenamefont {Quinn}, \citenamefont
  {Wolff},\ and\ \citenamefont {Pavlov}}]{Magn2}%
  \BibitemOpen
  \bibfield  {author} {\bibinfo {author} {\bibfnamefont {G.}~\bibnamefont
  {Kazakevich}}, \bibinfo {author} {\bibfnamefont {R.}~\bibnamefont {Johnson}},
  \bibinfo {author} {\bibfnamefont {G.}~\bibnamefont {Flanagan}}, \bibinfo
  {author} {\bibfnamefont {F.}~\bibnamefont {Marhauser}}, \bibinfo {author}
  {\bibfnamefont {V.}~\bibnamefont {Yakovlev}}, \bibinfo {author}
  {\bibfnamefont {B.}~\bibnamefont {Chase}}, \bibinfo {author} {\bibfnamefont
  {V.}~\bibnamefont {Lebedev}}, \bibinfo {author} {\bibfnamefont
  {S.}~\bibnamefont {Nagaitsev}}, \bibinfo {author} {\bibfnamefont
  {R.}~\bibnamefont {Pasquinelli}}, \bibinfo {author} {\bibfnamefont
  {N.}~\bibnamefont {Solyak}}, \bibinfo {author} {\bibfnamefont
  {K.}~\bibnamefont {Quinn}}, \bibinfo {author} {\bibfnamefont
  {D.}~\bibnamefont {Wolff}},\ and\ \bibinfo {author} {\bibfnamefont
  {V.}~\bibnamefont {Pavlov}},\ }\bibfield  {title} {\bibinfo {title}
  {High-power magnetron transmitter as an rf source for superconducting linear
  accelerators},\ }\href
  {https://doi.org/https://doi.org/10.1016/j.nima.2014.05.069} {\bibfield
  {journal} {\bibinfo  {journal} {Nuclear Instruments and Methods in Physics
  Research Section A: Accelerators, Spectrometers, Detectors and Associated
  Equipment}\ }\textbf {\bibinfo {volume} {760}},\ \bibinfo {pages} {19}
  (\bibinfo {year} {2014})}\BibitemShut {NoStop}%
\bibitem [{\citenamefont {Liu}\ \emph {et~al.}(2016)\citenamefont {Liu},
  \citenamefont {Huang}, \citenamefont {Liu}, \citenamefont {Huo},\ and\
  \citenamefont {Huang}}]{Magn3}%
  \BibitemOpen
  \bibfield  {author} {\bibinfo {author} {\bibfnamefont {C.}~\bibnamefont
  {Liu}}, \bibinfo {author} {\bibfnamefont {H.}~\bibnamefont {Huang}}, \bibinfo
  {author} {\bibfnamefont {Z.}~\bibnamefont {Liu}}, \bibinfo {author}
  {\bibfnamefont {F.}~\bibnamefont {Huo}},\ and\ \bibinfo {author}
  {\bibfnamefont {K.}~\bibnamefont {Huang}},\ }\bibfield  {title} {\bibinfo
  {title} {Experimental study on microwave power combining based on
  injection-locked 15-kw $s$ -band continuous-wave magnetrons},\ }\href
  {https://doi.org/10.1109/TPS.2016.2565564} {\bibfield  {journal} {\bibinfo
  {journal} {IEEE Transactions on Plasma Science}\ }\textbf {\bibinfo {volume}
  {44}},\ \bibinfo {pages} {1291} (\bibinfo {year} {2016})}\BibitemShut
  {NoStop}%
\bibitem [{\citenamefont {Kazakevich}\ \emph {et~al.}(2018)\citenamefont
  {Kazakevich}, \citenamefont {Johnson}, \citenamefont {Lebedev}, \citenamefont
  {Yakovlev},\ and\ \citenamefont {Pavlov}}]{Magn4}%
  \BibitemOpen
  \bibfield  {author} {\bibinfo {author} {\bibfnamefont {G.}~\bibnamefont
  {Kazakevich}}, \bibinfo {author} {\bibfnamefont {R.}~\bibnamefont {Johnson}},
  \bibinfo {author} {\bibfnamefont {V.}~\bibnamefont {Lebedev}}, \bibinfo
  {author} {\bibfnamefont {V.}~\bibnamefont {Yakovlev}},\ and\ \bibinfo
  {author} {\bibfnamefont {V.}~\bibnamefont {Pavlov}},\ }\bibfield  {title}
  {\bibinfo {title} {Resonant interaction of the electron beam with a
  synchronous wave in controlled magnetrons for high-current superconducting
  accelerators},\ }\href {https://doi.org/10.1103/PhysRevAccelBeams.21.062001}
  {\bibfield  {journal} {\bibinfo  {journal} {Phys. Rev. Accel. Beams}\
  }\textbf {\bibinfo {volume} {21}},\ \bibinfo {pages} {062001} (\bibinfo
  {year} {2018})}\BibitemShut {NoStop}%
\bibitem [{\citenamefont {Chen}\ \emph {et~al.}(2020)\citenamefont {Chen},
  \citenamefont {Yang}, \citenamefont {Shinohara},\ and\ \citenamefont
  {Liu}}]{Magn5}%
  \BibitemOpen
  \bibfield  {author} {\bibinfo {author} {\bibfnamefont {X.}~\bibnamefont
  {Chen}}, \bibinfo {author} {\bibfnamefont {B.}~\bibnamefont {Yang}}, \bibinfo
  {author} {\bibfnamefont {N.}~\bibnamefont {Shinohara}},\ and\ \bibinfo
  {author} {\bibfnamefont {C.}~\bibnamefont {Liu}},\ }\bibfield  {title}
  {\bibinfo {title} {Modeling and experiments of an injection-locked magnetron
  with various load reflection levels},\ }\href
  {https://doi.org/10.1109/TED.2020.3009901} {\bibfield  {journal} {\bibinfo
  {journal} {IEEE Transactions on Electron Devices}\ }\textbf {\bibinfo
  {volume} {67}},\ \bibinfo {pages} {3802} (\bibinfo {year}
  {2020})}\BibitemShut {NoStop}%
\bibitem [{\citenamefont {Li}\ \emph {et~al.}(2021)\citenamefont {Li},
  \citenamefont {Zhang}, \citenamefont {Zhou}, \citenamefont {Zhang},
  \citenamefont {Liu}, \citenamefont {Zhang},\ and\ \citenamefont
  {An}}]{Magn6}%
  \BibitemOpen
  \bibfield  {author} {\bibinfo {author} {\bibfnamefont {W.}~\bibnamefont
  {Li}}, \bibinfo {author} {\bibfnamefont {P.}~\bibnamefont {Zhang}}, \bibinfo
  {author} {\bibfnamefont {B.}~\bibnamefont {Zhou}}, \bibinfo {author}
  {\bibfnamefont {H.}~\bibnamefont {Zhang}}, \bibinfo {author} {\bibfnamefont
  {Y.}~\bibnamefont {Liu}}, \bibinfo {author} {\bibfnamefont {L.}~\bibnamefont
  {Zhang}},\ and\ \bibinfo {author} {\bibfnamefont {S.}~\bibnamefont {An}},\
  }\bibfield  {title} {\bibinfo {title} {Development of a 1497 {M}hz, 13.5 k{W}
  continuous-wave magnetron for superconducting radio frequency accelerator},\
  }\href {https://doi.org/10.1109/TPS.2020.3048982} {\bibfield  {journal}
  {\bibinfo  {journal} {IEEE Transactions on Plasma Science}\ }\textbf
  {\bibinfo {volume} {49}},\ \bibinfo {pages} {663} (\bibinfo {year}
  {2021})}\BibitemShut {NoStop}%
\bibitem [{\citenamefont {Ha}\ \emph {et~al.}(2022)\citenamefont {Ha},
  \citenamefont {Kim}, \citenamefont {Jang}, \citenamefont {Kim},\ and\
  \citenamefont {Han}}]{Magn7}%
  \BibitemOpen
  \bibfield  {author} {\bibinfo {author} {\bibfnamefont {C.-S.}\ \bibnamefont
  {Ha}}, \bibinfo {author} {\bibfnamefont {T.-H.}\ \bibnamefont {Kim}},
  \bibinfo {author} {\bibfnamefont {S.-R.}\ \bibnamefont {Jang}}, \bibinfo
  {author} {\bibfnamefont {J.-S.}\ \bibnamefont {Kim}},\ and\ \bibinfo {author}
  {\bibfnamefont {S.-T.}\ \bibnamefont {Han}},\ }\bibfield  {title} {\bibinfo
  {title} {Experimental study on the effect of the characteristics of a
  switching-mode power supply on a phase-locked magnetron},\ }\href
  {https://doi.org/10.1109/LED.2022.3156543} {\bibfield  {journal} {\bibinfo
  {journal} {IEEE Electron Device Letters}\ }\textbf {\bibinfo {volume} {43}},\
  \bibinfo {pages} {619} (\bibinfo {year} {2022})}\BibitemShut {NoStop}%
\bibitem [{\citenamefont {Wang}\ \emph {et~al.}(2022)\citenamefont {Wang},
  \citenamefont {Blum}, \citenamefont {Coriton}, \citenamefont {Jordan},
  \citenamefont {Moeller}, \citenamefont {Nelson}, \citenamefont {Overstreet},
  \citenamefont {Rimmer}, \citenamefont {Thackston}, \citenamefont {Vega},\
  and\ \citenamefont {Ziemyte}}]{Magn8}%
  \BibitemOpen
  \bibfield  {author} {\bibinfo {author} {\bibfnamefont {H.}~\bibnamefont
  {Wang}}, \bibinfo {author} {\bibfnamefont {J.}~\bibnamefont {Blum}}, \bibinfo
  {author} {\bibfnamefont {B.}~\bibnamefont {Coriton}}, \bibinfo {author}
  {\bibfnamefont {K.}~\bibnamefont {Jordan}}, \bibinfo {author} {\bibfnamefont
  {C.}~\bibnamefont {Moeller}}, \bibinfo {author} {\bibfnamefont
  {R.}~\bibnamefont {Nelson}}, \bibinfo {author} {\bibfnamefont
  {S.}~\bibnamefont {Overstreet}}, \bibinfo {author} {\bibfnamefont
  {R.}~\bibnamefont {Rimmer}}, \bibinfo {author} {\bibfnamefont
  {K.}~\bibnamefont {Thackston}}, \bibinfo {author} {\bibfnamefont
  {J.}~\bibnamefont {Vega}},\ and\ \bibinfo {author} {\bibfnamefont
  {G.}~\bibnamefont {Ziemyte}},\ }\bibfield  {title} {\bibinfo {title}
  {{Magnetron R$\&$D Progress for High Efficiency CW RF Sources of Industrial
  Accelerators}},\ }in\ \href {https://doi.org/10.18429/JACoW-NAPAC2022-WEZD3}
  {\emph {\bibinfo {booktitle} {Proc. NAPAC'22}}},\ \bibinfo {series and
  number} {\bibinfo {series} {International Particle Accelerator Conference}\
  No.~\bibinfo {number} {5}}\ (\bibinfo  {publisher} {JACoW Publishing, Geneva,
  Switzerland},\ \bibinfo {year} {2022})\ pp.\ \bibinfo {pages}
  {597--600}\BibitemShut {NoStop}%
\bibitem [{\citenamefont {de~Loos}\ and\ \citenamefont {van~der Geer}()}]{GPT}%
  \BibitemOpen
  \bibfield  {author} {\bibinfo {author} {\bibfnamefont {M.~J.}\ \bibnamefont
  {de~Loos}}\ and\ \bibinfo {author} {\bibfnamefont {S.~B.}\ \bibnamefont
  {van~der Geer}},\ }\bibfield  {title} {\bibinfo {title} {{General Particle
  Tracer: A New 3D Code for Accelerator and Beamline Design}},\ }in\ \href@noop
  {} {\emph {\bibinfo {booktitle} {Proc. EPAC'96}}}\ (\bibinfo  {publisher}
  {JACoW Publishing, Geneva, Switzerland})\ pp.\ \bibinfo {pages} {1245 --
  1247}\BibitemShut {NoStop}%
\bibitem [{\citenamefont {Pierini}\ \emph {et~al.}()\citenamefont {Pierini},
  \citenamefont {Barni}, \citenamefont {Bosotti}, \citenamefont {Ciovati},\
  and\ \citenamefont {Pagani}}]{buildcav}%
  \BibitemOpen
  \bibfield  {author} {\bibinfo {author} {\bibfnamefont {P.}~\bibnamefont
  {Pierini}}, \bibinfo {author} {\bibfnamefont {D.}~\bibnamefont {Barni}},
  \bibinfo {author} {\bibfnamefont {A.}~\bibnamefont {Bosotti}}, \bibinfo
  {author} {\bibfnamefont {G.}~\bibnamefont {Ciovati}},\ and\ \bibinfo {author}
  {\bibfnamefont {C.}~\bibnamefont {Pagani}},\ }\bibfield  {title} {\bibinfo
  {title} {{Cavity Design Tools and Applications to the TRASCO Project}},\ }in\
  \href {https://jacow.org/SRF99/papers/WEP004.pdf} {\emph {\bibinfo
  {booktitle} {Proc. SRF'99}}}\ (\bibinfo  {publisher} {JACoW Publishing,
  Geneva, Switzerland})\ pp.\ \bibinfo {pages} {380--383}\BibitemShut {NoStop}%
\bibitem [{\citenamefont {Billen}\ and\ \citenamefont {Young}(1996)}]{SF}%
  \BibitemOpen
  \bibfield  {author} {\bibinfo {author} {\bibfnamefont {J.~H.}\ \bibnamefont
  {Billen}}\ and\ \bibinfo {author} {\bibfnamefont {L.~M.}\ \bibnamefont
  {Young}},\ }\href@noop {} {\emph {\bibinfo {title} {Poisson Superfish}}},\
  \bibinfo {type} {Tech. Rep.}\ \bibinfo {number} {{LA-UR-96-1834}}\ (\bibinfo
  {institution} {{Los Alamos National Laboratory}},\ \bibinfo {year} {1996})\
  \bibinfo {note} {revised February 6, 2003}\BibitemShut {NoStop}%
\bibitem [{\citenamefont {Donoghue}\ \emph {et~al.}()\citenamefont {Donoghue},
  \citenamefont {Wu}, \citenamefont {Mammosser}, \citenamefont {Rimmer},
  \citenamefont {Stirbet}, \citenamefont {Phillips},\ and\ \citenamefont
  {Wang}}]{donoghue:srf05}%
  \BibitemOpen
  \bibfield  {author} {\bibinfo {author} {\bibfnamefont {E.}~\bibnamefont
  {Donoghue}}, \bibinfo {author} {\bibfnamefont {G.}~\bibnamefont {Wu}},
  \bibinfo {author} {\bibfnamefont {J.}~\bibnamefont {Mammosser}}, \bibinfo
  {author} {\bibfnamefont {R.}~\bibnamefont {Rimmer}}, \bibinfo {author}
  {\bibfnamefont {M.}~\bibnamefont {Stirbet}}, \bibinfo {author} {\bibfnamefont
  {L.}~\bibnamefont {Phillips}},\ and\ \bibinfo {author} {\bibfnamefont
  {H.}~\bibnamefont {Wang}},\ }\bibfield  {title} {\bibinfo {title} {{Studies
  of Electron Activities in SNS-Type Superconducting RF Cavities}},\ }in\ \href
  {https://jacow.org/SRF2005/papers/TUP67.pdf} {\emph {\bibinfo {booktitle}
  {Proc. SRF'05}}}\ (\bibinfo  {publisher} {JACoW Publishing, Geneva,
  Switzerland})\ pp.\ \bibinfo {pages} {402--405}\BibitemShut {NoStop}%
\bibitem [{\citenamefont {Xu}\ \emph {et~al.}(2012)\citenamefont {Xu},
  \citenamefont {Altinbas}, \citenamefont {Belomestnykh}, \citenamefont
  {Ben-Zvi}, \citenamefont {Cole}, \citenamefont {Deonarine}, \citenamefont
  {Falletta}, \citenamefont {Jamilkowski}, \citenamefont {Gassner},
  \citenamefont {Kankiya}, \citenamefont {Kayran}, \citenamefont {Laloudakis},
  \citenamefont {Masi}, \citenamefont {McIntyre}, \citenamefont {Pate},
  \citenamefont {Philips}, \citenamefont {Seda}, \citenamefont {Schultheiss},
  \citenamefont {Steszyn}, \citenamefont {Tallerico}, \citenamefont {Todd},
  \citenamefont {Weiss}, \citenamefont {Whitbeck},\ and\ \citenamefont
  {Zaltsman}}]{FPC}%
  \BibitemOpen
  \bibfield  {author} {\bibinfo {author} {\bibfnamefont {W.}~\bibnamefont
  {Xu}}, \bibinfo {author} {\bibfnamefont {Z.}~\bibnamefont {Altinbas}},
  \bibinfo {author} {\bibfnamefont {S.}~\bibnamefont {Belomestnykh}}, \bibinfo
  {author} {\bibfnamefont {I.}~\bibnamefont {Ben-Zvi}}, \bibinfo {author}
  {\bibfnamefont {M.}~\bibnamefont {Cole}}, \bibinfo {author} {\bibfnamefont
  {S.}~\bibnamefont {Deonarine}}, \bibinfo {author} {\bibfnamefont
  {M.}~\bibnamefont {Falletta}}, \bibinfo {author} {\bibfnamefont
  {J.}~\bibnamefont {Jamilkowski}}, \bibinfo {author} {\bibfnamefont
  {D.}~\bibnamefont {Gassner}}, \bibinfo {author} {\bibfnamefont
  {P.}~\bibnamefont {Kankiya}}, \bibinfo {author} {\bibfnamefont
  {D.}~\bibnamefont {Kayran}}, \bibinfo {author} {\bibfnamefont
  {N.}~\bibnamefont {Laloudakis}}, \bibinfo {author} {\bibfnamefont
  {L.}~\bibnamefont {Masi}}, \bibinfo {author} {\bibfnamefont {G.}~\bibnamefont
  {McIntyre}}, \bibinfo {author} {\bibfnamefont {D.}~\bibnamefont {Pate}},
  \bibinfo {author} {\bibfnamefont {D.}~\bibnamefont {Philips}}, \bibinfo
  {author} {\bibfnamefont {T.}~\bibnamefont {Seda}}, \bibinfo {author}
  {\bibfnamefont {T.}~\bibnamefont {Schultheiss}}, \bibinfo {author}
  {\bibfnamefont {A.}~\bibnamefont {Steszyn}}, \bibinfo {author} {\bibfnamefont
  {T.}~\bibnamefont {Tallerico}}, \bibinfo {author} {\bibfnamefont
  {R.}~\bibnamefont {Todd}}, \bibinfo {author} {\bibfnamefont {D.}~\bibnamefont
  {Weiss}}, \bibinfo {author} {\bibfnamefont {G.}~\bibnamefont {Whitbeck}},\
  and\ \bibinfo {author} {\bibfnamefont {A.}~\bibnamefont {Zaltsman}},\
  }\bibfield  {title} {\bibinfo {title} {Design, simulations, and conditioning
  of 500 k{W} fundamental power couplers for a superconducting rf gun},\ }\href
  {https://doi.org/10.1103/PhysRevSTAB.15.072001} {\bibfield  {journal}
  {\bibinfo  {journal} {Phys. Rev. ST Accel. Beams}\ }\textbf {\bibinfo
  {volume} {15}},\ \bibinfo {pages} {072001} (\bibinfo {year}
  {2012})}\BibitemShut {NoStop}%
\bibitem [{\citenamefont {{CST Studio Suite}}(2022)}]{CST}%
  \BibitemOpen
  \bibfield  {author} {\bibinfo {author} {\bibnamefont {{CST Studio Suite}}},\
  }\href@noop {} {\bibinfo {title} {3{D} electromagnetic software package}},\
  \bibinfo {howpublished}
  {\url{https://www.3ds.com/products-services/simulia/products/cst-studio-suite/}}
  (\bibinfo {year} {2022})\BibitemShut {NoStop}%
\bibitem [{\citenamefont {Marhauser}\ \emph {et~al.}()\citenamefont
  {Marhauser}, \citenamefont {Rimmer}, \citenamefont {Tian},\ and\
  \citenamefont {Wang}}]{marhauser:pac09}%
  \BibitemOpen
  \bibfield  {author} {\bibinfo {author} {\bibfnamefont {F.}~\bibnamefont
  {Marhauser}}, \bibinfo {author} {\bibfnamefont {R.~A.}\ \bibnamefont
  {Rimmer}}, \bibinfo {author} {\bibfnamefont {K.}~\bibnamefont {Tian}},\ and\
  \bibinfo {author} {\bibfnamefont {H.}~\bibnamefont {Wang}},\ }\bibfield
  {title} {\bibinfo {title} {{Enhanced Method for Cavity Impedance
  Calculations}},\ }in\ \href {https://jacow.org/PAC2009/papers/FR5PFP094.pdf}
  {\emph {\bibinfo {booktitle} {Proc. PAC'09}}}\ (\bibinfo  {publisher} {JACoW
  Publishing, Geneva, Switzerland})\ pp.\ \bibinfo {pages}
  {4523--4525}\BibitemShut {NoStop}%
\bibitem [{\citenamefont {Pudasaini}\ \emph {et~al.}(2019)\citenamefont
  {Pudasaini}, \citenamefont {Eremeev}, \citenamefont {Angle}, \citenamefont
  {Tuggle}, \citenamefont {Reece},\ and\ \citenamefont {Kelley}}]{Uttar_JVSTA}%
  \BibitemOpen
  \bibfield  {author} {\bibinfo {author} {\bibfnamefont {U.}~\bibnamefont
  {Pudasaini}}, \bibinfo {author} {\bibfnamefont {G.~V.}\ \bibnamefont
  {Eremeev}}, \bibinfo {author} {\bibfnamefont {J.~W.}\ \bibnamefont {Angle}},
  \bibinfo {author} {\bibfnamefont {J.}~\bibnamefont {Tuggle}}, \bibinfo
  {author} {\bibfnamefont {C.~E.}\ \bibnamefont {Reece}},\ and\ \bibinfo
  {author} {\bibfnamefont {M.~J.}\ \bibnamefont {Kelley}},\ }\bibfield  {title}
  {\bibinfo {title} {Growth of {N}b$_3${S}n coating in tin vapor-diffusion
  process},\ }\href {https://doi.org/10.1116/1.5113597} {\bibfield  {journal}
  {\bibinfo  {journal} {Journal of Vacuum Science \& Technology A}\ }\textbf
  {\bibinfo {volume} {37}},\ \bibinfo {pages} {051509} (\bibinfo {year}
  {2019})},\ \Eprint {https://arxiv.org/abs/https://doi.org/10.1116/1.5113597}
  {https://doi.org/10.1116/1.5113597} \BibitemShut {NoStop}%
\bibitem [{\citenamefont {Ciovati}\ \emph {et~al.}(2022)\citenamefont
  {Ciovati}, \citenamefont {Dhakal}, \citenamefont {Parajuli}, \citenamefont
  {Saeki},\ and\ \citenamefont {Pathiranage}}]{Ciovati_SRF21}%
  \BibitemOpen
  \bibfield  {author} {\bibinfo {author} {\bibfnamefont {G.}~\bibnamefont
  {Ciovati}}, \bibinfo {author} {\bibfnamefont {P.}~\bibnamefont {Dhakal}},
  \bibinfo {author} {\bibfnamefont {I.}~\bibnamefont {Parajuli}}, \bibinfo
  {author} {\bibfnamefont {T.}~\bibnamefont {Saeki}},\ and\ \bibinfo {author}
  {\bibfnamefont {M.~W.}\ \bibnamefont {Pathiranage}},\ }\bibfield  {title}
  {\bibinfo {title} {{Thermal Conductivity of Electroplated Copper Onto Bulk
  Niobium at Cryogenic Temperatures}},\ }in\ \href
  {https://doi.org/10.18429/JACoW-SRF2021-WEPFDV008} {\emph {\bibinfo
  {booktitle} {Proc. SRF'21}}},\ \bibinfo {series and number} {\bibinfo
  {series} {International Conference on RF Superconductivity}\ No.~\bibinfo
  {number} {20}}\ (\bibinfo  {publisher} {JACoW Publishing, Geneva,
  Switzerland},\ \bibinfo {year} {2022})\ pp.\ \bibinfo {pages}
  {576--580}\BibitemShut {NoStop}%
\bibitem [{\citenamefont {{Ansys Mechanical}}(2022)}]{ansys}%
  \BibitemOpen
  \bibfield  {author} {\bibinfo {author} {\bibnamefont {{Ansys Mechanical}}},\
  }\href@noop {} {\bibinfo {title} {3{D} engineering and designing software}},\
  \bibinfo {howpublished}
  {\url{https://www.ansys.com/products/structures/ansys-mechanical}} (\bibinfo
  {year} {2022})\BibitemShut {NoStop}%
\bibitem [{\citenamefont {Simon}\ \emph {et~al.}(1992)\citenamefont {Simon},
  \citenamefont {Drexler},\ and\ \citenamefont {Reed}}]{Cu_NIST}%
  \BibitemOpen
  \bibfield  {author} {\bibinfo {author} {\bibfnamefont {N.~J.}\ \bibnamefont
  {Simon}}, \bibinfo {author} {\bibfnamefont {E.~S.}\ \bibnamefont {Drexler}},\
  and\ \bibinfo {author} {\bibfnamefont {R.~P.}\ \bibnamefont {Reed}},\
  }\href@noop {} {\emph {\bibinfo {title} {Properties of copper and copper
  alloys at cryogenic temperatures}}},\ \bibinfo {type} {Tech. Rep.}\ \bibinfo
  {number} {{NIST Monograph 177}}\ (\bibinfo  {institution} {{National
  Institute of Standards and Technology}},\ \bibinfo {year} {1992})\BibitemShut
  {NoStop}%
\bibitem [{\citenamefont {Cody}\ and\ \citenamefont
  {Cohen}(1964)}]{Nb3Sn_thermcond}%
  \BibitemOpen
  \bibfield  {author} {\bibinfo {author} {\bibfnamefont {G.~D.}\ \bibnamefont
  {Cody}}\ and\ \bibinfo {author} {\bibfnamefont {R.~W.}\ \bibnamefont
  {Cohen}},\ }\bibfield  {title} {\bibinfo {title} {Thermal conductivity of
  {N}b$_3${S}n},\ }\href {https://doi.org/10.1103/RevModPhys.36.121} {\bibfield
   {journal} {\bibinfo  {journal} {Rev. Mod. Phys.}\ }\textbf {\bibinfo
  {volume} {36}},\ \bibinfo {pages} {121} (\bibinfo {year} {1964})}\BibitemShut
  {NoStop}%
\bibitem [{\citenamefont {Dillon}\ \emph {et~al.}(2017)\citenamefont {Dillon},
  \citenamefont {McCusker}, \citenamefont {Dyke}, \citenamefont {Isler},\ and\
  \citenamefont {Christiansen}}]{Dillon_2017}%
  \BibitemOpen
  \bibfield  {author} {\bibinfo {author} {\bibfnamefont {A.}~\bibnamefont
  {Dillon}}, \bibinfo {author} {\bibfnamefont {K.}~\bibnamefont {McCusker}},
  \bibinfo {author} {\bibfnamefont {J.~V.}\ \bibnamefont {Dyke}}, \bibinfo
  {author} {\bibfnamefont {B.}~\bibnamefont {Isler}},\ and\ \bibinfo {author}
  {\bibfnamefont {M.}~\bibnamefont {Christiansen}},\ }\bibfield  {title}
  {\bibinfo {title} {Thermal interface material characterization for cryogenic
  electronic packaging solutions},\ }\href
  {https://doi.org/10.1088/1757-899X/278/1/012054} {\bibfield  {journal}
  {\bibinfo  {journal} {IOP Conference Series: Materials Science and
  Engineering}\ }\textbf {\bibinfo {volume} {278}},\ \bibinfo {pages} {012054}
  (\bibinfo {year} {2017})}\BibitemShut {NoStop}%
\bibitem [{\citenamefont {Halbritter}(1970)}]{BCS}%
  \BibitemOpen
  \bibfield  {author} {\bibinfo {author} {\bibfnamefont {J.}~\bibnamefont
  {Halbritter}},\ }\href@noop {} {\emph {\bibinfo {title} {{FORTRAN-Program}
  for the computation of the surface impedance of superconductors}}},\ \bibinfo
  {type} {Tech. Rep.}\ \bibinfo {number} {{FZK 3/70-6}}\ (\bibinfo
  {institution} {{Forschungszentrum Karlsruhe}},\ \bibinfo {year}
  {1970})\BibitemShut {NoStop}%
\bibitem [{Note1()}]{Note1}%
  \BibitemOpen
  \bibinfo {note} {The power into the FPC was slightly higher than the target
  value because the $Q_{ext}$ of the FPC in the 3D model was not exactly at the
  target value.}\BibitemShut {Stop}%
\bibitem [{\citenamefont {Satogata}\ and\ \citenamefont
  {Zhang}(2018)}]{JLEIC1}%
  \BibitemOpen
  \bibfield  {author} {\bibinfo {author} {\bibfnamefont {T.}~\bibnamefont
  {Satogata}}\ and\ \bibinfo {author} {\bibfnamefont {Y.}~\bibnamefont {Zhang}}
  (\bibinfo {collaboration} {JLEIC Design Study}),\ }\bibfield  {title}
  {\bibinfo {title} {{JLEIC - A Polarized Electron-Ion Collider at Jefferson
  Lab}},\ }\href@noop {} {\bibfield  {journal} {\bibinfo  {journal} {ICFA Beam
  Dyn. Newslett.}\ }\textbf {\bibinfo {volume} {74}},\ \bibinfo {pages} {92}
  (\bibinfo {year} {2018})}\BibitemShut {NoStop}%
\bibitem [{\citenamefont {Rimmer}\ \emph {et~al.}()\citenamefont {Rimmer},
  \citenamefont {Clemens}, \citenamefont {Fors}, \citenamefont {Guo},
  \citenamefont {Hannon}, \citenamefont {Henry}, \citenamefont {Marhauser},
  \citenamefont {Turlington}, \citenamefont {Wang},\ and\ \citenamefont
  {Wang}}]{JLEIC2}%
  \BibitemOpen
  \bibfield  {author} {\bibinfo {author} {\bibfnamefont {R.~A.}\ \bibnamefont
  {Rimmer}}, \bibinfo {author} {\bibfnamefont {W.~A.}\ \bibnamefont {Clemens}},
  \bibinfo {author} {\bibfnamefont {F.}~\bibnamefont {Fors}}, \bibinfo {author}
  {\bibfnamefont {J.}~\bibnamefont {Guo}}, \bibinfo {author} {\bibfnamefont
  {F.~E.}\ \bibnamefont {Hannon}}, \bibinfo {author} {\bibfnamefont
  {J.}~\bibnamefont {Henry}}, \bibinfo {author} {\bibfnamefont
  {F.}~\bibnamefont {Marhauser}}, \bibinfo {author} {\bibfnamefont
  {L.}~\bibnamefont {Turlington}}, \bibinfo {author} {\bibfnamefont
  {H.}~\bibnamefont {Wang}},\ and\ \bibinfo {author} {\bibfnamefont
  {S.}~\bibnamefont {Wang}},\ }\bibfield  {title} {\bibinfo {title} {{952.6 MHz
  SRF Cavity Development for JLEIC}},\ }in\ \href
  {https://doi.org/10.18429/JACoW-IPAC2018-THPAL144} {\emph {\bibinfo
  {booktitle} {Proc. IPAC'18}}}\ (\bibinfo  {publisher} {JACoW Publishing,
  Geneva, Switzerland})\ pp.\ \bibinfo {pages} {3982--3985}\BibitemShut
  {NoStop}%
\bibitem [{\citenamefont {Marhauser}(2018)}]{Marhauser_FCC}%
  \BibitemOpen
  \bibfield  {author} {\bibinfo {author} {\bibfnamefont {F.}~\bibnamefont
  {Marhauser}},\ }\href
  {https://indico.cern.ch/event/656491/contributions/2932251/attachments/1629681/2597650/5_cell_Cavity_Marhauser.pdf}
  {\bibinfo {title} {{Recent Results on a Multi-Cell 802 MHz bulk Nb
  Cavity}}},\ \bibinfo {howpublished} {presented at FCC Week 2018} (\bibinfo
  {year} {9-13 April 2018})\BibitemShut {NoStop}%
\bibitem [{\citenamefont {Daly}\ \emph {et~al.}()\citenamefont {Daly},
  \citenamefont {Davis},\ and\ \citenamefont {Hicks}}]{daly:pac05}%
  \BibitemOpen
  \bibfield  {author} {\bibinfo {author} {\bibfnamefont {E.}~\bibnamefont
  {Daly}}, \bibinfo {author} {\bibfnamefont {G.~K.}\ \bibnamefont {Davis}},\
  and\ \bibinfo {author} {\bibfnamefont {W.~R.}\ \bibnamefont {Hicks}},\
  }\bibfield  {title} {\bibinfo {title} {{Testing of the New Tuner Design for
  the CEBAF 12 GeV Upgrade SRF Cavities}},\ }in\ \href
  {https://jacow.org/p05/papers/TPPT073.pdf} {\emph {\bibinfo {booktitle}
  {Proc. PAC'05}}}\ (\bibinfo  {publisher} {JACoW Publishing, Geneva,
  Switzerland})\ pp.\ \bibinfo {pages} {3910--3912}\BibitemShut {NoStop}%
\bibitem [{\citenamefont {Pischalnikov}\ \emph {et~al.}()\citenamefont
  {Pischalnikov}, \citenamefont {Cheban},\ and\ \citenamefont
  {Yun}}]{pischalnikov:ipac18}%
  \BibitemOpen
  \bibfield  {author} {\bibinfo {author} {\bibfnamefont {Y.~M.}\ \bibnamefont
  {Pischalnikov}}, \bibinfo {author} {\bibfnamefont {S.}~\bibnamefont
  {Cheban}},\ and\ \bibinfo {author} {\bibfnamefont {J.~C.}\ \bibnamefont
  {Yun}},\ }\bibfield  {title} {\bibinfo {title} {{Design of 650 MHz Tuner for
  PIP-II Project}},\ }in\ \href
  {https://doi.org/10.18429/JACoW-IPAC2018-WEPML002} {\emph {\bibinfo
  {booktitle} {Proc. IPAC'18}}}\ (\bibinfo  {publisher} {JACoW Publishing,
  Geneva, Switzerland})\ pp.\ \bibinfo {pages} {2671--2673}\BibitemShut
  {NoStop}%
\bibitem [{\citenamefont {Pischalnikov}\ \emph {et~al.}(2015)\citenamefont
  {Pischalnikov}, \citenamefont {Borissov}, \citenamefont {Gonin},
  \citenamefont {Holzbauer}, \citenamefont {Khabiboulline}, \citenamefont
  {Schappert}, \citenamefont {Smith},\ and\ \citenamefont
  {Yun}}]{Pischalnikov:IPAC2015}%
  \BibitemOpen
  \bibfield  {author} {\bibinfo {author} {\bibfnamefont {Y.}~\bibnamefont
  {Pischalnikov}}, \bibinfo {author} {\bibfnamefont {E.}~\bibnamefont
  {Borissov}}, \bibinfo {author} {\bibfnamefont {I.}~\bibnamefont {Gonin}},
  \bibinfo {author} {\bibfnamefont {J.}~\bibnamefont {Holzbauer}}, \bibinfo
  {author} {\bibfnamefont {T.}~\bibnamefont {Khabiboulline}}, \bibinfo {author}
  {\bibfnamefont {W.}~\bibnamefont {Schappert}}, \bibinfo {author}
  {\bibfnamefont {S.}~\bibnamefont {Smith}},\ and\ \bibinfo {author}
  {\bibfnamefont {J.}~\bibnamefont {Yun}},\ }\bibfield  {title} {\bibinfo
  {title} {{D}esign and {T}est of the {C}ompact {T}uner for {N}arrow
  {B}andwidth {SRF} {C}avities},\ }in\ \href
  {https://doi.org/https://doi.org/10.18429/JACoW-IPAC2015-WEPTY035} {\emph
  {\bibinfo {booktitle} {Proc. 6th International Particle Accelerator
  Conference (IPAC'15), Richmond, VA, USA, May 3-8, 2015}}},\ \bibinfo {series
  and number} {\bibinfo {series} {International Particle Accelerator
  Conference}\ No.~\bibinfo {number} {6}}\ (\bibinfo  {publisher} {JACoW},\
  \bibinfo {address} {Geneva, Switzerland},\ \bibinfo {year} {2015})\ pp.\
  \bibinfo {pages} {3352--3354},\ \bibinfo {note}
  {https://doi.org/10.18429/JACoW-IPAC2015-WEPTY035}\BibitemShut {NoStop}%
\bibitem [{\citenamefont {Padamsee}(2009)}]{Padamsee}%
  \BibitemOpen
  \bibfield  {author} {\bibinfo {author} {\bibfnamefont {H.}~\bibnamefont
  {Padamsee}},\ }\bibinfo {title} {Applications and operations},\ in\ \href
  {https://doi.org/10.1002/9783527627172.ch10} {\emph {\bibinfo {booktitle} {RF
  Superconductivity}}}\ (\bibinfo  {publisher} {John Wiley \& Sons, Ltd},\
  \bibinfo {year} {2009})\ Chap.~\bibinfo {chapter} {10}, pp.\ \bibinfo {pages}
  {313--332}\BibitemShut {NoStop}%
\bibitem [{\citenamefont {Corruccini}\ and\ \citenamefont
  {Gniewek}(1961)}]{Nb_CTE}%
  \BibitemOpen
  \bibfield  {author} {\bibinfo {author} {\bibfnamefont {R.~J.}\ \bibnamefont
  {Corruccini}}\ and\ \bibinfo {author} {\bibfnamefont {J.~J.}\ \bibnamefont
  {Gniewek}},\ }\href@noop {} {\emph {\bibinfo {title} {Thermal expansion of
  technical solids at low temperatures}}},\ \bibinfo {type} {Tech. Rep.}\
  \bibinfo {number} {{NBS Monograph 29}}\ (\bibinfo  {institution} {{National
  Bureau of Standards}},\ \bibinfo {year} {1961})\BibitemShut {NoStop}%
\bibitem [{\citenamefont {Poirier}\ \emph {et~al.}(1984)\citenamefont
  {Poirier}, \citenamefont {Plamondon}, \citenamefont {Cheeke},\ and\
  \citenamefont {Bussière}}]{Nb3Sn_elastic_constants}%
  \BibitemOpen
  \bibfield  {author} {\bibinfo {author} {\bibfnamefont {M.}~\bibnamefont
  {Poirier}}, \bibinfo {author} {\bibfnamefont {R.}~\bibnamefont {Plamondon}},
  \bibinfo {author} {\bibfnamefont {J.~D.~N.}\ \bibnamefont {Cheeke}},\ and\
  \bibinfo {author} {\bibfnamefont {J.~F.}\ \bibnamefont {Bussière}},\
  }\bibfield  {title} {\bibinfo {title} {Elastic constants of polycrystalline
  nb3sn between 4.2 and 300 k},\ }\href {https://doi.org/10.1063/1.333370}
  {\bibfield  {journal} {\bibinfo  {journal} {Journal of Applied Physics}\
  }\textbf {\bibinfo {volume} {55}},\ \bibinfo {pages} {3327} (\bibinfo {year}
  {1984})},\ \Eprint {https://arxiv.org/abs/https://doi.org/10.1063/1.333370}
  {https://doi.org/10.1063/1.333370} \BibitemShut {NoStop}%
\bibitem [{\citenamefont {Eremeev}\ \emph {et~al.}(2020)\citenamefont
  {Eremeev}, \citenamefont {Clemens}, \citenamefont {Macha}, \citenamefont
  {Reece}, \citenamefont {Valente-Feliciano}, \citenamefont {Williams},
  \citenamefont {Pudasaini},\ and\ \citenamefont {Kelley}}]{Eremeev_RSI}%
  \BibitemOpen
  \bibfield  {author} {\bibinfo {author} {\bibfnamefont {G.}~\bibnamefont
  {Eremeev}}, \bibinfo {author} {\bibfnamefont {W.}~\bibnamefont {Clemens}},
  \bibinfo {author} {\bibfnamefont {K.}~\bibnamefont {Macha}}, \bibinfo
  {author} {\bibfnamefont {C.~E.}\ \bibnamefont {Reece}}, \bibinfo {author}
  {\bibfnamefont {A.~M.}\ \bibnamefont {Valente-Feliciano}}, \bibinfo {author}
  {\bibfnamefont {S.}~\bibnamefont {Williams}}, \bibinfo {author}
  {\bibfnamefont {U.}~\bibnamefont {Pudasaini}},\ and\ \bibinfo {author}
  {\bibfnamefont {M.}~\bibnamefont {Kelley}},\ }\bibfield  {title} {\bibinfo
  {title} {{Nb$_3$Sn multicell cavity coating system at Jefferson Lab}},\
  }\href {https://doi.org/10.1063/1.5144490} {\bibfield  {journal} {\bibinfo
  {journal} {Review of Scientific Instruments}\ }\textbf {\bibinfo {volume}
  {91}},\ \bibinfo {pages} {073911} (\bibinfo {year} {2020})},\ \Eprint
  {https://arxiv.org/abs/https://doi.org/10.1063/1.5144490}
  {https://doi.org/10.1063/1.5144490} \BibitemShut {NoStop}%
\bibitem [{\citenamefont {Trenikhina}\ \emph {et~al.}(2017)\citenamefont
  {Trenikhina}, \citenamefont {Posen}, \citenamefont {Romanenko}, \citenamefont
  {Sardela}, \citenamefont {Zuo}, \citenamefont {Hall},\ and\ \citenamefont
  {Liepe}}]{Trenikhina_2018}%
  \BibitemOpen
  \bibfield  {author} {\bibinfo {author} {\bibfnamefont {Y.}~\bibnamefont
  {Trenikhina}}, \bibinfo {author} {\bibfnamefont {S.}~\bibnamefont {Posen}},
  \bibinfo {author} {\bibfnamefont {A.}~\bibnamefont {Romanenko}}, \bibinfo
  {author} {\bibfnamefont {M.}~\bibnamefont {Sardela}}, \bibinfo {author}
  {\bibfnamefont {J.-M.}\ \bibnamefont {Zuo}}, \bibinfo {author} {\bibfnamefont
  {D.~L.}\ \bibnamefont {Hall}},\ and\ \bibinfo {author} {\bibfnamefont
  {M.}~\bibnamefont {Liepe}},\ }\bibfield  {title} {\bibinfo {title}
  {{Performance-defining properties of Nb$_3$Sn coating in SRF cavities}},\
  }\href {https://doi.org/10.1088/1361-6668/aa9694} {\bibfield  {journal}
  {\bibinfo  {journal} {Superconductor Science and Technology}\ }\textbf
  {\bibinfo {volume} {31}},\ \bibinfo {pages} {015004} (\bibinfo {year}
  {2017})}\BibitemShut {NoStop}%
\bibitem [{\citenamefont {Pudasaini}\ \emph {et~al.}(2022)\citenamefont
  {Pudasaini}, \citenamefont {Reece},\ and\ \citenamefont
  {Tiskumara}}]{pudasaini:srf2021}%
  \BibitemOpen
  \bibfield  {author} {\bibinfo {author} {\bibfnamefont {U.}~\bibnamefont
  {Pudasaini}}, \bibinfo {author} {\bibfnamefont {C.}~\bibnamefont {Reece}},\
  and\ \bibinfo {author} {\bibfnamefont {J.}~\bibnamefont {Tiskumara}},\
  }\bibfield  {title} {\bibinfo {title} {{Managing Sn-Supply to Tune Surface
  Characteristics of Vapor-Diffusion Coating of Nb$_3$Sn}},\ }in\ \href
  {https://doi.org/10.18429/JACoW-SRF2021-TUPTEV013} {\emph {\bibinfo
  {booktitle} {Proc. SRF'21}}},\ \bibinfo {series and number} {\bibinfo
  {series} {International Conference on RF Superconductivity}\ No.~\bibinfo
  {number} {20}}\ (\bibinfo  {publisher} {JACoW Publishing, Geneva,
  Switzerland},\ \bibinfo {year} {2022})\ pp.\ \bibinfo {pages}
  {516--521}\BibitemShut {NoStop}%
\bibitem [{\citenamefont {Hall}\ \emph {et~al.}()\citenamefont {Hall},
  \citenamefont {Liepe}, \citenamefont {Liarte},\ and\ \citenamefont
  {Sethna}}]{Hall}%
  \BibitemOpen
  \bibfield  {author} {\bibinfo {author} {\bibfnamefont {D.}~\bibnamefont
  {Hall}}, \bibinfo {author} {\bibfnamefont {M.}~\bibnamefont {Liepe}},
  \bibinfo {author} {\bibfnamefont {D.}~\bibnamefont {Liarte}},\ and\ \bibinfo
  {author} {\bibfnamefont {J.~P.}\ \bibnamefont {Sethna}},\ }\bibfield  {title}
  {\bibinfo {title} {Impact of trapped magnetic flux and thermal gradients on
  the performance of {N}b$_3${S}n cavities},\ }in\ \href
  {https://doi.org/10.18429/JACoW-IPAC2017-MOPVA118} {\emph {\bibinfo
  {booktitle} {{Proc. 8th Int. Particle Accelerator Conf. (IPAC’17),
  Copenhagen, Denmark, May 2017}}}}\ (\bibinfo  {publisher} {JACoW Publishing,
  Geneva, Switzerland})\ pp.\ \bibinfo {pages} {1127--1129}\BibitemShut
  {NoStop}%
\bibitem [{\citenamefont {Turneaure}\ \emph {et~al.}(1991)\citenamefont
  {Turneaure}, \citenamefont {Halbritter},\ and\ \citenamefont
  {Schwettman}}]{Halb_JAP}%
  \BibitemOpen
  \bibfield  {author} {\bibinfo {author} {\bibfnamefont {J.~P.}\ \bibnamefont
  {Turneaure}}, \bibinfo {author} {\bibfnamefont {J.}~\bibnamefont
  {Halbritter}},\ and\ \bibinfo {author} {\bibfnamefont {H.~A.}\ \bibnamefont
  {Schwettman}},\ }\bibfield  {title} {\bibinfo {title} {{The surface impedance
  of superconductors and normal conductors: The Mattis-Bardeen theory}},\
  }\href {https://doi.org/10.1007/BF00618215} {\bibfield  {journal} {\bibinfo
  {journal} {Journal of Superconductivity}\ }\textbf {\bibinfo {volume} {4}},\
  \bibinfo {pages} {341} (\bibinfo {year} {1991})}\BibitemShut {NoStop}%
\bibitem [{\citenamefont {Hein}(1999)}]{Hein}%
  \BibitemOpen
  \bibfield  {author} {\bibinfo {author} {\bibfnamefont {M.}~\bibnamefont
  {Hein}},\ }\bibinfo {title} {Measurements of the surface impedance at linear
  response},\ in\ \href {https://doi.org/10.1007/BFb0111182} {\emph {\bibinfo
  {booktitle} {{High-Temperature-Superconductor Thin Films at Microwave
  Frequencies}}}}\ (\bibinfo  {publisher} {Springer Berlin},\ \bibinfo {year}
  {1999})\ Chap.~\bibinfo {chapter} {2}, pp.\ \bibinfo {pages}
  {43--102}\BibitemShut {NoStop}%
\bibitem [{\citenamefont {Trollier}\ \emph {et~al.}(2016)\citenamefont
  {Trollier}, \citenamefont {Tanchon}, \citenamefont {Lacapere}, \citenamefont
  {Renaud}, \citenamefont {Rey},\ and\ \citenamefont {Ravex}}]{Trollier}%
  \BibitemOpen
  \bibfield  {author} {\bibinfo {author} {\bibfnamefont {T.}~\bibnamefont
  {Trollier}}, \bibinfo {author} {\bibfnamefont {J.}~\bibnamefont {Tanchon}},
  \bibinfo {author} {\bibfnamefont {J.}~\bibnamefont {Lacapere}}, \bibinfo
  {author} {\bibfnamefont {P.}~\bibnamefont {Renaud}}, \bibinfo {author}
  {\bibfnamefont {J.}~\bibnamefont {Rey}},\ and\ \bibinfo {author}
  {\bibfnamefont {A.}~\bibnamefont {Ravex}},\ }\bibfield  {title} {\bibinfo
  {title} {Flexible thermal link assembly solutions for space applications},\
  }in\ \href {https://cryocooler.org/Cryocoolers-19} {\emph {\bibinfo
  {booktitle} {{Proc. Cryocoolers 19}}}}\ (\bibinfo  {publisher} {ICC Press,
  Boulder, CO},\ \bibinfo {year} {2016})\ pp.\ \bibinfo {pages}
  {595--603}\BibitemShut {NoStop}%
\bibitem [{\citenamefont {Cheng}\ \emph {et~al.}(2019)\citenamefont {Cheng},
  \citenamefont {Ciovati},\ and\ \citenamefont {Morrone}}]{cheng:srf2019}%
  \BibitemOpen
  \bibfield  {author} {\bibinfo {author} {\bibfnamefont {G.}~\bibnamefont
  {Cheng}}, \bibinfo {author} {\bibfnamefont {G.}~\bibnamefont {Ciovati}},\
  and\ \bibinfo {author} {\bibfnamefont {M.}~\bibnamefont {Morrone}},\
  }\bibfield  {title} {\bibinfo {title} {{Evaluation of Low Heat Conductivity
  RF Cables}},\ }in\ \href {https://doi.org/10.18429/JACoW-SRF2019-THP068}
  {\emph {\bibinfo {booktitle} {Proc. SRF'19}}},\ \bibinfo {series and number}
  {\bibinfo {series} {International Conference on RF Superconductivity}\
  No.~\bibinfo {number} {19}}\ (\bibinfo  {publisher} {JACoW Publishing,
  Geneva, Switzerland},\ \bibinfo {year} {2019})\ pp.\ \bibinfo {pages}
  {1045--1049},\ \bibinfo {note}
  {https://doi.org/10.18429/JACoW-SRF2019-THP068}\BibitemShut {NoStop}%
\bibitem [{\citenamefont {Powers}(2019)}]{powers:srf2019}%
  \BibitemOpen
  \bibfield  {author} {\bibinfo {author} {\bibfnamefont {T.}~\bibnamefont
  {Powers}},\ }\bibfield  {title} {\bibinfo {title} {{Practical Aspects of SRF
  Cavity Testing and Operations}}} (\bibinfo {year} {2019}),\ \bibinfo {note}
  {presented at SRF2019 in Dresden, Germany, unpublished}\BibitemShut {NoStop}%
\bibitem [{\citenamefont {{Van Harlingen}}(1982)}]{VANHARLINGEN1982}%
  \BibitemOpen
  \bibfield  {author} {\bibinfo {author} {\bibfnamefont {D.}~\bibnamefont {{Van
  Harlingen}}},\ }\bibfield  {title} {\bibinfo {title} {Thermoelectric effects
  in the superconducting state},\ }\href
  {https://doi.org/https://doi.org/10.1016/0378-4363(82)90195-4} {\bibfield
  {journal} {\bibinfo  {journal} {Physica B+C}\ }\textbf {\bibinfo {volume}
  {109-110}},\ \bibinfo {pages} {1710} (\bibinfo {year} {1982})},\ \bibinfo
  {note} {16th International Conference on Low Temperature Physics, Part
  3}\BibitemShut {NoStop}%
\bibitem [{\citenamefont {Shelly}\ \emph {et~al.}(2016)\citenamefont {Shelly},
  \citenamefont {Matrozova},\ and\ \citenamefont
  {Petrashov}}]{Thermoelectric_SC}%
  \BibitemOpen
  \bibfield  {author} {\bibinfo {author} {\bibfnamefont {C.~D.}\ \bibnamefont
  {Shelly}}, \bibinfo {author} {\bibfnamefont {E.~A.}\ \bibnamefont
  {Matrozova}},\ and\ \bibinfo {author} {\bibfnamefont {V.~T.}\ \bibnamefont
  {Petrashov}},\ }\bibfield  {title} {\bibinfo {title} {Resolving
  thermoelectric "paradox" in superconductors},\ }\href
  {https://doi.org/10.1126/sciadv.1501250} {\bibfield  {journal} {\bibinfo
  {journal} {Science Advances}\ }\textbf {\bibinfo {volume} {2}},\ \bibinfo
  {pages} {e1501250} (\bibinfo {year} {2016})},\ \Eprint
  {https://arxiv.org/abs/https://www.science.org/doi/pdf/10.1126/sciadv.1501250}
  {https://www.science.org/doi/pdf/10.1126/sciadv.1501250} \BibitemShut
  {NoStop}%
\bibitem [{\citenamefont {Kramer}\ \emph {et~al.}(2020)\citenamefont {Kramer},
  \citenamefont {Kugeler}, \citenamefont {K\"oszegi},\ and\ \citenamefont
  {Knobloch}}]{kugeler_PRAB2020}%
  \BibitemOpen
  \bibfield  {author} {\bibinfo {author} {\bibfnamefont {F.}~\bibnamefont
  {Kramer}}, \bibinfo {author} {\bibfnamefont {O.}~\bibnamefont {Kugeler}},
  \bibinfo {author} {\bibfnamefont {J.-M.}\ \bibnamefont {K\"oszegi}},\ and\
  \bibinfo {author} {\bibfnamefont {J.}~\bibnamefont {Knobloch}},\ }\bibfield
  {title} {\bibinfo {title} {Impact of geometry on flux trapping and the
  related surface resistance in a superconducting cavity},\ }\href
  {https://doi.org/10.1103/PhysRevAccelBeams.23.123101} {\bibfield  {journal}
  {\bibinfo  {journal} {Phys. Rev. Accel. Beams}\ }\textbf {\bibinfo {volume}
  {23}},\ \bibinfo {pages} {123101} (\bibinfo {year} {2020})}\BibitemShut
  {NoStop}%
\bibitem [{\citenamefont {Longuevergne}\ and\ \citenamefont
  {Miyazaki}(2021)}]{Miyazaki_PRAB2021}%
  \BibitemOpen
  \bibfield  {author} {\bibinfo {author} {\bibfnamefont {D.}~\bibnamefont
  {Longuevergne}}\ and\ \bibinfo {author} {\bibfnamefont {A.}~\bibnamefont
  {Miyazaki}},\ }\bibfield  {title} {\bibinfo {title} {Impact of geometry on
  the magnetic flux trapping of superconducting accelerating cavities},\ }\href
  {https://doi.org/10.1103/PhysRevAccelBeams.24.083101} {\bibfield  {journal}
  {\bibinfo  {journal} {Phys. Rev. Accel. Beams}\ }\textbf {\bibinfo {volume}
  {24}},\ \bibinfo {pages} {083101} (\bibinfo {year} {2021})}\BibitemShut
  {NoStop}%
\bibitem [{\citenamefont {Powers}\ \emph {et~al.}(2019)\citenamefont {Powers},
  \citenamefont {Brock},\ and\ \citenamefont {Davis}}]{Powers_microphonics}%
  \BibitemOpen
  \bibfield  {author} {\bibinfo {author} {\bibfnamefont {T.}~\bibnamefont
  {Powers}}, \bibinfo {author} {\bibfnamefont {N.}~\bibnamefont {Brock}},\ and\
  \bibinfo {author} {\bibfnamefont {G.}~\bibnamefont {Davis}},\ }\bibfield
  {title} {\bibinfo {title} {Microphonics testing of {LCLS} {II} cryomodules at
  {J}efferson {L}ab},\ }in\ \href
  {https://doi.org/doi:10.18429/JACoW-SRF2019-TUP034} {\emph {\bibinfo
  {booktitle} {Proc. SRF'19}}},\ \bibinfo {series and number} {\bibinfo
  {series} {International Conference on RF Superconductivity}\ No.~\bibinfo
  {number} {19}}\ (\bibinfo  {publisher} {JACoW Publishing, Geneva,
  Switzerland},\ \bibinfo {year} {2019})\ pp.\ \bibinfo {pages} {493--498},\
  \bibinfo {note} {https://doi.org/10.18429/JACoW-SRF2019-TUP034}\BibitemShut
  {NoStop}%
\bibitem [{\citenamefont {Bauer}\ \emph {et~al.}(2006)\citenamefont {Bauer},
  \citenamefont {Solyak}, \citenamefont {Ciovati}, \citenamefont {Eremeev},
  \citenamefont {Gurevich}, \citenamefont {Lilje},\ and\ \citenamefont
  {Visentin}}]{Bauer_PhysicaC}%
  \BibitemOpen
  \bibfield  {author} {\bibinfo {author} {\bibfnamefont {P.}~\bibnamefont
  {Bauer}}, \bibinfo {author} {\bibfnamefont {N.}~\bibnamefont {Solyak}},
  \bibinfo {author} {\bibfnamefont {G.}~\bibnamefont {Ciovati}}, \bibinfo
  {author} {\bibfnamefont {G.}~\bibnamefont {Eremeev}}, \bibinfo {author}
  {\bibfnamefont {A.}~\bibnamefont {Gurevich}}, \bibinfo {author}
  {\bibfnamefont {L.}~\bibnamefont {Lilje}},\ and\ \bibinfo {author}
  {\bibfnamefont {B.}~\bibnamefont {Visentin}},\ }\bibfield  {title} {\bibinfo
  {title} {Evidence for non-linear bcs resistance in srf cavities},\ }\href
  {https://doi.org/https://doi.org/10.1016/j.physc.2006.03.056} {\bibfield
  {journal} {\bibinfo  {journal} {Physica C: Superconductivity}\ }\textbf
  {\bibinfo {volume} {441}},\ \bibinfo {pages} {51} (\bibinfo {year} {2006})},\
  \bibinfo {note} {proceedings of the 12th International Workshop on RF
  Superconductivity}\BibitemShut {NoStop}%
\bibitem [{\citenamefont {Gurevich}(2012)}]{Gurevich_RAST}%
  \BibitemOpen
  \bibfield  {author} {\bibinfo {author} {\bibfnamefont {A.}~\bibnamefont
  {Gurevich}},\ }\bibfield  {title} {\bibinfo {title} {Superconducting
  radio-frequency fundamentals for particle accelerators},\ }\href
  {https://doi.org/10.1142/S1793626812300058} {\bibfield  {journal} {\bibinfo
  {journal} {Reviews of Accelerator Science and Technology}\ }\textbf {\bibinfo
  {volume} {05}},\ \bibinfo {pages} {119} (\bibinfo {year} {2012})},\ \Eprint
  {https://arxiv.org/abs/https://doi.org/10.1142/S1793626812300058}
  {https://doi.org/10.1142/S1793626812300058} \BibitemShut {NoStop}%
\bibitem [{\citenamefont {Van~Sciver}(1986)}]{VanSciver1986}%
  \BibitemOpen
  \bibfield  {author} {\bibinfo {author} {\bibfnamefont {S.~W.}\ \bibnamefont
  {Van~Sciver}},\ }\bibinfo {title} {Pool boiling he i heat transfer},\ in\
  \href {https://doi.org/10.1007/978-1-4899-0499-7_6} {\emph {\bibinfo
  {booktitle} {Helium Cryogenics}}}\ (\bibinfo  {publisher} {Springer US},\
  \bibinfo {address} {Boston, MA},\ \bibinfo {year} {1986})\ pp.\ \bibinfo
  {pages} {199--238}\BibitemShut {NoStop}%
\bibitem [{\citenamefont {Posen}\ and\ \citenamefont
  {Liepe}(2014)}]{Posen_PhysRevSTAB.17.112001}%
  \BibitemOpen
  \bibfield  {author} {\bibinfo {author} {\bibfnamefont {S.}~\bibnamefont
  {Posen}}\ and\ \bibinfo {author} {\bibfnamefont {M.}~\bibnamefont {Liepe}},\
  }\bibfield  {title} {\bibinfo {title} {Advances in development of
  ${\mathrm{nb}}_{3}\mathrm{Sn}$ superconducting radio-frequency cavities},\
  }\href {https://doi.org/10.1103/PhysRevSTAB.17.112001} {\bibfield  {journal}
  {\bibinfo  {journal} {Phys. Rev. ST Accel. Beams}\ }\textbf {\bibinfo
  {volume} {17}},\ \bibinfo {pages} {112001} (\bibinfo {year}
  {2014})}\BibitemShut {NoStop}%
\bibitem [{\citenamefont {Lee}\ \emph {et~al.}(2020)\citenamefont {Lee},
  \citenamefont {Mao}, \citenamefont {He}, \citenamefont {Sung}, \citenamefont
  {Spina}, \citenamefont {Baik}, \citenamefont {Hall}, \citenamefont {Liepe},
  \citenamefont {Seidman},\ and\ \citenamefont {Posen}}]{Nb3Sn_GB}%
  \BibitemOpen
  \bibfield  {author} {\bibinfo {author} {\bibfnamefont {J.}~\bibnamefont
  {Lee}}, \bibinfo {author} {\bibfnamefont {Z.}~\bibnamefont {Mao}}, \bibinfo
  {author} {\bibfnamefont {K.}~\bibnamefont {He}}, \bibinfo {author}
  {\bibfnamefont {Z.~H.}\ \bibnamefont {Sung}}, \bibinfo {author}
  {\bibfnamefont {T.}~\bibnamefont {Spina}}, \bibinfo {author} {\bibfnamefont
  {S.-I.}\ \bibnamefont {Baik}}, \bibinfo {author} {\bibfnamefont {D.~L.}\
  \bibnamefont {Hall}}, \bibinfo {author} {\bibfnamefont {M.}~\bibnamefont
  {Liepe}}, \bibinfo {author} {\bibfnamefont {D.~N.}\ \bibnamefont {Seidman}},\
  and\ \bibinfo {author} {\bibfnamefont {S.}~\bibnamefont {Posen}},\ }\bibfield
   {title} {\bibinfo {title} {Grain-boundary structure and segregation in nb3sn
  coatings on nb for high-performance superconducting radiofrequency cavity
  applications},\ }\href
  {https://doi.org/https://doi.org/10.1016/j.actamat.2020.01.055} {\bibfield
  {journal} {\bibinfo  {journal} {Acta Materialia}\ }\textbf {\bibinfo {volume}
  {188}},\ \bibinfo {pages} {155} (\bibinfo {year} {2020})}\BibitemShut
  {NoStop}%
\bibitem [{\citenamefont {Suenaga}\ and\ \citenamefont
  {Jansen}(1983)}]{Nb3Sn_GB2}%
  \BibitemOpen
  \bibfield  {author} {\bibinfo {author} {\bibfnamefont {M.}~\bibnamefont
  {Suenaga}}\ and\ \bibinfo {author} {\bibfnamefont {W.}~\bibnamefont
  {Jansen}},\ }\bibfield  {title} {\bibinfo {title} {Chemical compositions at
  and near the grain boundaries in bronze‐processed superconducting
  nb$_3$sn},\ }\href {https://doi.org/10.1063/1.94457} {\bibfield  {journal}
  {\bibinfo  {journal} {Applied Physics Letters}\ }\textbf {\bibinfo {volume}
  {43}},\ \bibinfo {pages} {791} (\bibinfo {year} {1983})},\ \Eprint
  {https://arxiv.org/abs/https://doi.org/10.1063/1.94457}
  {https://doi.org/10.1063/1.94457} \BibitemShut {NoStop}%
\bibitem [{\citenamefont {Sandim}\ \emph {et~al.}(2013)\citenamefont {Sandim},
  \citenamefont {Tytko}, \citenamefont {Kostka}, \citenamefont {Choi},
  \citenamefont {Awaji}, \citenamefont {Watanabe},\ and\ \citenamefont
  {Raabe}}]{Nb3Sn_GB3}%
  \BibitemOpen
  \bibfield  {author} {\bibinfo {author} {\bibfnamefont {M.~J.~R.}\
  \bibnamefont {Sandim}}, \bibinfo {author} {\bibfnamefont {D.}~\bibnamefont
  {Tytko}}, \bibinfo {author} {\bibfnamefont {A.}~\bibnamefont {Kostka}},
  \bibinfo {author} {\bibfnamefont {P.}~\bibnamefont {Choi}}, \bibinfo {author}
  {\bibfnamefont {S.}~\bibnamefont {Awaji}}, \bibinfo {author} {\bibfnamefont
  {K.}~\bibnamefont {Watanabe}},\ and\ \bibinfo {author} {\bibfnamefont
  {D.}~\bibnamefont {Raabe}},\ }\bibfield  {title} {\bibinfo {title} {Grain
  boundary segregation in a bronze-route nb3sn superconducting wire studied by
  atom probe tomography},\ }\href
  {https://doi.org/10.1088/0953-2048/26/5/055008} {\bibfield  {journal}
  {\bibinfo  {journal} {Superconductor Science and Technology}\ }\textbf
  {\bibinfo {volume} {26}},\ \bibinfo {pages} {055008} (\bibinfo {year}
  {2013})}\BibitemShut {NoStop}%
\bibitem [{\citenamefont {Lee}\ \emph {et~al.}(2018)\citenamefont {Lee},
  \citenamefont {Posen}, \citenamefont {Mao}, \citenamefont {Trenikhina},
  \citenamefont {He}, \citenamefont {Hall}, \citenamefont {Liepe},\ and\
  \citenamefont {Seidman}}]{Nb3Sn_GB4}%
  \BibitemOpen
  \bibfield  {author} {\bibinfo {author} {\bibfnamefont {J.}~\bibnamefont
  {Lee}}, \bibinfo {author} {\bibfnamefont {S.}~\bibnamefont {Posen}}, \bibinfo
  {author} {\bibfnamefont {Z.}~\bibnamefont {Mao}}, \bibinfo {author}
  {\bibfnamefont {Y.}~\bibnamefont {Trenikhina}}, \bibinfo {author}
  {\bibfnamefont {K.}~\bibnamefont {He}}, \bibinfo {author} {\bibfnamefont
  {D.~L.}\ \bibnamefont {Hall}}, \bibinfo {author} {\bibfnamefont
  {M.}~\bibnamefont {Liepe}},\ and\ \bibinfo {author} {\bibfnamefont {D.~N.}\
  \bibnamefont {Seidman}},\ }\bibfield  {title} {\bibinfo {title}
  {{Atomic-scale analyses of Nb$_3$Sn on Nb prepared by vapor diffusion for
  superconducting radiofrequency cavity applications: a correlative study}},\
  }\href {https://doi.org/10.1088/1361-6668/aaf268} {\bibfield  {journal}
  {\bibinfo  {journal} {Superconductor Science and Technology}\ }\textbf
  {\bibinfo {volume} {32}},\ \bibinfo {pages} {024001} (\bibinfo {year}
  {2018})}\BibitemShut {NoStop}%
\bibitem [{\citenamefont {Hylton}\ and\ \citenamefont
  {Beasley}(1989)}]{GB_diss_1}%
  \BibitemOpen
  \bibfield  {author} {\bibinfo {author} {\bibfnamefont {T.~L.}\ \bibnamefont
  {Hylton}}\ and\ \bibinfo {author} {\bibfnamefont {M.~R.}\ \bibnamefont
  {Beasley}},\ }\bibfield  {title} {\bibinfo {title} {Effect of grain
  boundaries on magnetic field penetration in polycrystalline
  superconductors},\ }\href {https://doi.org/10.1103/PhysRevB.39.9042}
  {\bibfield  {journal} {\bibinfo  {journal} {Phys. Rev. B}\ }\textbf {\bibinfo
  {volume} {39}},\ \bibinfo {pages} {9042} (\bibinfo {year}
  {1989})}\BibitemShut {NoStop}%
\bibitem [{\citenamefont {McDonald}\ and\ \citenamefont
  {Clem}(1997)}]{GB_diss_2}%
  \BibitemOpen
  \bibfield  {author} {\bibinfo {author} {\bibfnamefont {J.}~\bibnamefont
  {McDonald}}\ and\ \bibinfo {author} {\bibfnamefont {J.~R.}\ \bibnamefont
  {Clem}},\ }\bibfield  {title} {\bibinfo {title} {Microwave response and
  surface impedance of weak links},\ }\href
  {https://doi.org/10.1103/PhysRevB.56.14723} {\bibfield  {journal} {\bibinfo
  {journal} {Phys. Rev. B}\ }\textbf {\bibinfo {volume} {56}},\ \bibinfo
  {pages} {14723} (\bibinfo {year} {1997})}\BibitemShut {NoStop}%
\bibitem [{\citenamefont {Sheikhzada}\ and\ \citenamefont
  {Gurevich}(2017)}]{GB_diss_3}%
  \BibitemOpen
  \bibfield  {author} {\bibinfo {author} {\bibfnamefont {A.}~\bibnamefont
  {Sheikhzada}}\ and\ \bibinfo {author} {\bibfnamefont {A.}~\bibnamefont
  {Gurevich}},\ }\bibfield  {title} {\bibinfo {title} {Dynamic transition of
  vortices into phase slips and generation of vortex-antivortex pairs in thin
  film josephson junctions under dc and ac currents},\ }\href
  {https://doi.org/10.1103/PhysRevB.95.214507} {\bibfield  {journal} {\bibinfo
  {journal} {Phys. Rev. B}\ }\textbf {\bibinfo {volume} {95}},\ \bibinfo
  {pages} {214507} (\bibinfo {year} {2017})}\BibitemShut {NoStop}%
\bibitem [{\citenamefont {Godeke}(2006)}]{Godeke_2006}%
  \BibitemOpen
  \bibfield  {author} {\bibinfo {author} {\bibfnamefont {A.}~\bibnamefont
  {Godeke}},\ }\bibfield  {title} {\bibinfo {title} {{A review of the
  properties of Nb$_3$Sn and their variation with A15 composition, morphology
  and strain state}},\ }\href {https://doi.org/10.1088/0953-2048/19/8/R02}
  {\bibfield  {journal} {\bibinfo  {journal} {Superconductor Science and
  Technology}\ }\textbf {\bibinfo {volume} {19}},\ \bibinfo {pages} {R68}
  (\bibinfo {year} {2006})}\BibitemShut {NoStop}%
\bibitem [{\citenamefont {Marzi}\ \emph {et~al.}(2013)\citenamefont {Marzi},
  \citenamefont {Morici}, \citenamefont {Muzzi}, \citenamefont {della Corte},\
  and\ \citenamefont {Nardelli}}]{DeMarzi_2013}%
  \BibitemOpen
  \bibfield  {author} {\bibinfo {author} {\bibfnamefont {G.~D.}\ \bibnamefont
  {Marzi}}, \bibinfo {author} {\bibfnamefont {L.}~\bibnamefont {Morici}},
  \bibinfo {author} {\bibfnamefont {L.}~\bibnamefont {Muzzi}}, \bibinfo
  {author} {\bibfnamefont {A.}~\bibnamefont {della Corte}},\ and\ \bibinfo
  {author} {\bibfnamefont {M.~B.}\ \bibnamefont {Nardelli}},\ }\bibfield
  {title} {\bibinfo {title} {Strain sensitivity and superconducting properties
  of {N}b$_3${S}n from first principles calculations},\ }\href
  {https://doi.org/10.1088/0953-8984/25/13/135702} {\bibfield  {journal}
  {\bibinfo  {journal} {Journal of Physics: Condensed Matter}\ }\textbf
  {\bibinfo {volume} {25}},\ \bibinfo {pages} {135702} (\bibinfo {year}
  {2013})}\BibitemShut {NoStop}%
\bibitem [{\citenamefont {Ding}\ and\ \citenamefont {Gao}(2021)}]{Ding_2021}%
  \BibitemOpen
  \bibfield  {author} {\bibinfo {author} {\bibfnamefont {H.}~\bibnamefont
  {Ding}}\ and\ \bibinfo {author} {\bibfnamefont {Y.}~\bibnamefont {Gao}},\
  }\bibfield  {title} {\bibinfo {title} {Analysis of the strain dependence of
  the superconducting critical properties of single-crystal and polycrystalline
  {N}b$_3${S}n},\ }\href {https://doi.org/10.1088/1361-6668/abfaef} {\bibfield
  {journal} {\bibinfo  {journal} {Superconductor Science and Technology}\
  }\textbf {\bibinfo {volume} {34}},\ \bibinfo {pages} {075006} (\bibinfo
  {year} {2021})}\BibitemShut {NoStop}%
\bibitem [{\citenamefont {Gurevich}(2017)}]{Gurevich_2017}%
  \BibitemOpen
  \bibfield  {author} {\bibinfo {author} {\bibfnamefont {A.}~\bibnamefont
  {Gurevich}},\ }\bibfield  {title} {\bibinfo {title} {Theory of rf
  superconductivity for resonant cavities},\ }\href
  {https://doi.org/10.1088/1361-6668/30/3/034004} {\bibfield  {journal}
  {\bibinfo  {journal} {Superconductor Science and Technology}\ }\textbf
  {\bibinfo {volume} {30}},\ \bibinfo {pages} {034004} (\bibinfo {year}
  {2017})}\BibitemShut {NoStop}%
\bibitem [{\citenamefont {Makita}\ \emph {et~al.}(2022)\citenamefont {Makita},
  \citenamefont {Sundahl}, \citenamefont {Ciovati}, \citenamefont {Eom},\ and\
  \citenamefont {Gurevich}}]{Junki_Nb3Sn}%
  \BibitemOpen
  \bibfield  {author} {\bibinfo {author} {\bibfnamefont {J.}~\bibnamefont
  {Makita}}, \bibinfo {author} {\bibfnamefont {C.}~\bibnamefont {Sundahl}},
  \bibinfo {author} {\bibfnamefont {G.}~\bibnamefont {Ciovati}}, \bibinfo
  {author} {\bibfnamefont {C.~B.}\ \bibnamefont {Eom}},\ and\ \bibinfo {author}
  {\bibfnamefont {A.}~\bibnamefont {Gurevich}},\ }\bibfield  {title} {\bibinfo
  {title} {{Nonlinear Meissner effect in ${\mathrm{Nb}}_{3}\mathrm{Sn}$
  coplanar resonators}},\ }\href
  {https://doi.org/10.1103/PhysRevResearch.4.013156} {\bibfield  {journal}
  {\bibinfo  {journal} {Phys. Rev. Res.}\ }\textbf {\bibinfo {volume} {4}},\
  \bibinfo {pages} {013156} (\bibinfo {year} {2022})}\BibitemShut {NoStop}%
\bibitem [{\citenamefont {Pudasaini}\ \emph {et~al.}(2020)\citenamefont
  {Pudasaini}, \citenamefont {Eremeev}, \citenamefont {Reece}, \citenamefont
  {Tuggle},\ and\ \citenamefont {Kelley}}]{Pudasaini_2020}%
  \BibitemOpen
  \bibfield  {author} {\bibinfo {author} {\bibfnamefont {U.}~\bibnamefont
  {Pudasaini}}, \bibinfo {author} {\bibfnamefont {G.~V.}\ \bibnamefont
  {Eremeev}}, \bibinfo {author} {\bibfnamefont {C.~E.}\ \bibnamefont {Reece}},
  \bibinfo {author} {\bibfnamefont {J.}~\bibnamefont {Tuggle}},\ and\ \bibinfo
  {author} {\bibfnamefont {M.~J.}\ \bibnamefont {Kelley}},\ }\bibfield  {title}
  {\bibinfo {title} {{Analysis of RF losses and material characterization of
  samples removed from a Nb$_3$Sn-coated superconducting RF cavity}},\ }\href
  {https://doi.org/10.1088/1361-6668/ab75a8} {\bibfield  {journal} {\bibinfo
  {journal} {Superconductor Science and Technology}\ }\textbf {\bibinfo
  {volume} {33}},\ \bibinfo {pages} {045012} (\bibinfo {year}
  {2020})}\BibitemShut {NoStop}%
\bibitem [{\citenamefont {Eremeev}\ \emph {et~al.}(2019)\citenamefont
  {Eremeev}, \citenamefont {Crahen}, \citenamefont {Henry}, \citenamefont
  {Marhauser}, \citenamefont {Pudasaini},\ and\ \citenamefont
  {Reece}}]{eremeev:srf2019-mop015}%
  \BibitemOpen
  \bibfield  {author} {\bibinfo {author} {\bibfnamefont {G.}~\bibnamefont
  {Eremeev}}, \bibinfo {author} {\bibfnamefont {W.}~\bibnamefont {Crahen}},
  \bibinfo {author} {\bibfnamefont {J.}~\bibnamefont {Henry}}, \bibinfo
  {author} {\bibfnamefont {F.}~\bibnamefont {Marhauser}}, \bibinfo {author}
  {\bibfnamefont {U.}~\bibnamefont {Pudasaini}},\ and\ \bibinfo {author}
  {\bibfnamefont {C.}~\bibnamefont {Reece}},\ }\bibfield  {title} {\bibinfo
  {title} {{RF Performance Sensitivity to Tuning of Nb$_3$Sn Coated CEBAF
  Cavities}},\ }in\ \href {https://doi.org/10.18429/JACoW-SRF2019-MOP015}
  {\emph {\bibinfo {booktitle} {Proc. SRF'19}}},\ \bibinfo {series and number}
  {\bibinfo {series} {International Conference on RF Superconductivity}\
  No.~\bibinfo {number} {19}}\ (\bibinfo  {publisher} {JACoW Publishing,
  Geneva, Switzerland},\ \bibinfo {year} {2019})\ pp.\ \bibinfo {pages}
  {55--59},\ \bibinfo {note}
  {https://doi.org/10.18429/JACoW-SRF2019-MOP015}\BibitemShut {NoStop}%
\bibitem [{\citenamefont {Tomaru}\ \emph {et~al.}(2004)\citenamefont {Tomaru},
  \citenamefont {Suzuki}, \citenamefont {Haruyama}, \citenamefont {Shintomi},
  \citenamefont {Yamamoto}, \citenamefont {Koyama},\ and\ \citenamefont
  {Li}}]{CCR_vibration}%
  \BibitemOpen
  \bibfield  {author} {\bibinfo {author} {\bibfnamefont {T.}~\bibnamefont
  {Tomaru}}, \bibinfo {author} {\bibfnamefont {T.}~\bibnamefont {Suzuki}},
  \bibinfo {author} {\bibfnamefont {T.}~\bibnamefont {Haruyama}}, \bibinfo
  {author} {\bibfnamefont {T.}~\bibnamefont {Shintomi}}, \bibinfo {author}
  {\bibfnamefont {A.}~\bibnamefont {Yamamoto}}, \bibinfo {author}
  {\bibfnamefont {T.}~\bibnamefont {Koyama}},\ and\ \bibinfo {author}
  {\bibfnamefont {R.}~\bibnamefont {Li}},\ }\bibfield  {title} {\bibinfo
  {title} {Vibration analysis of cryocoolers},\ }\href
  {https://doi.org/https://doi.org/10.1016/j.cryogenics.2004.02.003} {\bibfield
   {journal} {\bibinfo  {journal} {Cryogenics}\ }\textbf {\bibinfo {volume}
  {44}},\ \bibinfo {pages} {309} (\bibinfo {year} {2004})}\BibitemShut
  {NoStop}%
\bibitem [{\citenamefont {Bachimanchi}\ \emph {et~al.}(2015)\citenamefont
  {Bachimanchi}, \citenamefont {Allison}, \citenamefont {Daly}, \citenamefont
  {Drury}, \citenamefont {Hovater}, \citenamefont {Lahti}, \citenamefont
  {Mounts}, \citenamefont {Nelson},\ and\ \citenamefont
  {Plawski}}]{Bachimanchi:IPAC2015-THXB1}%
  \BibitemOpen
  \bibfield  {author} {\bibinfo {author} {\bibfnamefont {R.}~\bibnamefont
  {Bachimanchi}}, \bibinfo {author} {\bibfnamefont {T.}~\bibnamefont
  {Allison}}, \bibinfo {author} {\bibfnamefont {E.}~\bibnamefont {Daly}},
  \bibinfo {author} {\bibfnamefont {M.}~\bibnamefont {Drury}}, \bibinfo
  {author} {\bibfnamefont {C.}~\bibnamefont {Hovater}}, \bibinfo {author}
  {\bibfnamefont {G.}~\bibnamefont {Lahti}}, \bibinfo {author} {\bibfnamefont
  {C.}~\bibnamefont {Mounts}}, \bibinfo {author} {\bibfnamefont
  {R.}~\bibnamefont {Nelson}},\ and\ \bibinfo {author} {\bibfnamefont {T.~E.}\
  \bibnamefont {Plawski}},\ }\bibfield  {title} {\bibinfo {title} {{CEBAF}
  {SRF} {P}erformance during {I}nitial 12 {G}e{V} {C}ommissioning},\ }in\ \href
  {https://doi.org/https://doi.org/10.18429/JACoW-IPAC2015-THXB1} {\emph
  {\bibinfo {booktitle} {Proc. 6th International Particle Accelerator
  Conference (IPAC'15), Richmond, VA, USA, May 3-8, 2015}}},\ \bibinfo {series
  and number} {\bibinfo {series} {International Particle Accelerator
  Conference}\ No.~\bibinfo {number} {6}}\ (\bibinfo  {publisher} {JACoW},\
  \bibinfo {address} {Geneva, Switzerland},\ \bibinfo {year} {2015})\ pp.\
  \bibinfo {pages} {3638--3642},\ \bibinfo {note}
  {https://doi.org/10.18429/JACoW-IPAC2015-THXB1}\BibitemShut {NoStop}%
\bibitem [{\citenamefont {{Model PT420}}()}]{PT}%
  \BibitemOpen
  \bibfield  {author} {\bibinfo {author} {\bibnamefont {{Model PT420}}},\
  }\href@noop {} {\bibinfo {title} {{Cryomech, Syracuse, NY, USA}}},\ \bibinfo
  {howpublished} {\url{https://www.cryomech.com/products/pt420/}},\ \bibinfo
  {note} {accessed: 2023-01-10}\BibitemShut {NoStop}%
\bibitem [{\citenamefont {{Model RDE-418D4}}()}]{GM}%
  \BibitemOpen
  \bibfield  {author} {\bibinfo {author} {\bibnamefont {{Model RDE-418D4}}},\
  }\href@noop {} {\bibinfo {title} {{Sumitomo (SHI) Cryogenics of America, Inc,
  Allentown, PA, USA}}},\ \bibinfo {howpublished}
  {\url{https://www.shicryogenics.com/product/rde-418d4-4k-cryocooler-series/}},\
  \bibinfo {note} {accessed: 2023-01-10}\BibitemShut {NoStop}%
\bibitem [{\citenamefont {Woodcraft}(2005)}]{Thermal_cond_Al}%
  \BibitemOpen
  \bibfield  {author} {\bibinfo {author} {\bibfnamefont {A.~L.}\ \bibnamefont
  {Woodcraft}},\ }\bibfield  {title} {\bibinfo {title} {Recommended values for
  the thermal conductivity of aluminium of different purities in the cryogenic
  to room temperature range, and a comparison with copper},\ }\href
  {https://doi.org/https://doi.org/10.1016/j.cryogenics.2005.06.008} {\bibfield
   {journal} {\bibinfo  {journal} {Cryogenics}\ }\textbf {\bibinfo {volume}
  {45}},\ \bibinfo {pages} {626} (\bibinfo {year} {2005})}\BibitemShut
  {NoStop}%
\bibitem [{\citenamefont {Dhuley}\ \emph {et~al.}(2019)\citenamefont {Dhuley},
  \citenamefont {Kostin}, \citenamefont {Prokofiev}, \citenamefont {Geelhoed},
  \citenamefont {Nicol}, \citenamefont {Posen}, \citenamefont {Thangaraj},
  \citenamefont {Kroc},\ and\ \citenamefont {Kephart}}]{Dhuley_TLA}%
  \BibitemOpen
  \bibfield  {author} {\bibinfo {author} {\bibfnamefont {R.~C.}\ \bibnamefont
  {Dhuley}}, \bibinfo {author} {\bibfnamefont {R.}~\bibnamefont {Kostin}},
  \bibinfo {author} {\bibfnamefont {O.}~\bibnamefont {Prokofiev}}, \bibinfo
  {author} {\bibfnamefont {M.~I.}\ \bibnamefont {Geelhoed}}, \bibinfo {author}
  {\bibfnamefont {T.~H.}\ \bibnamefont {Nicol}}, \bibinfo {author}
  {\bibfnamefont {S.}~\bibnamefont {Posen}}, \bibinfo {author} {\bibfnamefont
  {J.~C.~T.}\ \bibnamefont {Thangaraj}}, \bibinfo {author} {\bibfnamefont
  {T.~K.}\ \bibnamefont {Kroc}},\ and\ \bibinfo {author} {\bibfnamefont
  {R.~D.}\ \bibnamefont {Kephart}},\ }\bibfield  {title} {\bibinfo {title}
  {Thermal link design for conduction cooling of srf cavities using
  cryocoolers},\ }\href {https://doi.org/10.1109/TASC.2019.2901252} {\bibfield
  {journal} {\bibinfo  {journal} {IEEE Transactions on Applied
  Superconductivity}\ }\textbf {\bibinfo {volume} {29}},\ \bibinfo {pages} {1}
  (\bibinfo {year} {2019})}\BibitemShut {NoStop}%
\bibitem [{\citenamefont {Dhuley}\ \emph {et~al.}(2018)\citenamefont {Dhuley},
  \citenamefont {Geelhoed},\ and\ \citenamefont
  {Thangaraj}}]{Dhuley_thermalRes}%
  \BibitemOpen
  \bibfield  {author} {\bibinfo {author} {\bibfnamefont {R.}~\bibnamefont
  {Dhuley}}, \bibinfo {author} {\bibfnamefont {M.}~\bibnamefont {Geelhoed}},\
  and\ \bibinfo {author} {\bibfnamefont {J.}~\bibnamefont {Thangaraj}},\
  }\bibfield  {title} {\bibinfo {title} {Thermal resistance of pressed contacts
  of aluminum and niobium at liquid helium temperatures},\ }\href
  {https://doi.org/https://doi.org/10.1016/j.cryogenics.2018.06.003} {\bibfield
   {journal} {\bibinfo  {journal} {Cryogenics}\ }\textbf {\bibinfo {volume}
  {93}},\ \bibinfo {pages} {86} (\bibinfo {year} {2018})}\BibitemShut {NoStop}%
\bibitem [{\citenamefont {Kerr}\ and\ \citenamefont {Horne}(1991)}]{JointTR1}%
  \BibitemOpen
  \bibfield  {author} {\bibinfo {author} {\bibfnamefont {A.~R.}\ \bibnamefont
  {Kerr}}\ and\ \bibinfo {author} {\bibfnamefont {N.}~\bibnamefont {Horne}},\
  }\href@noop {} {\emph {\bibinfo {title} {The low temperature thermal
  resistance of high purity copper and bolted copper joints.}}},\ \bibinfo
  {type} {Tech. Rep.}\ \bibinfo {number} {EDTN 163}\ (\bibinfo  {institution}
  {{National Radio Astronomy Observatory - Electronic Division}},\ \bibinfo
  {year} {1991})\BibitemShut {NoStop}%
\bibitem [{\citenamefont {Kerr}\ and\ \citenamefont {Groves}(2006)}]{JointTR2}%
  \BibitemOpen
  \bibfield  {author} {\bibinfo {author} {\bibfnamefont {A.~R.}\ \bibnamefont
  {Kerr}}\ and\ \bibinfo {author} {\bibfnamefont {R.}~\bibnamefont {Groves}},\
  }\href@noop {} {\emph {\bibinfo {title} {Measurements of Copper Heat Straps
  Near 4 {K} With and Without {Apiezon-N} Grease}}},\ \bibinfo {type} {Tech.
  Rep.}\ \bibinfo {number} {EDTN 204}\ (\bibinfo  {institution} {{National
  Radio Astronomy Observatory - Electronic Division}},\ \bibinfo {year}
  {2006})\BibitemShut {NoStop}%
\bibitem [{\citenamefont {Salerno}\ \emph {et~al.}(1994)\citenamefont
  {Salerno}, \citenamefont {Kittel},\ and\ \citenamefont {Spivak}}]{JointTR3}%
  \BibitemOpen
  \bibfield  {author} {\bibinfo {author} {\bibfnamefont {L.}~\bibnamefont
  {Salerno}}, \bibinfo {author} {\bibfnamefont {P.}~\bibnamefont {Kittel}},\
  and\ \bibinfo {author} {\bibfnamefont {A.}~\bibnamefont {Spivak}},\
  }\bibfield  {title} {\bibinfo {title} {Thermal conductance of pressed
  metallic contacts augmented with indium foil or {A}piezon grease at liquid
  helium temperatures},\ }\href
  {https://doi.org/https://doi.org/10.1016/0011-2275(94)90142-2} {\bibfield
  {journal} {\bibinfo  {journal} {Cryogenics}\ }\textbf {\bibinfo {volume}
  {34}},\ \bibinfo {pages} {649} (\bibinfo {year} {1994})}\BibitemShut
  {NoStop}%
\bibitem [{\citenamefont {Dhuley}(2019)}]{JointTR4}%
  \BibitemOpen
  \bibfield  {author} {\bibinfo {author} {\bibfnamefont {R.}~\bibnamefont
  {Dhuley}},\ }\bibfield  {title} {\bibinfo {title} {Pressed copper and
  gold-plated copper contacts at low temperatures – a review of thermal
  contact resistance},\ }\href
  {https://doi.org/https://doi.org/10.1016/j.cryogenics.2019.06.008} {\bibfield
   {journal} {\bibinfo  {journal} {Cryogenics}\ }\textbf {\bibinfo {volume}
  {101}},\ \bibinfo {pages} {111} (\bibinfo {year} {2019})}\BibitemShut
  {NoStop}%
\bibitem [{\citenamefont {Withanage}\ \emph {et~al.}(2021)\citenamefont
  {Withanage}, \citenamefont {Juliao},\ and\ \citenamefont
  {Cooley}}]{Withanage_2021}%
  \BibitemOpen
  \bibfield  {author} {\bibinfo {author} {\bibfnamefont {W.~K.}\ \bibnamefont
  {Withanage}}, \bibinfo {author} {\bibfnamefont {A.}~\bibnamefont {Juliao}},\
  and\ \bibinfo {author} {\bibfnamefont {L.~D.}\ \bibnamefont {Cooley}},\
  }\bibfield  {title} {\bibinfo {title} {Rapid {N}b$_3${S}n film growth by
  sputtering {N}b on hot bronze},\ }\href
  {https://doi.org/10.1088/1361-6668/abf66f} {\bibfield  {journal} {\bibinfo
  {journal} {Superconductor Science and Technology}\ }\textbf {\bibinfo
  {volume} {34}},\ \bibinfo {pages} {06LT01} (\bibinfo {year}
  {2021})}\BibitemShut {NoStop}%
\bibitem [{\citenamefont {Ilyina}\ \emph {et~al.}(2019)\citenamefont {Ilyina},
  \citenamefont {Rosaz}, \citenamefont {Descarrega}, \citenamefont
  {Vollenberg}, \citenamefont {Lunt}, \citenamefont {Leaux}, \citenamefont
  {Calatroni}, \citenamefont {Venturini-Delsolaro},\ and\ \citenamefont
  {Taborelli}}]{Ilyina_2019}%
  \BibitemOpen
  \bibfield  {author} {\bibinfo {author} {\bibfnamefont {E.~A.}\ \bibnamefont
  {Ilyina}}, \bibinfo {author} {\bibfnamefont {G.}~\bibnamefont {Rosaz}},
  \bibinfo {author} {\bibfnamefont {J.~B.}\ \bibnamefont {Descarrega}},
  \bibinfo {author} {\bibfnamefont {W.}~\bibnamefont {Vollenberg}}, \bibinfo
  {author} {\bibfnamefont {A.~J.~G.}\ \bibnamefont {Lunt}}, \bibinfo {author}
  {\bibfnamefont {F.}~\bibnamefont {Leaux}}, \bibinfo {author} {\bibfnamefont
  {S.}~\bibnamefont {Calatroni}}, \bibinfo {author} {\bibfnamefont
  {W.}~\bibnamefont {Venturini-Delsolaro}},\ and\ \bibinfo {author}
  {\bibfnamefont {M.}~\bibnamefont {Taborelli}},\ }\bibfield  {title} {\bibinfo
  {title} {Development of sputtered nb$_3$sn films on copper substrates for
  superconducting radiofrequency applications},\ }\href
  {https://doi.org/10.1088/1361-6668/aaf61f} {\bibfield  {journal} {\bibinfo
  {journal} {Superconductor Science and Technology}\ }\textbf {\bibinfo
  {volume} {32}},\ \bibinfo {pages} {035002} (\bibinfo {year}
  {2019})}\BibitemShut {NoStop}%
\bibitem [{\citenamefont {Sun}\ \emph {et~al.}(2022)\citenamefont {Sun},
  \citenamefont {Arias}, \citenamefont {Baraissov}, \citenamefont {Dobson},
  \citenamefont {Gaitan}, \citenamefont {Ge}, \citenamefont {Howard},
  \citenamefont {Kelley}, \citenamefont {Liepe}, \citenamefont {Muller},
  \citenamefont {Oseroff}, \citenamefont {Porter}, \citenamefont {Sethna},\
  and\ \citenamefont {Sitaraman}}]{sun:srf2021-weotev03}%
  \BibitemOpen
  \bibfield  {author} {\bibinfo {author} {\bibfnamefont {Z.}~\bibnamefont
  {Sun}}, \bibinfo {author} {\bibfnamefont {T.}~\bibnamefont {Arias}}, \bibinfo
  {author} {\bibfnamefont {Z.}~\bibnamefont {Baraissov}}, \bibinfo {author}
  {\bibfnamefont {K.}~\bibnamefont {Dobson}}, \bibinfo {author} {\bibfnamefont
  {G.}~\bibnamefont {Gaitan}}, \bibinfo {author} {\bibfnamefont
  {M.}~\bibnamefont {Ge}}, \bibinfo {author} {\bibfnamefont {K.}~\bibnamefont
  {Howard}}, \bibinfo {author} {\bibfnamefont {M.}~\bibnamefont {Kelley}},
  \bibinfo {author} {\bibfnamefont {M.}~\bibnamefont {Liepe}}, \bibinfo
  {author} {\bibfnamefont {D.}~\bibnamefont {Muller}}, \bibinfo {author}
  {\bibfnamefont {T.}~\bibnamefont {Oseroff}}, \bibinfo {author} {\bibfnamefont
  {R.}~\bibnamefont {Porter}}, \bibinfo {author} {\bibfnamefont
  {J.}~\bibnamefont {Sethna}},\ and\ \bibinfo {author} {\bibfnamefont
  {N.}~\bibnamefont {Sitaraman}},\ }\bibfield  {title} {\bibinfo {title}
  {{Toward Stoichiometric and Low-Surface-Roughness Nb$_3$Sn Thin Films via
  Direct Electrochemical Deposition}},\ }in\ \href
  {https://doi.org/10.18429/JACoW-SRF2021-WEOTEV03} {\emph {\bibinfo
  {booktitle} {Proc. SRF'21}}},\ \bibinfo {series and number} {\bibinfo
  {series} {International Conference on RF Superconductivity}\ No.~\bibinfo
  {number} {20}}\ (\bibinfo  {publisher} {JACoW Publishing, Geneva,
  Switzerland},\ \bibinfo {year} {2022})\ pp.\ \bibinfo {pages}
  {710--713}\BibitemShut {NoStop}%
\bibitem [{\citenamefont {Ge}\ \emph {et~al.}(2019)\citenamefont {Ge},
  \citenamefont {Arrieta}, \citenamefont {Gruber}, \citenamefont {Kaufman},
  \citenamefont {Liepe}, \citenamefont {Maniscalco}, \citenamefont {McNeal},
  \citenamefont {Oseroff}, \citenamefont {Porter},\ and\ \citenamefont
  {Sun}}]{ge:srf2019-tufub8}%
  \BibitemOpen
  \bibfield  {author} {\bibinfo {author} {\bibfnamefont {M.}~\bibnamefont
  {Ge}}, \bibinfo {author} {\bibfnamefont {V.}~\bibnamefont {Arrieta}},
  \bibinfo {author} {\bibfnamefont {T.}~\bibnamefont {Gruber}}, \bibinfo
  {author} {\bibfnamefont {J.}~\bibnamefont {Kaufman}}, \bibinfo {author}
  {\bibfnamefont {M.}~\bibnamefont {Liepe}}, \bibinfo {author} {\bibfnamefont
  {J.}~\bibnamefont {Maniscalco}}, \bibinfo {author} {\bibfnamefont
  {S.}~\bibnamefont {McNeal}}, \bibinfo {author} {\bibfnamefont
  {T.}~\bibnamefont {Oseroff}}, \bibinfo {author} {\bibfnamefont
  {R.}~\bibnamefont {Porter}},\ and\ \bibinfo {author} {\bibfnamefont
  {Z.}~\bibnamefont {Sun}},\ }\bibfield  {title} {\bibinfo {title} {{CVD Coated
  Copper Substrate SRF Cavity Research at Cornell University}},\ }in\ \href
  {https://doi.org/10.18429/JACoW-SRF2019-TUFUB8} {\emph {\bibinfo {booktitle}
  {Proc. SRF'19}}},\ \bibinfo {series and number} {\bibinfo {series}
  {International Conference on RF Superconductivity}\ No.~\bibinfo {number}
  {19}}\ (\bibinfo  {publisher} {JACoW Publishing, Geneva, Switzerland},\
  \bibinfo {year} {2019})\ pp.\ \bibinfo {pages} {381--386},\ \bibinfo {note}
  {https://doi.org/10.18429/JACoW-SRF2019-TUFUB8}\BibitemShut {NoStop}%
\end{thebibliography}%

\end{document}